\documentclass[10pt,twoside]{article}
\usepackage[english,russianb]{babel}
\usepackage{droid}
\usepackage{famems}
\usepackage{cmap}
\usepackage{graphics}
\usepackage{graphicx}
\usepackage{graphpap}
\usepackage{amsmath,amstext,amsfonts,amssymb,amsthm}
\usepackage{mathtools}
\usepackage{braket}
\usepackage{dsfont}
\usepackage[mathscr]{eucal}
\usepackage{array}
\usepackage{multicol}
\usepackage{multirow}
\usepackage{multicol}
\usepackage{tabularx}
\usepackage{xcolor}
\usepackage{color}
\usepackage{colortbl}
\usepackage{bibunits}
\usepackage{ifthen}
\usepackage{epstopdf}
\usepackage[normalem]{ulem}
\usepackage{rotating}
\usepackage{hhline}
\usepackage{pdflscape}
\usepackage{fontawesome}
\usepackage{scrextend}
\usepackage{arydshln}
\usepackage{tikz}
\usepackage{ctable}
\usepackage{pict2e}
\usepackage{setspace}
\usepackage{url}
\usepackage[breaklinks]{hyperref}
\usepackage{hyperref}
\hypersetup{
    colorlinks,
    citecolor=black,
    filecolor=black,
    linkcolor=black,
    urlcolor=
teal
}

\definecolor{MyAqua0}{RGB}{0,255,255}
\definecolor{MyGreen}{rgb}{0,1,0}
\definecolor{MyOrangeBayes}{rgb}{1,0.24,0.24}
\definecolor{MyYellow50}{RGB}{255,255,128}
\definecolor{MyYellow0}{RGB}{255,255,0}
\definecolor{MyYellow75}{RGB}{255,255,192}
\definecolor{MyYellow85}{RGB}{255,255,240}
\definecolor{MyAqua50}{RGB}{127,255,255}
\definecolor{MyAqua75}{RGB}{192,255,255}
\definecolor{MyAqua85}{RGB}{224,255,255}
\definecolor{MyYellow}{rgb}{1,1,0}
\definecolor{MyYellow75}{rgb}{1,1,0.75}
\definecolor{MyRed95}{rgb}{1,0.95,0.95}
\definecolor{MyRed85}{rgb}{1,0.85,0.85}
\definecolor{MyRed75}{rgb}{1,0.75,0.75}
\definecolor{MyRed50}{rgb}{1,0.50,0.50}
\definecolor{MyRed25}{rgb}{1,0.25,0.25}
\definecolor{MyRed0}{rgb}{1,0,0}
\definecolor{MyRed}{rgb}{1,0,0}
\definecolor{MyBlue}{rgb}{0,0,1}
\definecolor{MyGreen0}{RGB}{0,255,0}
\definecolor{MyGreenD}{RGB}{0,160,0}
\definecolor{MyGreen20}{rgb}{0.20,1,0.20}
\definecolor{MyGreen25}{rgb}{0.25,1,0.25}
\definecolor{MyGreen35}{rgb}{0.35,1,0.35}
\definecolor{MyGreen50}{RGB}{128,255,128}
\definecolor{MyGreen60}{rgb}{0.60,1,0.60}
\definecolor{MyGreen70}{rgb}{0.70,1,0.70}
\definecolor{MyGreen75}{rgb}{0.75,1,0.75}
\definecolor{MyBlue40}{rgb}{0.40,0.40,1}
\definecolor{MyBlue60}{rgb}{0.60,0.60,1}
\definecolor{MyBlue75}{rgb}{0.75,0.75,1}
\definecolor{MyBlue85}{rgb}{0.85,0.85,1}
\definecolor{MyGray}{rgb}{0.95,0.95,0.95}
\definecolor{MyWhite}{rgb}{1,1,1}
\definecolor{MyBlack}{rgb}{0,0,0}
\definecolor{MyProbability}{rgb}{1.00,0.25,0.25}
\definecolor{MyConfidency}{rgb}{1.00,0.25,1.00}
\definecolor{MyEventology}{rgb}{0.25,1.00,1.00}
\definecolor{MyOrange}{rgb}{1,0.56,0.25}
\definecolor{MyOrangeB}{rgb}{1,0.65,0.45}
\definecolor{MyMagentaP}{rgb}{1.00,0.45,0.65}
\definecolor{MyMagenta}{rgb}{1.00,0,1.00}
\definecolor{MyMagenta0}{rgb}{1.00,0.75,1.00}
\definecolor{MyAqua}{rgb}{0,0.75,0.75}

\makeatletter
\renewcommand{\paragraph}{\@startsection{paragraph}{4}{0ex}%
   {-3.25ex plus -1ex minus -0.2ex}%
   {1.5ex plus 0.2ex}%
   {\normalfont\normalsize\tt}}
\makeatother

\begin{document}
\newcounter{ctrwar}\setcounter{ctrwar}{0} 
\newcounter{ctrdef}\setcounter{ctrdef}{0}
\newcounter{ctrdefpre}\setcounter{ctrdefpre}{0}
\newcounter{ctrTh}\setcounter{ctrTh}{0}
\newcounter{ctrnot}\setcounter{ctrnot}{0}
\newcounter{ctrATT}\setcounter{ctrATT}{0} 
\newcounter{ctrcor}\setcounter{ctrcor}{0}
\newcounter{ctrAx}\setcounter{ctrAx}{0}
\newcounter{ctrexa}\setcounter{ctrexa}{0}
\newcounter{ctrlem}\setcounter{ctrlem}{0}
\newcounter{ctrPRO}\setcounter{ctrPRO}{0}
\newcounter{ctrrem}\setcounter{ctrrem}{0}
\newcounter{ctrass}\setcounter{ctrass}{0}
\newcounter{ctrmem}\setcounter{ctrmem}{0}

\newcommand{\bfPB}{\mbox{\protect\reflectbox{\bf P}\hspace{-0.4em}{\bf B}}}
\newcommand{\bfEl}{\mbox{\protect\reflectbox{$\mathbf E$}}}
\newcommand{\bfEr}{\mathbf{E}}
\newcommand{\bfEE}{\mbox{\protect\reflectbox{$\mathbf E$}\hspace{-0.5em}{$\mathbf E$}}}
\newcommand{\bfphi}{\mbox{\boldmath$\varphi$}}
\newcommand{\bfPhi}{\mbox{\boldmath$\Phi$}}
\newcommand{\bfPhii}{\mbox{\scriptsize\boldmath$\Phi$}}
\newcommand{\rS}{\reflectbox{\bf S}}
\newcommand{\rsS}{\,\reflectbox{\scriptsize\bf S}}
\newcommand{\rssS}{\,\reflectbox{\tiny\bf S}}
\newcommand{\reS}{\reflectbox{S}}
\newcommand{\resS}{\,\reflectbox{\scriptsize S}}
\newcommand{\ressS}{\,\reflectbox{\tiny S}}
\newcommand{\scrER}{\mathscr E\!\mathscr R}
\newcommand{\ER}{\mathscr E\!\mathscr R}

\numberwithin{equation}{section}

\renewcommand{\figurename}{\scriptsize  Figure}
\renewcommand{\tablename}{\scriptsize  Table}
\renewcommand\refname{\scriptsize References}
\renewcommand\contentsname{\scriptsize Contents}
\setcounter{page}{92}

\titleOneAuthorenOne{
The logic of uncertainty as a logic of experience and chance\\[-2pt] and the co$\sim$event-based Bayes' theorem

}{Oleg Yu. Vorobyev}{
Institute of mathematics and computer science\\
Siberian Federal University \\
Krasnoyarsk \\
\tiny
\url{mailto:oleg.yu.vorobyev@gmail.com}\\
\url{http://www.sfu-kras.academia.edu/OlegVorobyev}\\
\url{http://olegvorobyev.academia.edu}}

\label{vorobyev10en}

\colontitleen{Oleg Yu Vorobyev. The logic of uncertainty as a logic of experience and chance}

\footavten{O.Yu.Vorobyev}

\setcounter{footnote}{0} \setcounter{equation}{0} \setcounter{figure}{0} \setcounter{table}{0} \setcounter{section}{0}

\vspace{-22pt}

\begin{abstracten}
\emph{The logic of uncertainty is not the logic of experience and as well as it is not the logic of chance. It is the logic of experience and chance.
Experience and chance are two inseparable poles. These are two dual reflections of one essence, which is called co$\sim$event. The theory of
experience and chance is the theory of co$\sim$events. To study the co$\sim$events, it is not enough to study the experience and to study the chance.
For this, it is necessary to study the experience and chance as a single entire, a co$\sim$event. In other words, it is necessary to study their
interaction within a co$\sim$event. The new co$\sim$event axiomatics and the theory of co$\sim$events following from it were created precisely for
these purposes. In this work, I am going to demonstrate the effectiveness of the new theory of co$\sim$events in a studying the logic of uncertainty.
I will do this by the example of a co$\sim$event splitting of the logic of the Bayesian scheme, which has a long history of fierce debates between
Bayesianists and frequentists. I hope the logic of the theory of experience and chance will make its modest contribution to the application of these
old dual debaters.}
\end{abstracten}\\[5pt]
\begin{keywordsen}
\emph{Eventology, event, probability, probability theory, Kolmogorov's axiomatics, experience, chance, cause, consequence, co$\sim$event, set of
co$\sim$events, bra-event, set of bra-events, ket-event, set of ket-events, believability, certainty, believability theory, certainty theory, theory
of co$\sim$events, theory of experience and chance, co$\sim$event dualism, co$\sim$event axiomatics, logic of uncertainty, logic of experience and
chance, logic of cause and consequence, logic of the past and the future, Bayesian scheme.}
\end{keywordsen}\\[5pt]
\begin{MSCen}
60A05, 60A10, 60A86, 62A01, 62A86, 62H10, 62H11, 62H12, 68T01, 68T27, 81P05, 81P10, 91B08, 91B10, 91B12, 91B14, 91B30, 91B42, 91B80, 93B07, 94D05
\end{MSCen}

\hfill {\scriptsize\em The statements of science are not of what is true and what is not true,}\\[-22pt]

\hfill {\scriptsize\em but statements of what is known with different degrees of \emph{certainty}.}\\[-20pt]

\hfill {\scriptsize Richard Feynman.}

\vspace{-0.5cm}

\section{``Collider'' splitting the classical Bayes' scheme by the theory of co$\sim$events}

The axiomatics of the theory of experience and chance  \cite{Vorobyev2016famems2} is fundamentally new with the new semantics\footnote{I consider the
new co$\sim$event-based semantic is very important for understanding this paper. For this reason, I'm going to highlight and mark especially the most
key places in the text with new semantics for some ``become familiar'' terms of the classical probability theory and statistics.} of terms and
statements. This axiomatics is not too complicated. However, any attempts to apply the new theory in various fields, which are a long-lived fiefdom
of probability theory and mathematical statistics, face difficulties in translating from a probabilistic language into the dual language of the new
theory. This translation reminds me of an investigation of the splitting of previously unsplittable notions of \emph{probability} and \emph{event} by
means of a \emph{co$\sim$event-based ``collider''} that is provided with sufficient power of the new theory to split what was previously unsplittable
and to get instead of one probability two measures: \emph{probability} and \emph{believability}, and instead of one event two: a \emph{ket-event} and
a \emph{bra-event}. Here I intend to split the logic of uncertainty by the example of splitting the logic and the interpretation of Bayes' scheme by
this \emph{co$\sim$event-based ``collider''}, i.e., by the co$\sim$event-based axiomatics of the new theory. In order to present in all details this
process of co$\sim$event-based splitting of the previously unsplittable logic of uncertainty in Bayes' scheme, I will first present a simpler, crude
and naive procedure for the co$\sim$event-based splitting, and then I will go into its refinement. I will carry out this co$\sim$event-based
splitting of the Bayesian scheme in two stages. In the first stage\footnote{See details in Tables \ref{dual_analogies},
\ref{dual_analogies_abbreviations}, and \ref{dual_analogies_ostensibly}.}, I am going to first compare three formally possible splitting methods to
choose the most suitable of them. In the second stage (see paragraph \ref{refinement}), I'm going to continue the co$\sim$event-based splitting.
However, now I will split the chosen splitting method itself in order to construct the final formula, which I would call the
\textbf{co$\sim$event-based Bayes' theorem}.

\clearpage
\begin{table}[h!]
\vspace*{-1cm}
  \begin{center}
    \begin{tabular}{p{8cm}|p{8cm}}
      \multicolumn{2}{p{16cm}}{\begin{center}\bf Certainty theory\end{center}}\\[-0.75cm]
      \multicolumn{2}{p{16cm}}{ } \\ \hline
      &\\[-0.75cm]
      \begin{center}\bf Probability theory\end{center}
      &
      \begin{center}\bf Believability theory\end{center}
      \\[-0.45cm] \hline
      \multicolumn{2}{p{16cm}}{ } \\[-0.45cm]
      \multicolumn{2}{p{16cm}}{\centering\scriptsize\bf  In pondering an observable chance feature of Nature, and\\ an accumulated own experience, a scientist after all}
      \\  \hline
      \scriptsize of the random experiment observes the chance as a limited array of \textbf{random observations}.
      &
      \scriptsize of the experienced experiment accumulates his/her observer' experience as a limit array of \textbf{accumulated hypotheses}.
      \\ \hline
      \multicolumn{2}{p{16cm}}{ } \\[-0.45cm]
      \multicolumn{2}{p{16cm}}{\centering\scriptsize\bf Suppose we list them and denote them by the symbols}\\  \hline
      & \\[-0.85cm]
      \scriptsize
\begin{eqnarray}\label{observations}
      &&\ket{X_1};\ket{X_2}; \ldots;\ket{X_N}
\end{eqnarray}
      where $\ket{X_j}$ for $X_j\in\rS^{\frak X_{\!\mathscr R}}$ is an abbreviation for the ket-event $\ket{\textsf{ter}(X_j/\!\!/\frak X_{\!\mathscr R})}\subseteq \ket{\Omega}$; &
      \scriptsize
\begin{eqnarray}\label{hypotheses}
      &&\bra{x_1};\bra{x_2}; \ldots;\bra{x_M}
\end{eqnarray}
      where $\bra{x_i}\subseteq \bra{\Omega}$ is a bra-event for $x_i\in\frak X_{\!\mathscr R}$.
      \\ \hline
      \multicolumn{2}{p{16cm}}{ } \\[-0.45cm]
      \multicolumn{2}{p{16cm}}{\centering\scriptsize\bf Presumably, each of these} \\  \hline
      \scriptsize \textbf{random observations} has some degree of \textbf{probability} that it will happen, or else it would not be in the listing.
      &
      \scriptsize \textbf{experienced hypotheses} has some degree of \textbf{believability}\footnote{In the theory of experience and chance the new strict mathematical term ``believability'' is a neologism VS the old polysemantic terms: ``credibility'', ``confidence'', ``likelihood'' and so on.} that it was experienced by the observer, or else it would not have been in the listing.
      \\ \hline
      \multicolumn{2}{p{16cm}}{ } \\[-0.45cm]
      \multicolumn{2}{p{16cm}}{\centering\scriptsize\bf Since the list is meant to be exhaustive of} \\  \hline
      \scriptsize
 the chances that the scientist expected, a total \textbf{probability} of all chances of the experienced-random experiment would be distributed among these $N$ \textbf{random observations}. Norming the total \textbf{probability} to a unit of 1, relative \textbf{probabilities} of these random observations are expressed by the \textbf{probabilities}
\begin{eqnarray}\label{priorP}
&&\mathbf P(\ket{X_1}); \mathbf P(\ket{X_2}); \ldots; \mathbf P(\ket{X_N});
\end{eqnarray}
which sum to 1 where $\mathbf P(\ket{X_j})=\bfPhi(\braket{\Omega|X_j}), j=1,\ldots,N$.
So they are expressed by the \textbf{certainties}
\begin{eqnarray}\label{priorPhiP}
&&\bfPhi(\braket{\Omega|X_1}); \bfPhi(\braket{\Omega|X_2}); \ldots; \bfPhi(\braket{\Omega|X_N});
\end{eqnarray}      &
      \scriptsize
 the hypotheses that the scientist experienced, a total \textbf{believability} of all hypotheses in the experienced-random experiment would be distributed among these $M$ \textbf{experienced hypotheses}. Norming the total \textbf{believability} to a unit of 1, relative \textbf{believabilities} of these experienced hypotheses are expressed by the \textbf{believabilities}
\begin{eqnarray}\label{priorB}
&&\mathbf B(\bra{x_1}); \mathbf B(\bra{x_2}); \ldots; \mathbf B(\bra{x_M});
\end{eqnarray}
which sum to 1 where $\mathbf B(\bra{x_i})=\bfPhi(\braket{x_i|\Omega}), i=1,\ldots,M$.
So they are expressed by the \textbf{certainties}
\begin{eqnarray}\label{priorPhiB}
&&\bfPhi(\braket{x_1|\Omega}); \bfPhi(\braket{x_2|\Omega}); \ldots; \bfPhi(\braket{x_M|\Omega});
\end{eqnarray}
      \\ \hline
      \multicolumn{2}{p{16cm}}{ } \\[-0.45cm]
      \multicolumn{2}{p{16cm}}{\centering\scriptsize\bf Empirical science is oriented towards collecting new experimental or historical observations that can inform us about} \\  \hline
      \scriptsize
 the chances of these \textbf{random observations} (\ref{observations}). Suppose we denote the numerical data that summarises the result of an experienced random experiment, i.e. some co$\sim$event, by a letter $\mathscr R$. Co$\sim$event-based Bayes' scheme provides a logical mathematical framework for calibrating how the co$\sim$event $\mathscr R$ should be used to inform us about a new understanding of the relative \textbf{certainties} given this new data. Technically, it yields numerical values for the conditional \textbf{certainties}
\begin{equation}\label{posteriorPhiP}
\bfPhi(\braket{\Omega|X_1}|\mathscr R); \bfPhi(\braket{\Omega|X_2}|\mathscr R); \ldots; \bfPhi(\braket{\Omega|X_N}|\mathscr R);
\end{equation}
that are based on their initial information sources augmented by the new co$\sim$event, $\mathscr R$. The numerical differences between the \textbf{certainties} (\ref{priorPhiP}) and the conditional \textbf{certainties} (\ref{posteriorPhiP}) display what one can rightfully learn about our assessment of this co$\sim$event. This assessment, in the long and the short of statistical theory, is the goal of scientific experimentation and data analysis. Because these distributions can represent uncertain observable features of Nature before and after co$\sim$event, $\mathscr R$, the former is commonly called a ``prior'' \textbf{certainties} (\ref{priorPhiP}), while the latter is called the ``posterior'' \textbf{certainties}
(\ref{posteriorPhiP}) for the \textbf{random observations} (\ref{observations}).
      &
      \scriptsize
 the experience of these \textbf{accumulated hypotheses} (\ref{hypotheses}). Suppose we denote the numerical data that summarises the result of an experienced random experiment, i.e. some co$\sim$event, by a letter $\mathscr R$. Co$\sim$event-based Bayes' scheme provides a logical mathematical framework for calibrating how the co$\sim$event $\mathscr R$ should be used to inform us about a new understanding of the relative \textbf{certainties} given this new data. Technically, it yields numerical values for the conditional \textbf{certainties}
\begin{equation}\label{posteriorPhiB}
\bfPhi(\braket{x_1|\Omega}|\mathscr R); \bfPhi(\braket{x_2|\Omega}|\mathscr R); \ldots; \bfPhi(\braket{x_M|\Omega}|\mathscr R);
\end{equation}
that are based on their initial information sources augmented by the new co$\sim$event, $\mathscr R$. The numerical differences between the \textbf{certainties} (\ref{priorPhiB}) and the conditional \textbf{certainties} (\ref{posteriorPhiB}) display what one can rightfully learn about our assessment of the \textbf{accumulated hypotheses} from this co$\sim$event. This assessment, in the long and the short of statistical theory, is the goal of scientific experimentation and data analysis. Because these distributions can represent uncertain opinions before and after co$\sim$event, $\mathscr R$, the former is commonly called a ``prior'' \textbf{certainties} (\ref{priorPhiB}), while the latter is called the ``posterior'' \textbf{certainties} (\ref{posteriorPhiB}) for the \textbf{accumulated hypotheses} (\ref{hypotheses}).
      \\ \hline
      \multicolumn{2}{p{16cm}}{ } \\[-0.45cm]
      \multicolumn{2}{p{16cm}}{\centering\scriptsize\bf The Bayes' scheme in the context of the theory of experience and chance\\ requires that we assess the certainty of the co$\sim$event $\mathscr R$ that occurs, given each of the relevant} \\  \hline
      \scriptsize
\textbf{random observations}. These assessments are represented by the numbers
\begin{equation}\label{posteriorPhiP2}
\bfPhi(\mathscr R|\braket{\Omega|X_1}); \bfPhi(\mathscr R|\braket{\Omega|X_2}); \ldots; \bfPhi(\mathscr R|\braket{\Omega|X_N})
\end{equation}
where $\bfPhi(\mathscr R|\braket{\Omega|X_j})=\mathbf B(\bra{X_j}), j=1,\ldots,N$.
The relative sizes of these numbers for the various \textbf{random observations} based on the same data are commonly referred to as the \textbf{certainties} for the \textbf{random observations}. Using these certainties for the observed data on the basis of each of the \textbf{random observations}, co$\sim$event-based Bayes' scheme allows the computation of each of the ``posterior'' \textbf{certainties} (\ref{posteriorPhiP}) according to the computational formula
\begin{equation}\label{BayesFormulaPhiP}
\bfPhi(\braket{\Omega|X_j}|\mathscr R) = \frac{\bfPhi(\mathscr R|\braket{\Omega|X_j})\bfPhi(\braket{\Omega|X_j})}
{\displaystyle\sum_{k=1}^N\bfPhi(\mathscr R|\braket{\Omega|X_k})\bfPhi(\braket{\Omega|X_k})}
\end{equation}
for each \textbf{random observation} $\ket{X_j}$ and ``prior'' \textbf{certainties} (\ref{priorPhiP})
where $\bfPhi(\braket{\Omega|X_j}|\mathscr R)=\mathbf B(\bra{X_j})$.
      &
      \scriptsize
\textbf{accumulated hypotheses}. These assessments are represented by the numbers
\begin{equation}\label{posteriorPhiB2}
\bfPhi(\mathscr R|\braket{x_1|\Omega}); \bfPhi(\mathscr R|\braket{x_2|\Omega}); \ldots; \bfPhi(\mathscr R|\braket{x_M|\Omega}).
\end{equation}
where $\bfPhi(\mathscr R|\braket{x_i|\Omega})=\mathbf P(\ket{x_i}), i=1,\ldots,M$.
The relative sizes of these numbers for the various \textbf{accumulated hypotheses} based on the same data are commonly referred to as the \textbf{certainties} for the \textbf{accumulated hypotheses}. Using these certainties for the observed data on the basis of each of the \textbf{accumulated hypotheses}, co$\sim$event-based Bayes' scheme allows the computation of each of the ``posterior'' \textbf{certainties} (\ref{posteriorPhiB}) according to the computational formula called the \textbf{co$\sim$event-based Bayes theorem}:
\begin{equation}\label{BayesFormulaPhiB}
\bfPhi(\braket{x_i|\Omega}|\mathscr R) = \frac{\bfPhi(\mathscr R|\braket{x_i|\Omega})\bfPhi(\braket{x_i|\Omega})}
{\displaystyle\sum_{k=1}^M\bfPhi(\mathscr R|\braket{x_k|\Omega})\bfPhi(\braket{x_k|\Omega})}
\end{equation}
for each \textbf{accumulated hypothesis} $\bra{x_i}$ and ``prior'' \textbf{certainties} (\ref{priorPhiB})
where $\bfPhi(\braket{x_i|\Omega}|\mathscr R)=\mathbf P(\ket{x_i})$.
      \\ \hline
    \end{tabular}
  \end{center}
  \caption{A logic of the dual co$\sim$event-based analogies of Bayes' scheme in the theory of experience and chance.\label{dual_analogies}}
\end{table}

\clearpage
\begin{table}[h!]
\vspace*{-0.5cm}
  \begin{center}
    \begin{tabular}{p{8cm}|p{8cm}}
      \multicolumn{2}{p{16cm}}{\begin{center}\bf Certainty bra-ket theory\end{center}}\\[-0.75cm]
      \multicolumn{2}{p{16cm}}{ } \\ \hline
      &\\[-0.75cm]
      \begin{center}\bf Probability ket-theory\end{center}
      &
      \begin{center}\bf Believability bra-theory\end{center}
      \\[-0.45cm] \hline
      \multicolumn{2}{p{16cm}}{ } \\[-0.45cm]
      \multicolumn{2}{p{16cm}}{\centering\scriptsize\bf  Rewrite co$\sim$event-based Bayes' ket- and bra-formulas (\ref{BayesFormulaPhiP}, \ref{BayesFormulaPhiB})\protect\\ by applying the following obvious equalities and useful abbreviations}
      \\  \hline
      &\\[-0.75cm]
      \scriptsize
\begin{eqnarray}\label{ket-abbreviations}
      \mbox{prior:}& &\bfPhi(\mathscr R|\braket{\Omega|X_{\!j}})=b(X_{\!j}),\\
      \mbox{prior:}& &\bfPhi(\braket{\Omega|X_{\!j}})=p(X_{\!j}),\\
      \mbox{posterior:}& &\bfPhi(\braket{\Omega|X_{\!j}}|\mathscr R)=p^{\mbox{\tiny post}}(X_{\!j}).
\end{eqnarray}
      &
      \scriptsize
\begin{eqnarray}\label{ket-abbreviations}
      \mbox{prior:}& &\bfPhi(\mathscr R|\braket{x_{\!i}|\Omega})=p_{x_{\!i}},\\
      \mbox{prior:}& &\bfPhi(\braket{x_{\!i}|\Omega})=b_{x_{\!i}},\\
      \mbox{posterior:}& &\bfPhi(\braket{x_{\!i}|\Omega}|\mathscr R)=b^{\mbox{\tiny post}}_{x_{\!i}}.
\end{eqnarray}
\\[-0.45cm] \hline
      &\\[-1.21cm]
      \begin{minipage}{8.0cm}
      \vspace{0.80cm}
      \colorbox{MyRed0}
      {\scriptsize\bf  \hspace*{0.26cm}Co$\sim$event-based Bayes' ket-formula that follows from (\ref{BayesFormulaPhiP})\hspace*{0.27cm}}
      \end{minipage}
      &
      \begin{minipage}{8.0cm}
      \vspace{0.80cm}
      \colorbox{MyAqua0}
      {\scriptsize\bf  \hspace*{0.48cm}Co$\sim$event-based Bayes' theorem that follows from (\ref{BayesFormulaPhiB})\hspace*{0.48cm}}
      \end{minipage}
      \\[-0.05cm] \hline
      &\\[-0.75cm]
      \vspace{0.04cm}
      \colorbox{MyRed0}
      {
      \begin{minipage}{7.75cm}
      \scriptsize
\begin{eqnarray}\label{ket-abbreviations}
\begin{cases}
p^{\mbox{\tiny post}}(X_{\!j})&\displaystyle =p(X_j)\frac{b(X_j)}{\bfPhi(\mathscr R)},\\
&\\
p^{\mbox{\tiny post}}_{x_{\!i}}&\displaystyle =\sum_{x_{\!i}\in X_{\!j}}p^{\mbox{\tiny post}}(X_{\!j}).
\end{cases}
\end{eqnarray}
      \end{minipage}
      }
      &
      \vspace{0.08cm}
      \colorbox{MyAqua0}
      {
      \begin{minipage}{7.75cm}
      \scriptsize
\begin{eqnarray}\label{bra-abbreviations}
\begin{cases}
b^{\mbox{\tiny post}}_{x_{\!i}}&\displaystyle =b_{x_{\!i}}\frac{p_{x_{\!i}}}{\bfPhi(\mathscr R)},\\
&\\
b^{\mbox{\tiny post}}(X_{\!j})&\displaystyle =\sum_{x_{\!i}\in X_{\!j}}b^{\mbox{\tiny post}}_{x_{\!i}}.
\end{cases}
\end{eqnarray}
      \end{minipage}
      }
      \\[-0.0cm]  \hline
    \end{tabular}
  \end{center}
  \caption{A logic of the dual co$\sim$event-based analogies of Bayes' scheme in the theory of experience and chance in abbreviations.\label{dual_analogies_abbreviations}}
\end{table}

\begin{table}[h!]
  \begin{center}
    \begin{tabular}{p{8.5cm}|p{8.5cm}}
      \multicolumn{2}{p{17cm}}{\begin{center}\bf Certainty bra-ket theory\end{center}}\\[-0.75cm]
      \multicolumn{2}{p{17cm}}{ }\\[-0.15cm] \hline
      \multicolumn{2}{p{17cm}}{ }\\[-0.45cm]
      \multicolumn{2}{p{17cm}}{\centering\scriptsize\bf In pondering a chance of Nature, and an accumulated own experience, a scientist after all}\\
      \multicolumn{2}{p{17cm}}{ }\\[-0.45cm] \hline
      \multicolumn{2}{p{17cm}}{ }\\[-0.45cm]
            \multicolumn{2}{p{17cm}}{\scriptsize
      of the experienced-random experiment observes the chance and accumulates his/her observer' experience
      as a limited array of ``\textbf{accumulated hypotheses}'' $\times$ ``\textbf{random observations}''.
   }\\ \hline

    \multicolumn{2}{p{17cm}}{ }\\[-0.45cm]
      \multicolumn{2}{p{17cm}}{\centering\scriptsize\bf Suppose we list them and denote them by the symbols}\\
      \multicolumn{2}{p{17cm}}{ }\\[-0.45cm] \hline
      \multicolumn{2}{p{17cm}}{ }\\[-0.45cm]
            \multicolumn{2}{p{17cm}}{\scriptsize
      $\braket{x_i|X_j}$, an abbreviation for the elementary co$\sim$event $\braket{x_i|\textsf{ter}(X_j/\!\!/\frak X_{\!\mathscr R})}\subseteq \braket{\Omega|\Omega}$, $i=1,\ldots,M, j=1,\ldots,N$, for $x_i\in\frak X_{\!\mathscr R}$, $X_j\in\rS^{\frak X_{\!\mathscr R}}$;}\\ \hline

      \multicolumn{2}{p{17cm}}{ }\\[-0.45cm]
      \multicolumn{2}{p{17cm}}{\centering\scriptsize\bf
      Since the list is meant to be exhaustive of
      }\\
      \multicolumn{2}{p{17cm}}{ }\\[-0.45cm] \hline
      \multicolumn{2}{p{17cm}}{ }\\[-0.45cm]
      \multicolumn{2}{p{17cm}}{\scriptsize
   the elementary co$\sim$events that the scientist observed and accumulated, a total certainty of all elementary co$\sim$events of the experienced-random experiment would be distributed among these $M\times N$ elementary co$\sim$events $\braket{x_i|X_j}$. Norming the total certainty to a unit of 1, relative certainties of these elementary co$\sim$events are expressed by the certainties
$\varphi_{x_{\!i}}(X_{\!j})=\bfPhi(\braket{x_i|X_j}), i=1,\ldots,M, j=1,\ldots,N$.
      }\\ \hline

      \multicolumn{2}{p{17cm}}{ }\\[-0.45cm]
      \multicolumn{2}{p{17cm}}{\centering\scriptsize\bf
Empirical science is oriented towards collecting new experimental or historical co$\sim$events that can inform us about
      }\\
      \multicolumn{2}{p{17cm}}{ }\\[-0.45cm] \hline
      \multicolumn{2}{p{17cm}}{ }\\[-0.45cm]
      \multicolumn{2}{p{17cm}}{\scriptsize
 these co$\sim$events. Suppose we denote the numerical data that summarises the result of an experienced-random experiment, i.e. some co$\sim$event, by a letter $\mathscr R$. Co$\sim$event-based Bayes' scheme provides a logical mathematical framework for calibrating how the co$\sim$event $\mathscr R$ should be used to inform us about a new understanding of the relative certainties given this new data. Technically, it yields numerical values for the conditional certainties
$\bfPhi(\braket{x_i|X_j}|\mathscr R)$ that are based on their initial information sources augmented by the new co$\sim$event, $\mathscr R$. The
numerical differences between the initial certainties and the conditional certainties display what one can rightfully learn about our assessment of
this co$\sim$event. This assessment, in the long and the short of statistical theory, is the goal of scientific experienced-random experimentation
and data analysis. Because these distributions can represent the uncertainty of Nature before and after co$\sim$event, $\mathscr R$, the former is
commonly called a ``prior'' certainties, while the latter is called the ``posterior'' certainties of the elementary co$\sim$events.
      }\\ \hline

      \multicolumn{2}{p{17cm}}{ }\\[-0.45cm]
      \multicolumn{2}{p{17cm}}{\centering\scriptsize\bf
  The Bayes' scheme in the context of the theory of experience and chance\\
  requires that we assess the certainty of the co$\sim$event $\mathscr R$
  that occurs, given each of the relevant
      }\\
      \multicolumn{2}{p{17cm}}{ }\\[-0.45cm] \hline
      \multicolumn{2}{p{17cm}}{ }\\[-0.45cm]
      \multicolumn{2}{p{17cm}}{\scriptsize
elementary co$\sim$events. These assessments are represented by the numbers $\bfPhi(\mathscr R|\braket{x_i|X_j})$. The relative sizes of these
numbers for the various elementary co$\sim$events based on the same data are commonly referred to as the certainties for the elementary
co$\sim$events. Using these certainties for the observed data on the basis of each of the co$\sim$events, co$\sim$event-based Bayes' scheme allows
the computation of each of the ``posterior'' certainties according to the computational formula
\begin{equation}\label{BayesFormulaPhi}
\bfPhi(\braket{x_{\!i}|X_{\!j}}|\mathscr R) = \frac{\bfPhi(\mathscr R|\braket{x_{\!i}|X_{\!j}})\bfPhi(\braket{x_{\!i}|X_{\!j}})}
{\bfPhi(\mathscr R)}
\end{equation}
for each elementary co$\sim$event $\braket{x_i|X_j}$ and their ``prior'' certainties $\varphi_{x_{\!i}}(X_{\!j})$ where
\begin{eqnarray}
\mbox{prior certainty:} &\bfPhi(\braket{x_{\!i}|X_{\!j}})&=\varphi_{x_{\!i}}(X_{\!j})=b_{x_{\!i}} p(X_{\!j}),\\
\mbox{posterior certainty:} &\bfPhi(\braket{x_{\!i}|X_{\!j}}|\mathscr R)&=\varphi^{\mbox{\tiny post}}_{x_{\!i}}(X_{\!j})=[b_{x_{\!i}} p(X_{\!j})]^{\mbox{\tiny post}},\\
\nonumber&&\\
&\bfPhi(\mathscr R|\braket{x_{\!i}|X_{\!j}})=\mathbf 1_{X_{\!j}}(x_{\!i}) &=
\begin{cases}
1, & x_{\!i}\in X_{\!j},\\
0, & \mbox{otherwise}.
\end{cases}
\end{eqnarray}
      }\\[-0.35cm] \hline

      \multicolumn{2}{p{17cm}}{ }\\[-1.21cm]
      \multicolumn{2}{p{17cm}}{
      \begin{minipage}{17.0cm}
      \centering
      \vspace{0.80cm}
      \colorbox{MyRed0}
      {\scriptsize\bf  \hspace*{2.50cm}Hence united co$\sim$event-based Bayes' bra-ket formula (\ref{BayesFormulaPhi}) can be rewritten in the short form:\hspace*{2.70cm}}
      \end{minipage}
      }\\
      \multicolumn{2}{p{17cm}}{ }\\[-0.45cm] \hline
      \multicolumn{2}{p{17cm}}{ }\\[-0.85cm]
      \multicolumn{2}{p{17cm}}{
      \vspace{0.15cm}
      \colorbox{MyRed0}
      {
      \begin{minipage}{16.75cm}
      \scriptsize
\begin{eqnarray}
\varphi^{\mbox{\tiny post}}_{x_{\!i}}(X_{\!j})=
\begin{cases}
\displaystyle
\frac{\varphi_{x_{\!i}}(X_{\!j})}{\bfPhi(\mathscr R)}, & x_{\!i}\in X_{\!j},\\
&\\
0                                                  , & \mbox{otherwise}.
\end{cases}
\end{eqnarray}
      \end{minipage}
      }
}
\\[-0.0cm] \hline
    \end{tabular}
  \end{center}
  \caption{A logic of the united co$\sim$event-based analog of Bayes' scheme in the theory of experience and chance.\label{dual_analogies_ostensibly}}
\end{table}

\clearpage
\section{Co$\sim$event-based Bayes' theorem for a believability measure in the context of the theory of experience and chance}

Taking up this matter, primarily, I believed that as a result of this dual splitting I will get two dual theorems that are analogous to Bayes'
scheme. The first is about ``posterior believability'' of an accumulated hypothesis, and the second is about ``posterior probability'' of a random
observation. Moreover, I assumed I would get one general theorem on the ``posterior certainty'' of the co$\sim$event. However, this time my
``formal'' intuition turned out to be powerless. A brutal reality has revealed many more interesting things than all my ordinary assumptions.

The Tables \ref{dual_analogies} and \ref{dual_analogies_abbreviations} compare the logic of the dual analogies of the classical Bayes' scheme
formally in the contexts of probability theory and believability theory. It remains for me to note that within the framework of the theory of
certainties ostensibly another analogy it unites both dual analogies may seem possible (see Table \ref{dual_analogies_ostensibly}).

The logic of uncertainty it follows from the theory of experience and chance allows us to draw a completely definite conclusion about the three
analogies of Bayes' scheme (\ref{BayesFormulaPhiP}, \ref{BayesFormulaPhi}, \ref{bra-abbreviations}). Of the three listed analogies, I can consider
only the bra-formula (\ref{bra-abbreviations}), or what is the same, the bra-formula (\ref{BayesFormulaPhiB}), as a more or less suitable contender
for the title of the \textbf{co$\sim$event-based Bayes' theorem}. Why?

Firstly, the ket-formula (\ref{BayesFormulaPhiP}) ``puts the cart before the horse'', ``puts the chance before the experience'', ``puts the sequence
before the cause'', and ``puts the past before the future''. It considers that the a posteriori probability of the chance observation as an outcome
of a random experiment depends on the a priori believability of the observer' experience. This is nonsense.

Secondly, the bra-ket formula (\ref{BayesFormulaPhi}) inherits all the sins of Bayes' formula interpreted within the framework of probability theory,
comparing the probabilities of co$\sim$events from different spaces --- of the original and conditional spaces.

And thirdly, only the bra-formula (\ref{bra-abbreviations}) is free from these contradictions, calculating the a posteriori believability in the
experience of the observer through his/her a priori believability and through the a priori probabilities of his/her chance observations.

At the same time, the third formula (\ref{bra-abbreviations}, \ref{BayesFormulaPhiB}) reflects only the most general co$\sim$event-based logic of
uncertainty in the Bayes' scheme. Although the third formula is true in principle it is too rude and naive from the co$\sim$event-based point of
view. We now turn to the co$\sim$event-based refinement of these formulas.

\subsection{Co$\sim$event-based refinement of the Bayes' scheme\label{refinement}}

To refine the formulas (\ref{bra-abbreviations}, \ref{BayesFormulaPhiB}) in the co$\sim$event-based Bayesian scheme, we need to sacrifice the brevity
of the notation. This is a needful measure. Since three co$\sim$event actors, not one, are involved in this scheme, and they act in the same bra-ket
space $\braket{\Omega|\Omega}$.

In pondering a chance of Nature, and an accumulated own experience, a scientist after all of the experienced-random experiment observes the chance
and accumulates his/her observer' experience as two co$\sim$events:
\begin{eqnarray}\label{circ_times}
\mathscr R&=&\mbox{``\emph{Reality} (chance, or random observations)'' }\subseteq\braket{\Omega|\Omega},\\
\mathscr H&=&\mbox{``\emph{Hypotheses} (accumulated experience)'' }\subseteq\braket{\Omega|\Omega}.
\end{eqnarray}
From these co$\sim$events we will construct one more useful co$\sim$event
\begin{equation}\label{M}
\mathscr M = (\mathscr H \Delta\, \mathscr R)^c = \mbox{``\emph{Match Hypotheses with Reality}'' }\subseteq\braket{\Omega|\Omega}.
\end{equation}

Having at our disposal these three co$\sim$event actors, $\mathscr R, \mathscr H$, and $\mathscr M$, we can fully describe what is occurring in the
co$\sim$event-based Bayesian scheme.

What does the expression ``the co$\sim$event $\mathscr M$ is at our disposal'' in the theory of experience and chance mean?

This means that we know the probability, believability, and certainty distributions of the co$\sim$event $\mathscr M$:
\begin{eqnarray}
\label{P}
\mathbf P(\ket{X}) &=& p(X/\!\!/\frak X_{\!\mathscr M}), \ \ \ X\in\rS^{\frak X_{\!\mathscr M}}, \\
\label{B}
\mathbf B(\bra{x}) &=& b_{x}, \ \ \ x\in\frak X_{\!\mathscr M},\\
\label{Phi}
\bfPhi(\braket{x|X}) &=& \varphi_x(X/\!\!/\frak X_{\!\mathscr M}), \ \ \
(x,X)\in\frak X_{\!\mathscr M} \times \rS^{\frak X_{\!\mathscr M}},
\end{eqnarray}
defined on the bra-ket space $\braket{\Omega|\Omega}$, which has the labelling $\braket{\frak X_{\!\mathscr M}|\rS^{\frak X_{\!\mathscr M}}}$
generated by $\mathscr M$. Similar information is known to us about events $\mathscr H$ and $\mathscr R$. We know the probability, believability, and
certainty distributions of the co$\sim$event $\mathscr H$:
\begin{eqnarray}
\label{P}
\mathbf P(\ket{X}) &=& p(X/\!\!/\frak X_{\!\mathscr H}), \ \ \ X\in\rS^{\frak X_{\!\mathscr H}}, \\
\label{B}
\mathbf B(\bra{x}) &=& b_{x}, \ \ \ x\in\frak X_{\!\mathscr H},\\
\label{Phi}
\bfPhi(\braket{x|X}) &=& \varphi_x(X/\!\!/\frak X_{\!\mathscr H}), \ \ \
(x,X)\in\frak X_{\!\mathscr H} \times \rS^{\frak X_{\!\mathscr H}},
\end{eqnarray}
defined on the bra-ket space $\braket{\Omega|\Omega}$, which has the labelling $\braket{\frak X_{\!\mathscr H}|\rS^{\frak X_{\!\mathscr H}}}$
generated by $\mathscr H$. And we know the probability, believability, and certainty distributions of the co$\sim$event $\mathscr R$:
\begin{eqnarray}
\label{P}
\mathbf P(\ket{X}) &=& p(X/\!\!/\frak X_{\!\mathscr R}), \ \ \ X\in\rS^{\frak X_{\!\mathscr R}}, \\
\label{B}
\mathbf B(\bra{x}) &=& b_{x}, \ \ \ x\in\frak X_{\!\mathscr R},\\
\label{Phi}
\bfPhi(\braket{x|X}) &=& \varphi_x(X/\!\!/\frak X_{\!\mathscr R}), \ \ \
(x,X)\in\frak X_{\!\mathscr R} \times \rS^{\frak X_{\!\mathscr R}},
\end{eqnarray}
defined on the bra-ket space $\braket{\Omega|\Omega}$, which has the labelling $\braket{\frak X_{\!\mathscr R}|\rS^{\frak X_{\!\mathscr R}}}$
generated by $\mathscr R$.

Now we see that to refine the mathematical description of the co$\sim$event-based Bayesian scheme, we must describe three co$\sim$events, which are
defined on the same bra-ket space $\braket{\Omega|\Omega}$. Here the most important difficulty is they generate three different labellings
$\braket{\frak X_{\!\mathscr H}|\rS^{\frak X_{\!\mathscr H}}}$, $\braket{\frak X_{\!\mathscr R}|\rS^{\frak X_{\!\mathscr R}}}$, and $\braket{\frak
X_{\!\mathscr M}|\rS^{\frak X_{\!\mathscr M}}}$ of this space.

The labelling $\braket{\frak X_{\!\mathscr H}|\rS^{\frak X_{\!\mathscr H}}}$ is generated by the co$\sim$event $\mathscr H$ that describes results of
the preliminary experienced-random experiment in which ``an observer for each observation puts forward one or another subset of hypotheses about the
observation''. For example, ``the observer preliminary checkups of the set of patients and for each patient puts forward the one or another subset of
hypotheses about her/his diagnosis''.

The labelling $\braket{\frak X_{\!\mathscr R}|\rS^{\frak X_{\!\mathscr R}}}$ is generated by the co$\sim$event $\mathscr R$ that describes results of
the experienced-random experiment in which ``an observer for each observation takes a real decision based on complete information about the
observation''. For example, ``the observer for each patient determines a real diagnosis based on a complete medical examination of the patient''.

The labelling $\braket{\frak X_{\!\mathscr M}|\rS^{\frak X_{\!\mathscr M}}}$ is generated by the co$\sim$event $\mathscr M =(\mathscr
H\Delta\,\mathscr M)^c$ that is a co$\sim$event-valued function of $\mathscr H$ and $\mathscr R$. In other words, the co$\sim$event $\mathscr M$ is
defined by two experienced-random experiments as a result of which two co$\sim$events occur:  $\mathscr H$ (``Hypotheses'') and $\mathscr R$
(``Reality''). The labelling $\braket{\frak X_{\!\mathscr M}|\rS^{\frak X_{\!\mathscr M}}}$ is defined as a Minkowski intersection of two previous
labellings
\begin{equation}\label{M-intersection}
\braket{\frak X_{\!\mathscr M}|\rS^{\frak X_{\!\mathscr M}}} = \braket{\frak X_{\!\mathscr H}|\rS^{\frak X_{\!\mathscr H}}} (\cap) \braket{\frak X_{\!\mathscr R}|\rS^{\frak X_{\!\mathscr R}}}
\end{equation}
where
\begin{eqnarray}\label{M-intersection12}
\bra{\frak X_{\!\mathscr M}} &=& \bra{\frak X_{\!\mathscr H}} (\cap) \bra{\frak X_{\!\mathscr R}}
=\Big\{\bra{x^{\!\times}} \cap \bra{x^{\!\circ}}\colon x^{\!\times}\in\frak X_{\!\mathscr H}, x^{\!\circ}\in\frak X_{\!\mathscr R}, \bra{x^{\!\times}} \cap \bra{x^{\!\circ}}\ne\varnothing_{\bra{\Omega}} \Big\},\\
\ket{\rS^{\frak X_{\!\mathscr M}}} &=& \ket{\rS^{\frak X_{\!\mathscr H}}} (\cap) \ket{\rS^{\frak X_{\!\mathscr H}}}
=\Big\{\ket{X^{\!\times}} \cap \ket{X^{\!\circ}}\colon X^{\!\times}\in \rS^{\frak X_{\!\mathscr H}}, X^{\!\circ}\in \rS^{\frak X_{\!\mathscr R}}, \ket{X^{\!\times}} \cap \ket{X^{\!\circ}}\ne\emptyset_{\ket{\Omega}}\Big\}.
\end{eqnarray}

In our case of co$\sim$event-based Bayesian scheme we have equality for all $x\in \frak X_{\mathscr M}$
\begin{equation}\label{bra-equality}
\bra{x}=\bra{x^{\!\times}}=\bra{x^{\!\circ}}
\end{equation}
from which it follows that
\begin{equation}\label{bra-equality}
\bra{\frak X_{\!\mathscr M}}=\bra{\frak X_{\!\mathscr H}}=\bra{\frak X_{\!\mathscr R}}.
\end{equation}

\section{Co$\sim$event-based Bayes' theorem}

The observer twice conducts a series of $N$ observations of the same set of corresponding $N$ objects. First, the observer conducts \emph{preliminary
observations} $\{\ket{X^{\!\times}_1},\ldots,\ket{X^{\!\times}_{N}}\}$ of all objects to put forward the corresponding subsets of hypotheses from the
set of $M$ hypotheses $\{\braket{x_1|x^{\!\times}_1},\ldots,\braket{x_{M}|x^{\!\times}_{M}}\}$ about the results of \emph{exhaustive observations and
examinations} $\{\ket{X^{\!\circ}_1},\ldots,\ket{X^{\!\circ}_{N}}\}$ of the $N$ objects, as a result of which the co$\sim$event
\begin{equation}\label{HHH}
\mathscr H= \mbox{ ``\emph{Hypotheses}'' }=\braket{x_1|x^{\!\times}_1}+\ldots+\braket{x_{M}|x^{\!\times}_{M}}\subseteq\braket{\Omega|\Omega}
\end{equation}
occurs. Then the observer conducts \emph{exhaustive observations and examinations} $\{\ket{X^{\!\circ}_1},\ldots,\ket{X^{\!\circ}_{N}}\}$ of the $N$
objects, as a result of which the co$\sim$event
\begin{equation}\label{RRR}
\mathscr R= \mbox{ ``\emph{Reality}'' } =\braket{x_1|x^{\!\circ}_1}+\ldots+\braket{x_{M}|x^{\!\circ}_{M}}\subseteq\braket{\Omega|\Omega}
\end{equation}
occurs and defines the corresponding subsets of the set of $M$ real conclusions
$\{\braket{x_1|x^{\!\circ}_1},\ldots,\braket{x_{M}|x^{\!\circ}_{M}}\}$. After comparing the events $\mathscr H$ and $\mathscr R$ the co$\sim$event
\begin{equation}\label{MMM}
\mathscr M=(\mathscr H\Delta\,\mathscr R)^c= \mbox{ ``\emph{Match Hypotheses with Reality}'' } =
\braket{x_1|(x^{\!\times}_1\Delta\,x^{\!\circ}_1)^c}+\ldots+\braket{x_{M}|(x^{\!\times}_1\Delta\,x^{\!\circ}_M)^c}
\end{equation}
occurs, which I am going to use as the basis in a formulation of the co$\sim$event-based Bayes' theorem\footnote{The figure \ref{10x10Bayes} shows
Venn diagrams of three co$\sim$events from the co$\sim$event-based Bayes' scheme: $\mathscr H, \mathscr R$, and $\mathscr M$ for $M=N=10$, obtained
as the results of two experienced-random experiments: of hypothetic where the co$\sim$event $\mathscr H$ occurs and of real where the co$\sim$event
$\mathscr R$ occurs.}.

\texttt{\indent Theorem \!\refstepcounter{ctrTh}\arabic{ctrTh}\,\label{CoeventBayesTheorem}\itshape\footnotesize (co$\sim$event-based Bayes
theorem)\!.} \emph{Let $\mathscr H=$ ``\textbf{Hypotheses}'', $\mathscr R=$ ``\textbf{Reality}'', and  $\mathscr M=(\mathscr H\Delta\,\mathscr R)^c=$
``\textbf{Match Hypotheses with Reality}'' be co$\sim$events defined on the bra-ket space $\braket{\Omega|\Omega}$ from the certainty space
$(\braket{\Omega|\Omega},\braket{\mathscr A|\mathscr A},\bfPhi)=\braket{\Omega,\mathscr A, \mathbf B|\Omega, \mathscr A, \mathbf P}$ with
believability $\mathbf B$, probability $\mathbf P$, and certainty $\bfPhi = \mathbf B \times \mathbf P$. Let $\braket{\frak X_{\mathscr H}|\rS^{\frak
X_{\mathscr H}}}$, $\braket{\frak X_{\mathscr R}|\rS^{\frak X_{\mathscr R}}}$, and $\braket{\frak X_{\mathscr M}|\rS^{\frak X_{\mathscr M}}}$ be
labellings generated by the co$\sim$events $\mathscr H, \mathscr R$, and $\mathscr M\subseteq\braket{\Omega|\Omega}$ respectively where
\begin{eqnarray}\label{ÑoeventBayesTheoremWhere1}
\braket{\frak X_{\!\mathscr M}|\rS^{\frak X_{\!\mathscr M}}} &=&
\braket{\frak X_{\!\mathscr H}|\rS^{\frak X_{\!\mathscr H}}} (\cap)
\braket{\frak X_{\!\mathscr R}|\rS^{\frak X_{\!\mathscr R}}},\\
\bra{\frak X_{\!\mathscr M}} &=& \bra{\frak X_{\!\mathscr H}} (\cap) \bra{\frak X_{\!\mathscr R}}
=\Big\{\bra{x^{\!\times}} \cap \bra{x^{\!\circ}}\colon x^{\!\times}\in\frak X_{\!\mathscr H}, x^{\!\circ}\in\frak X_{\!\mathscr R}, \bra{x^{\!\times}} \cap \bra{x^{\!\circ}}\ne\varnothing_{\bra{\Omega}} \Big\},
\end{eqnarray}
\begin{eqnarray}\label{ÑoeventBayesTheoremWhere2}
\ket{\rS^{\frak X_{\!\mathscr M}}} &=& \ket{\rS^{\frak X_{\!\mathscr H}}} (\cap) \ket{\rS^{\frak X_{\!\mathscr H}}}
=\Big\{\ket{X^{\!\times}} \cap \ket{X^{\!\circ}}\colon X^{\!\times}\in \rS^{\frak X_{\!\mathscr H}}, X^{\!\circ}\in \rS^{\frak X_{\!\mathscr R}}, \ket{X^{\!\times}} \cap \ket{X^{\!\circ}}\ne\emptyset_{\ket{\Omega}}\Big\},\\
\bra{\frak X_{\!\mathscr M}}&=&\bra{\frak X_{\!\mathscr H}}=\bra{\frak X_{\!\mathscr R}}.
\end{eqnarray}
Let also $b_{x^{\!\times}}=\mathbf B(\bra{x^{\!\times}})$ and $b_{x^{\!\times}}^{\mbox{\tiny post}}=\mathbf B^{\mbox{\tiny
post}}(\bra{x^{\!\times}})$ be respectively \textbf{a priori} and \textbf{a posteriori} believability of the hypothesis $\bra{x}, x\in\frak
X_{\mathscr M}$, and $\mu(x^{\!\times},x^{\!\circ})=\mathbf P((\ket{x^{\!\times}}\Delta\,\ket{x^{\!\circ}})^c)$ be a probability of match the
ket-event $\ket{x^{\!\times}}$ with the ket-event $\ket{x^{\!\circ}}\subseteq\ket{\Omega}$. At last, let \textbf{a posteriori} believability of
bra-event $\bra{x^{\!\times}}\subseteq\bra{\Omega}$ under the condition that $\mathscr M$ has occurred is equal to the fraction of certainty of the
sub-co$\sim$event $\braket{x^{\!\times}|(x^{\!\times}\Delta\,x^{\circ})^c}\subseteq\mathscr M$ in the certainty of the co$\sim$event $\mathscr M$:
\begin{equation}\label{CoeventBayesAssumption}
b_{x^{\!\times}}^{\mbox{\tiny post}}=\frac{\bfPhi(\braket{x^{\!\times}|(x^{\!\times}\Delta\,x^{\circ})^c})}{\bfPhi(\mathscr M)}.
\end{equation}
Then \textbf{a posteriori} and \textbf{a priori} believability distributions $\{b_{x^{\!\times}}\colon x\in \frak X_{\mathscr M} \}$ and
$\{b_{x^{\!\times}}^{\mbox{\tiny post}}\colon x\in \frak X_{\mathscr M}\}$ are related by the co$\sim$event-based Bayesian formula for $x\in \frak
X_{\mathscr M}$:}
\begin{eqnarray}
\label{CoeventBayesTheoremFormula}
b^{\mbox{\tiny post}}_{x^{\!\times}} &=& b_{x^{\!\times}}
\frac{\mu(x^{\!\times},x^{\!\circ})}
{\displaystyle \sum_{x\in\frak X_{\!\mathscr M}} b_{x^{\!\times}} \mu(x^{\!\times},x^{\!\circ})}.
\end{eqnarray}

\textsf{Proof.} We use as the template the third co$\sim$event formula (\ref{bra-abbreviations}, \ref{BayesFormulaPhiB}) of the Bayesian scheme
chosen by us from Tables \ref{dual_analogies}, \ref{dual_analogies_abbreviations}, and \ref{dual_analogies_ostensibly} in order to rewrite it for the
event $\mathscr M$ in the following way:
\begin{eqnarray}\label{CoeventBayesTheoremProof}
\bfPhi(\braket{x^{\!\times}|\Omega}|\mathscr M) &=&
\frac{\bfPhi(\mathscr M\mid\braket{x^{\!\times}|\Omega})\bfPhi(\braket{x^{\!\times}|\Omega})}{\bfPhi(\mathscr M)}.
\end{eqnarray}
Now it remains for us to notice the following equalities:
\begin{eqnarray}\label{CoeventBayesTheoremProof1}
\label{bx}
\bfPhi(\braket{x^{\!\times}|\Omega}) &=&
\mathbf B(\bra{x^{\!\times}})=b_{x^{\!\times}},\\
\label{Munderbx}
\bfPhi(\mathscr M\mid\braket{x^{\!\times}|\Omega}) &=&
\mathbf P((\ket{x^{\!\times}}\Delta\,\ket{x^{\!\circ}})^c)=\mu(x^{\!\times},x^{\!\circ}),\\
\label{bxpost}
\bfPhi(\braket{x^{\!\times}|\Omega}|\mathscr M) &=&
\mathbf B^{\mbox{\tiny post}}(\bra{x^{\!\times}})=b_{x^{\!\times}}^{\mbox{\tiny post}}
\end{eqnarray}
in order for the theorem to be proved. Indeed, we have the equality (\ref{bx}) because
\begin{eqnarray}\label{CoeventBayesTheoremProof2}
\nonumber
\bfPhi(\braket{x^{\!\times}|\Omega}) &=& \mathbf B(\bra{x^{\!\times}})\mathbf P(\ket{\Omega})\\
\nonumber
&=&\mathbf B(\bra{x^{\!\times}})=b_{x^{\!\times}},
\end{eqnarray}
the equality (\ref{Munderbx}) because
\begin{eqnarray}\label{CoeventBayesTheoremProof3}
\nonumber
\bfPhi(\mathscr M\mid\braket{x^{\!\times}|\Omega}) &=&
\frac{\bfPhi(\mathscr M\cap\braket{x^{\!\times}|\Omega})}{\bfPhi(\braket{x^{\!\times}|\Omega})}=
\frac{\bfPhi(\braket{x^{\!\times}|(x^{\!\times}\Delta\,x^{\circ})^c})}{\bfPhi(\braket{x^{\!\times}|\Omega})}\\
&=&
\frac{\mathbf B(\bra{x^{\!\times}})\mathbf P((\ket{x^{\!\times}}\Delta\,\ket{x^{\!\circ}})^c)}{\mathbf B(\bra{x^{\!\times}})}\\
&=&
\mathbf P((\ket{x^{\!\times}}\Delta\,\ket{x^{\!\circ}})^c)=\mu(x^{\!\times},x^{\!\circ})
\end{eqnarray}
due to the fact that for ket-events $\ket{(x^{\!\times}\Delta\,x^{\circ})^c}=(\ket{x^{\!\times}}\Delta\,\ket{x^{\circ}})^c$. At last, the equality
(\ref{bxpost}) follows from our main co$\sim$event-based Bayes assumption (\ref{CoeventBayesAssumption}) because
\begin{eqnarray}\label{CoeventBayesAssumptionProof}
\nonumber
\bfPhi(\braket{x^{\!\times}|\Omega}\mid \mathscr M) &=&
\frac{\bfPhi(\braket{x^{\!\times}|\Omega}\cap \mathscr M)}{\bfPhi(\mathscr M)}=
\frac{\bfPhi(\braket{x^{\!\times}|(x^{\!\times}\Delta\,x^{\circ})^c})}{\bfPhi(\mathscr M)}\\
&=&
\mathbf B^{\mbox{\tiny post}}(\bra{x^{\!\times}})=b_{x^{\!\times}}^{\mbox{\tiny post}}.
\end{eqnarray}
Finally, from (\ref{bx}), (\ref{Munderbx}), and (\ref{bxpost}) it follows that
\begin{eqnarray}
\label{CoeventBayesTheoremProof4}
\mathbf B^{\mbox{\tiny post}}(\bra{x^{\!\times}}) &=&
\mathbf B(\bra{x^{\!\times}})
\frac{\mathbf P((\ket{x^{\!\times}}\Delta\,\ket{x^{\!\circ}})^c)}
{\displaystyle\sum_{x\in\frak X_{\mathscr R}}\mathbf B(\bra{x^{\!\times}})\mathbf P((\ket{x^{\!\times}}\Delta\,\ket{x^{\!\circ}})^c)}
= b_{x^{\!\times}}
\frac{\mu(x^{\!\times},x^{\!\circ})}
{\displaystyle \sum_{x\in\frak X_{\!\mathscr M}} b_{x^{\!\times}} \mu(x^{\!\times},x^{\!\circ})}
\end{eqnarray}
where
\begin{eqnarray}
\nonumber
\mu(x^{\!\times},x^{\!\circ})&=&\mathbf P((\ket{x^{\!\times}}\Delta \ket{x^{\!\circ}})^c)=1-p_{x^{\!\times}}-p_{x^{\!\circ}}+2p_{x^{\!\times}\!x^{\circ}},\\
\nonumber
p_{x^{\!\times}} &=& \mathbf P(\ket{x^{\!\times}}),\\
\nonumber
p_{x^{\circ}} &=& \mathbf P(\ket{x^{\!\circ}}),\\
\nonumber
p_{x^{\!\times}\!x^{\circ}} &=& \mathbf P(\ket{x^{\!\times}} \cap \ket{x^{\!\circ}}).
\end{eqnarray}
The theorem is proved.

\texttt{\indent Corollary \!\refstepcounter{ctrcor}\arabic{ctrcor}\,\label{cor-posteriori-certainty}\itshape\footnotesize (on a posteriori certainty
of co$\sim$event $\mathscr M$)\!.} The posterior believability distribution $\{b_x^{\mbox{\tiny post}}\colon x\in\frak X_{\!\mathscr M}\}$ calculated
by the co$\sim$event-based Bayes' theorem (\ref{CoeventBayesTheorem}) defines the posterior certainty of the co$\sim$event $\mathscr M=$ ``Match
Hypotheses with Reality'' by the following formula:
\begin{equation}\label{PosteriorCertainty}
\bfPhi^{\mbox{\tiny post}}(\mathscr M) = \sum_{x\in\frak X_{\!\mathscr M}} \sum_{x\in X\in\rsS^{\frak X_{\!\mathscr R}}} b_x^{\mbox{\tiny post}} p(X/\!\!/\frak X_{\!\mathscr R}).
\end{equation}
\textsf{Proof.} The corollary is valid because $\varphi^{\mbox{\tiny post}}_x(X) = b_x^{\mbox{\tiny post}} p(X/\!\!/\frak X_{\!\mathscr M})$ for
$x\in\frak X_{\!\mathscr M}$ and $X\in\rS^{\frak X_{\!\mathscr M}}$ and because the certainty measure $\bfPhi$ is additive.

\section{An example: ``Doctor and Patients with a headache''}

Suppose you are a doctor, and a patient comes to you, complaining of a headache. Further, suppose that there are two yours hypotheses for why people
get headaches, they can have a headache due to a brain tumor: $\bra{x}\subseteq\bra{\Omega}$, or they can have a headache due to a cold:
$\bra{y}\subseteq\bra{\Omega}$, $\bra{x}+\bra{y}=\bra{\Omega}$. The diagnosis of people can be of four varieties, they might have a brain tumor:
$\ket{\{x\}}\subseteq\ket{\Omega}$, they might have a cold: $\ket{\{y\}}\subseteq\ket{\Omega}$, they might have a brain tumor and a cold:
$\ket{\{x,y\}}\subseteq\ket{\Omega}$, and, finally, they may have neither a brain tumor nor a cold: $\ket{\emptyset}\subseteq\ket{\Omega}$ where
$\ket{\{x\}}+\ket{\{y\}}+\ket{\{x,y\}}+\ket{\emptyset}=\ket{\Omega}$. A brain tumor always causes a headache, but exceedingly few people have a brain
tumor. In contrast, a headache is rarely a symptom of cold, but most people manage to catch a cold every single year.

Let us assume that after your preliminary checkup of the 200 patients your hypotheses formed the statistics of the co$\sim$event $\mathscr
H=$``headaches''$=\protect\braket{x|\{x^{\times}\}}+\protect\braket{\Omega|\{x^{\times},y^{\times}\}}+\protect\braket{y|\{y^{\times}\}}+\protect\braket{y|\emptyset^{\times}}\subseteq
\protect\braket{\Omega|\Omega}$. The statistics of hypothetic diagnoses allows you to evaluate the probabilities of hypothetic diagnoses of the 200
patients as follows:
\begin{equation}\label{hypotheses}
p(\{x^{\times}\})=\mathbf P(\ket{\{x^{\times}\}}), \ \ \ p(\{y^{\times}\})=\mathbf P(\ket{\{y^{\times}\}}), \ \ \ p(\{x^{\times},y^{\times}\})=\mathbf P(\ket{\{x^{\times},y^{\times}\}}), \ \ \  p(\emptyset^{\times})=P(\ket{\emptyset^{\times}}).
\end{equation}
Then after a complete medical examination of the same 200 patients, the real statistics of the co$\sim$event $\mathscr R=$
``headaches''$=\protect\braket{x|\{x^{\circ}\}}+\protect\braket{\Omega|\{x^{\circ},y^{\circ}\}}+\protect\braket{y|\{y^{\circ}\}}+\protect\braket{y|\emptyset^{\circ}}\subseteq
\protect\braket{\Omega|\Omega}$ turns out at your disposal. The real statistics allows you to evaluate the probabilities of real diagnoses of the 200
patients as follows:
\begin{equation}\label{real}
p(\{x^{\circ}\})=\mathbf P(\ket{\{x^{\circ}\}}), \ \ \ p(\{y^{\circ}\})=\mathbf P(\ket{\{y^{\circ}\}}), \ \ \ p(\{x^{\circ},y^{\circ}\})=\mathbf P(\ket{\{x^{\circ},y^{\circ}\}}), \ \ \ p(\emptyset^{\circ})=\mathbf P(\ket{\emptyset^{\circ}}).
\end{equation}
Provided that the co$\sim$event $\mathscr M=(\mathscr H \Delta\,\mathscr R)^c=$ ``match hypotheses with real diagnoses'' occurred and in the absence
of other information on believabilities of yours hypotheses, i.e. $\mathbf B(\bra{x})=b_x=1/2, \mathbf B(\bra{y})=b_y=1/2$, do you think it is more
believably that a headache is caused by a tumor: $\bra{x}$, or by a cold: $\bra{y}$? In other words, do you think how the believability distribution
of your hypotheses could be after a complete medical examination of the 200 patients? The answer to this question you can get with the help of the
co$\sim$event-based Bayes' theorem (\ref{CoeventBayesTheoremFormula}).

In the terminology of the example ``Doctor and Patients'' the labelling $\braket{\frak X_{\!\mathscr H}|\rS^{\frak X_{\!\mathscr H}}}$ is generated
by the co$\sim$event $\mathscr H$ that describes results of the experienced-random experiment ``preliminary checkup of the 200 patients and assigning
their diagnosis to the first or the second hypothesis''. The labelling $\braket{\frak X_{\!\mathscr R}|\rS^{\frak X_{\!\mathscr R}}}$ is generated by
the co$\sim$event $\mathscr R$ that describes results of the experienced-random experiment ``assigning diagnosis to the first or the second type
after a complete medical examination of the same 200 patients''. The labelling $\braket{\frak X_{\!\mathscr M}|\rS^{\frak X_{\!\mathscr M}}}$ is
generated by the co$\sim$event $\mathscr M =(\mathscr H\Delta\,\mathscr R)^c$ that is a co$\sim$event-valued function of $\mathscr H$ and $\mathscr
R$. In other words, the co$\sim$event $\mathscr R$ is defined by two experienced-random experiments ``checkup of the 200 patients and assigning their
diagnosis to the first or the second hypothesis'' ($\mathscr H$) and ``assigning diagnosis to the first or the second type after a complete medical
examination of the same 200 patients'' ($\mathscr R$). And the labelling $\braket{\frak X_{\!\mathscr M}|\rS^{\frak X_{\!\mathscr M}}}$ is defined as
a Minkowski intersection of two previous labellings
\begin{equation}\label{M-intersection}
\braket{\frak X_{\!\mathscr M}|\rS^{\frak X_{\!\mathscr M}}} = \braket{\frak X_{\!\mathscr H}|\rS^{\frak X_{\!\mathscr H}}} (\cap) \braket{\frak X_{\!\mathscr R}|\rS^{\frak X_{\!\mathscr R}}}
\end{equation}
where
\begin{eqnarray}\label{M-intersection12}
\bra{\frak X_{\!\mathscr M}} &=& \bra{\frak X_{\!\mathscr H}} (\cap) \bra{\frak X_{\!\mathscr R}}
=\Big\{\bra{x_{\!\times}} \cap \bra{x_{\!\circ}}\colon \bra{x_{\!\times}}\in\bra{\frak X_{\!\times}}, \bra{x_{\!\circ}}\in\bra{\frak X_{\!\circ}}, \bra{x_{\!\times}} \cap \bra{x_{\!\circ}}\ne\varnothing_{\bra{\Omega}} \Big\}\\
\ket{\rS^{\frak X_{\!\mathscr M}}} &=& \ket{\rS^{\frak X_{\!\mathscr H}}} (\cap) \ket{\rS^{\frak X_{\!\mathscr R}}}
=\Big\{\ket{X^{\!\times}} \cap \ket{X^{\!\circ}}\colon \ket{X^{\!\times}}\in \ket{\rS^{\frak X_{\!\mathscr H}}}, \ket{X^{\!\circ}}\in \ket{\rS^{\frak X_{\!\mathscr R}}}, \ket{X^{\!\times}} \cap \ket{X^{\!\circ}}\ne\emptyset_{\ket{\Omega}}\Big\}.
\end{eqnarray}

In our case of co$\sim$event-based Bayesian scheme we have equality $\bra{x}=\bra{x_{\times}}=\bra{x_{\circ}}$ for all $x\in \frak X_{\mathscr R}$
from which it follows that
\begin{equation}\label{bra-equality}
\bra{\frak X_{\!\mathscr M}}=\bra{\frak X_{\!\mathscr H}}=\bra{\frak X_{\!\mathscr R}}.
\end{equation}

We'll continue with a few brief examples, illustrating the co$\sim$event-based Bayes' theorem (see Theorem \ref{CoeventBayesTheorem}). Let's consider
examples in which the hypothesis statistics of the co$\sim$event $\mathscr H$ will be the same (see Table \ref{example1_priorTABxFinal}), and the
real statistics of the co$\sim$event $\mathscr R$ will be presented in three variants (see Table \ref{example1_TABFinal}, at the top).
Correspondingly, the statistics of the co$\sim$event $\mathscr M=(\mathscr H\Delta\,\mathscr R)^c$ will be presented also in the three variants (see
Table \ref{example1_TABFinal} at the bottom).

In the first example (Tables \ref{example1_priorTABxFinal}, \ref{example1_TABFinal}) \emph{with \textbf{a posteriori} certainty one}\footnote{Within
the theory of co$\sim$events, the semantics of texts about the measurement of uncertainty acquires a new and precise meaning. This is the novelty of
the semantics of terms, which is fundamentally different from the semantics of terms in Kolmogorov's theory of probability and is intended to
interpret the relationship between the three measures of this theory: \emph{believability, probability and certainty}.} the posteriori believability
distribution of your hypotheses is uniform:
\begin{eqnarray}
\label{example1_posterior_x}
b^{\mbox{\tiny post}}_x&=&b_x\frac{p_x}{\bfPhi(\mathscr M)}=1/2,\\[-3pt]
\label{example1_posterior_y}
b^{\mbox{\tiny post}}_y&=&b_y\frac{p_y}{\bfPhi(\mathscr M)}=1/2,\\[-3pt]
\label{example1_prior_phi}
\bfPhi(\mathscr M)&=&\sum_{z\in\{x,y\}}\sum_{Z\in\{\{x\},\{x,y\},\{y\},\emptyset\}}\varphi_z(Z)=1,\\[-3pt]
\label{example1_posterior_phi}
\bfPhi^{\mbox{\tiny post}}(\mathscr M)&=&\sum_{z\in\{x,y\}}\sum_{Z\in\{\{x\},\{x,y\},\{y\},\emptyset\}}\varphi^{\mbox{\tiny post}}_z(Z)=1.
\end{eqnarray}

In the second example (Tables \ref{example1_priorTABxFinal}, \ref{example2_TABFinal}) by virtue of the fact that $\bfPhi(\mathscr M)=0$, the
co$\sim$event-based Bayes' theorem is not applicable and the posterior believability distribution and the posterior certainty of $\mathscr M$ remain
to be undefined.

\begin{table}[!h]
\centering
\begin{tabular}{r|l|l|l|l|l|l|}
 &           &\hspace*{-0.40cm}&\footnotesize b.tumor: $\ket{\{x^{\!\times}\}}$&\footnotesize b.tumor $\&$ cold: $\ket{\{x^{\!\times},y^{\!\times}\}}$&\footnotesize cold: $\ket{\{y^{\!\times}\}}$&\footnotesize nothing: $\ket{\emptyset^{\!\times}}$\\ \hline
 &           &\hspace*{-0.40cm}&\hfill $p(\{x^{\!\times}\})$ &\hfill $p(\{x^{\!\times},y^{\!\times}\})$ &\hfill $p(\{y^{\!\times}\})$&\hfill $p(\emptyset^{\!\times})$ \\
 &           &\hspace*{-0.40cm}&\hfill$8/200$ &\hfill$2/200$ &\hfill$150/200$&\hfill$40/200$ \\ \hline
 &&\hspace*{-0.40cm}&&&&\\[-0.40cm] \hline
\footnotesize b.tumor: $\bra{x}$&\hfill $b_x$ &\hspace*{-0.40cm}&\hfill $\varphi_x(\{x^{\!\times}\})$&\hfill $\varphi_x(\{x^{\!\times},y^{\!\times}\})$&& \\
                           & $1/2$ &\hspace*{-0.40cm}&\hfill $4/200$&\hfill $1/200$&\hfill 0&\hfill 0 \\ \hline
\footnotesize cold: $\bra{y}$&\hfill $b_y$ &\hspace*{-0.40cm}&                        &\hfill $\varphi_y(\{x^{\!\times},y^{\!\times}\})$& \hfill$\varphi_y(\{y^{\!\times}\})$&\\
                           &\hfill $1/2$ &\hspace*{-0.40cm}&\hfill 0&\hfill $1/200$&\hfill $75/200$&\hfill 0\\ \hline
\end{tabular}
\caption{A priori. Venn diagram of the co$\sim$event $\mathscr H=$ ``Hypotheses''$=\protect\braket{x|\{x^{\!\times}\}}+\protect\braket{\Omega|\{x^{\!\times},y^{\!\times}\}}+\protect\braket{y|\{y^{\!\times}\}}\subseteq \protect\braket{\Omega|\Omega}$ with the labelling $\protect\braket{\frak X_{\!\mathscr H}|\protect\rsS^{\frak X_{\!\mathscr H}}}=\protect\braket{\{x,y\}|\{\{x^{\!\times}\},\{x^{\!\times},y^{\!\times}\},\{y^{\!\times}\},\emptyset^{\!\times}\}}$,
with probabilities of ket-events: $p_{x^{\!\times}}=10/200$, $p_{y^{\!\times}}=0.152/200$, \emph{a priori} believabilities of bra-events: $b_x=1/2, b_y=1/2$, and with certainty $\bfPhi(\mathscr H)=81/200$.\label{example1_priorTABxFinal}}
\end{table}

\begin{table}[!h]
\centering
\begin{tabular}{r|l|l|l|l|l|l|}
 &           &\hspace*{-0.40cm}&\footnotesize b.tumor: $\ket{\{x^{\circ}\}}$&\footnotesize b.tumor $\&$ cold: $\ket{\{x^{\circ},y^{\circ}\}}$&\footnotesize cold: $\ket{\{y^{\circ}\}}$&\footnotesize nothing: $\ket{\emptyset^{\circ}}$\\ \hline
 &           &\hspace*{-0.40cm}&\hfill $p(\{x^{\circ}\})$ &\hfill $p(\{x^{\circ},y^{\circ}\})$ &\hfill $p(\{y^{\circ}\})$&\hfill $p(\emptyset^{\circ})$ \\
 &           &\hspace*{-0.40cm}&\hfill$8/200$ &\hfill$2/200$ &\hfill$150/200$&\hfill$40/200$ \\ \hline
 &&\hspace*{-0.40cm}&&&&\\[-0.40cm] \hline
\footnotesize b.tumor: $\bra{x}$&\hfill $b_x$ &\hspace*{-0.40cm}&\hfill $\varphi_x(\{x^{\circ}\})$&\hfill $\varphi_x(\{x^{\circ},y^{\circ}\})$&& \\
                           & $1/2$ &\hspace*{-0.40cm}&\hfill $4/200$&\hfill $1/200$&\hfill 0&\hfill 0 \\ \hline
\footnotesize cold: $\bra{y}$&\hfill $b_y$ &\hspace*{-0.40cm}&                        &\hfill $\varphi_y(\{x^{\circ},y^{\circ}\})$& \hfill$\varphi_y(\{y^{\circ}\})$&\\
                           &\hfill $1/2$ &\hspace*{-0.40cm}&\hfill 0&\hfill $1/200$&\hfill $75/200$&\hfill 0\\ \hline
\end{tabular}

\vspace{10pt}

\begin{tabular}{r|l|l|l|l|l|l|}
 &           &\hspace*{-0.40cm}&\footnotesize b.tumor: \ \ $\ket{\{x\}}$&\footnotesize b.tumor $\&$ cold: \ \ $\ket{\{x,y\}}$&\footnotesize cold: \ \ $\ket{\{y\}}$&\footnotesize nothing: \ \ $\ket{\emptyset}$\\ \hline
 &           &\hspace*{-0.40cm}&\hfill $p(\{x\})$ &\hfill $p(\{x,y\})$ &\hfill $p(\{y\})$&\hfill $p(\emptyset)$ \\
 &           &\hspace*{-0.40cm}&\hfill$8/200$ &\hfill$2/200$ &\hfill$150/200$&\hfill$40/200$ \\ \hline
 &&\hspace*{-0.40cm}&&&&\\[-0.40cm] \hline
\footnotesize b.tumor: $\bra{x}$&\hfill $b_x$ &\hspace*{-0.40cm}&\hfill $\varphi_x(\{x\})$&\hfill $\varphi_x(\{x,y\})$&\hfill $\varphi_x(\{y\})$&\hfill $\varphi_x(\emptyset)$ \\
                           &\hfill $1/2$ &\hspace*{-0.40cm}&\hfill $4/200$&\hfill $1/200$&\hfill $75/200$&\hfill $20/200$ \\ \hline
\footnotesize cold: $\bra{y}$&\hfill $b_y$ &\hspace*{-0.40cm}&\hfill $\varphi_y(\{x\})$&\hfill $\varphi_y(\{x,y\})$& \hfill$\varphi_y(\{y\})$&\hfill $\varphi_y(\emptyset)$\\
                           &\hfill $1/2$ &\hspace*{-0.40cm}&\hfill $4/200$&\hfill $1/200$&\hfill $75/200$&\hfill $20/200$\\ \hline
 &&\hspace*{-0.40cm}&&&&\\[-0.40cm] \hline
\footnotesize b.tumor: $\bra{x}$&\hfill $b^{\mbox{\tiny post}}_x$ &\hspace*{-0.40cm}&\hfill $\varphi^{\mbox{\tiny post}}_x(\{x\})$&\hfill $\varphi^{\mbox{\tiny post}}_x(\{x,y\})$&\hfill $\varphi^{\mbox{\tiny post}}_x(\{y\})$&\hfill $\varphi^{\mbox{\tiny post}}_x(\emptyset)$ \\
                           &\hfill $1/2$ &\hspace*{-0.40cm}&\hfill $4/200$&\hfill $1/200$&\hfill $75/200$&\hfill $20/200$ \\ \hline
\footnotesize cold: $\bra{y}$&\hfill $b^{\mbox{\tiny post}}_y$ &\hspace*{-0.40cm}&\hfill $\varphi^{\mbox{\tiny post}}_y(\{x\})$&\hfill $\varphi^{\mbox{\tiny post}}_y(\{x,y\})$& \hfill$\varphi^{\mbox{\tiny post}}_y(\{y\})$&\hfill $\varphi^{\mbox{\tiny post}}_y(\emptyset)$\\
                           &\hfill $1/2$ &\hspace*{-0.40cm}&\hfill $4/200$&\hfill $1/200$&\hfill $75/200$&\hfill $20/200$\\ \hline
\end{tabular}

\caption{\textbf{Example 1, at the top: A priori}. Venn diagram of the co$\sim$event $\mathscr R=$
``Reality''$=\protect\braket{x|\{x^{\circ}\}}+\protect\braket{\Omega|\{x^{\circ},y^{\circ}\}}+\protect\braket{y|\{y^{\circ}\}}\subseteq
\protect\braket{\Omega|\Omega}$ with the labelling $\protect\braket{\frak X_{\!\mathscr R}|\protect\rsS^{\frak X_{\!\mathscr
R}}}=\protect\braket{\{x,y\}|\{\{x^{\circ}\},\{x^{\circ},y^{\circ}\},\{y^{\circ}\},\emptyset\}}$, with real statistics of ket-events:
$p_{x^{\!\circ}}=10/200$, $p_{y^{\!\circ}}=152/200$, \emph{a priori} believabilities of bra-events: $b_x=1/2, b_y=1/2$, and with certainty
$\bfPhi(\mathscr R)=81/200$. \textbf{At the bottom: A posteriori}. Venn diagram of the co$\sim$event $\mathscr M=(\mathscr H\Delta\,\mathscr R)^c=$
``Match Hypotheses with Reality''$=\protect\braket{x|\{x\}}+\protect\braket{\Omega|\{x,y\}}+\protect\braket{y|\{y\}}\subseteq
\protect\braket{\Omega|\Omega}$ with the labelling $\protect\braket{\frak X_{\!\mathscr M}|\protect\rsS^{\frak X_{\!\mathscr
M}}}=\protect\braket{\{x,y\}|\{\{x\},\{x,y\},\{y\},\emptyset\}}$, with real statistics of ket-events: $p_x=1$, $p_y=1$, \emph{a posteriori}
believabilities of bra-events: $b^{\mbox{\tiny post}}_x=1/2, b^{\mbox{\tiny post}}_y=1/2$, and \emph{a posteriori} certainty $\bfPhi^{\mbox{\tiny
post}}(\mathscr M)=1$.\label{example1_TABFinal}}
\end{table}

\begin{table}[h!]

\vspace{-0.0cm}

\hspace{0.5cm}
\begin{minipage}{80mm}
\tiny
\begin{equation}\nonumber
\begin{split}
\mbox{\normalsize Example 1: \hspace{2.5cm}} \frak X_{\!\mathscr{R}} &= \{x_{1},x_{2}\},\\
\ket{X_{1}}       &=
\ket{X_1^{\!\times}} \cap \ket{X_1^{\circ}},\\[-1pt]
\ket{X_{2}}       &=
\ket{X_2^{\!\times}} \cap \ket{X_2^{\circ}},\\[-1pt]
\ket{X_{3}}       &=
\ket{X_2^{\!\times}} \cap \ket{X_3^{\circ}},\\[-1pt]
\ket{X_{4}}       &=
\ket{X_4^{\!\times}} \cap \ket{X_4^{\circ}}.
\end{split}
\end{equation}
\end{minipage}

\vspace{-1.0cm}

\hspace{1.20cm}
\begin{minipage}{80mm}
\normalsize $\left\{b_{x_{1}}^{\mbox{\tiny post}},b_{x_{2}}^{\mbox{\tiny post}}\right\}=\{1/2, \ 1/2\}$
\end{minipage}

\vspace{0.5cm}

\vspace{-25mm}

\setlength{\tabcolsep}{1pt}        
\renewcommand{\arraystretch}{0.45} 
\centering
\hspace{8.65cm}$\overbrace{\hspace{6.75cm}}^{\mbox{\normalsize$\ket{\Omega}$}}$            

\hspace{8.70cm}$\overbrace{\hspace{1.3cm}}^{\ket{X^{\circ}_1}}
                \overbrace{\hspace{1.3cm}}^{\ket{X^{\circ}_2}}
                \overbrace{\hspace{3.3cm}}^{\ket{X^{\circ}_3}}
                \overbrace{\hspace{0.65cm}}^{\ket{X^{\circ}_4}}$

\hspace{6.84cm}\begin{minipage}{0.50cm}
\hspace*{0.30cm}$\bra{\Omega}\!\!\left\{\begin{matrix}\vspace{1.2cm}\end{matrix}\right.$ 
\end{minipage}
\hspace{0.15cm}
\begin{minipage}{0.50cm}
\hspace*{0.15cm}$\bra{x_{1}}\!\left\{\begin{matrix}\vspace{0.47cm}\end{matrix}\right.$\\       
\hspace*{0.15cm}$\bra{x_{2}}\!\left\{\begin{matrix}\vspace{0.47cm}\end{matrix}\right.$
\end{minipage}
\hspace{0.30cm}
\begin{tabular}{|p{0.60cm}p{0.60cm}|p{0.60cm}p{0.60cm}|p{0.60cm}p{0.60cm}p{0.60cm}p{0.60cm}p{0.60cm}|p{0.60cm}|} \hline
  &&&&&&&&& \\[-0.15cm]
  \begin{minipage}{0.50cm}\colorbox{MyAqua0}{\large\hspace{-0.03cm}$\bigcirc$\hspace{-0.08cm}}\end{minipage}&
  \begin{minipage}{0.50cm}\colorbox{MyAqua0}{\large\hspace{-0.03cm}$\bigcirc$\hspace{-0.08cm}}\end{minipage}&
  \begin{minipage}{0.50cm}\colorbox{MyAqua0}{\large\hspace{-0.03cm}$\bigcirc$\hspace{-0.08cm}}\end{minipage}&
  \begin{minipage}{0.50cm}\colorbox{MyAqua0}{\large\hspace{-0.03cm}$\bigcirc$\hspace{-0.08cm}}\end{minipage}&
  \begin{minipage}{0.50cm}\colorbox{MyWhite}{$\begin{matrix}\vspace{0.1cm}&\vspace{0.1cm}\end{matrix}$}\end{minipage}&
  \begin{minipage}{0.50cm}\colorbox{MyWhite}{$\begin{matrix}\vspace{0.1cm}&\vspace{0.1cm}\end{matrix}$}\end{minipage}&
  \begin{minipage}{0.50cm}\colorbox{MyWhite}{$\begin{matrix}\vspace{0.1cm}&\vspace{0.1cm}\end{matrix}$}\end{minipage}&
  \begin{minipage}{0.50cm}\colorbox{MyWhite}{$\begin{matrix}\vspace{0.1cm}&\vspace{0.1cm}\end{matrix}$}\end{minipage}&
  \begin{minipage}{0.50cm}\colorbox{MyWhite}{$\begin{matrix}\vspace{0.1cm}&\vspace{0.1cm}\end{matrix}$}\end{minipage}&
  \begin{minipage}{0.50cm}\colorbox{MyWhite}{$\begin{matrix}\vspace{0.1cm}&\vspace{0.1cm}\end{matrix}$}\end{minipage}\\
  &&&&&&&&& \\[-0.15cm] \hline
  &&&&&&&&& \\[-0.15cm]
  \begin{minipage}{0.50cm}\colorbox{MyWhite}{$\begin{matrix}\vspace{0.1cm}&\vspace{0.1cm}\end{matrix}$}\end{minipage}&
  \begin{minipage}{0.50cm}\colorbox{MyWhite}{$\begin{matrix}\vspace{0.1cm}&\vspace{0.1cm}\end{matrix}$}\end{minipage}&
  \begin{minipage}{0.50cm}\colorbox{MyAqua0}{\large\hspace{-0.03cm}$\bigcirc$\hspace{-0.08cm}}\end{minipage}&
  \begin{minipage}{0.50cm}\colorbox{MyAqua0}{\large\hspace{-0.03cm}$\bigcirc$\hspace{-0.08cm}}\end{minipage}&
  \begin{minipage}{0.50cm}\colorbox{MyAqua0}{\large\hspace{-0.03cm}$\bigcirc$\hspace{-0.08cm}}\end{minipage}&
  \begin{minipage}{0.50cm}\colorbox{MyAqua0}{\large\hspace{-0.03cm}$\bigcirc$\hspace{-0.08cm}}\end{minipage}&
  \begin{minipage}{0.50cm}\colorbox{MyAqua0}{\large\hspace{-0.03cm}$\bigcirc$\hspace{-0.08cm}}\end{minipage}&
  \begin{minipage}{0.50cm}\colorbox{MyAqua0}{\large\hspace{-0.03cm}$\bigcirc$\hspace{-0.08cm}}\end{minipage}&
  \begin{minipage}{0.50cm}\colorbox{MyAqua0}{\large\hspace{-0.03cm}$\bigcirc$\hspace{-0.08cm}}\end{minipage}&
  \begin{minipage}{0.50cm}\colorbox{MyWhite}{$\begin{matrix}\vspace{0.1cm}&\vspace{0.1cm}\end{matrix}$}\end{minipage}\\
  &&&&&&&&& \\[-0.15cm] \hline
\end{tabular}

\vspace{0.0cm}
\hspace{0.85cm}$\overbrace{\hspace{6.7cm}}^{\mbox{\normalsize$\ket{\Omega}$}}$  \hspace{0.78cm}  $\overbrace{\hspace{6.7cm}}^{\mbox{\normalsize$\ket{\Omega}$}}$  

\hspace{0.85cm}$\overbrace{\hspace{1.3cm}}^{\ket{X^{\times}_1}}
                \overbrace{\hspace{1.3cm}}^{\ket{X^{\times}_2}}
                \overbrace{\hspace{3.3cm}}^{\ket{X^{\times}_3}}
                \overbrace{\hspace{0.65cm}}^{\ket{X^{\times}_4}}$
\hspace{0.89cm}$\overbrace{\hspace{1.3cm}}^{\ket{X_1}}
                \overbrace{\hspace{1.3cm}}^{\ket{X_2}}
                \overbrace{\hspace{3.3cm}}^{\ket{X_3}}
                \overbrace{\hspace{0.65cm}}^{\ket{X_4}}$

\hspace{-0.82cm}\begin{minipage}{0.50cm} $\bra{\Omega}\!\!\left\{\begin{matrix}\vspace{1.2cm}\end{matrix}\right.$
\end{minipage}
\hspace{0.15cm}
\begin{minipage}{0.50cm}
\hspace*{0.00cm}$\bra{x_{1}}\!\left\{\begin{matrix}\vspace{0.49cm}\end{matrix}\right.$\\       
\hspace*{0.00cm}$\bra{x_{2}}\!\left\{\begin{matrix}\vspace{0.49cm}\end{matrix}\right.$
\end{minipage}
\hspace{0.15cm}
\begin{tabular}{|p{0.60cm}p{0.60cm}|p{0.60cm}p{0.60cm}|p{0.60cm}p{0.60cm}p{0.60cm}p{0.60cm}p{0.60cm}|p{0.60cm}|} \hline
  &&&&&&&&& \\[-0.15cm]
  \begin{minipage}{0.50cm}\colorbox{MyGreen}{\LARGE\hspace{-0.03cm}$\times$\hspace{-0.08cm}}\end{minipage}&
  \begin{minipage}{0.50cm}\colorbox{MyGreen}{\LARGE\hspace{-0.03cm}$\times$\hspace{-0.08cm}}\end{minipage}&
  \begin{minipage}{0.50cm}\colorbox{MyGreen}{\LARGE\hspace{-0.03cm}$\times$\hspace{-0.08cm}}\end{minipage}&
  \begin{minipage}{0.50cm}\colorbox{MyGreen}{\LARGE\hspace{-0.03cm}$\times$\hspace{-0.08cm}}\end{minipage}&
  \begin{minipage}{0.50cm}\colorbox{MyWhite}{$\begin{matrix}\vspace{0.1cm}&\vspace{0.1cm}\end{matrix}$}\end{minipage}&
  \begin{minipage}{0.50cm}\colorbox{MyWhite}{$\begin{matrix}\vspace{0.1cm}&\vspace{0.1cm}\end{matrix}$}\end{minipage}&
  \begin{minipage}{0.50cm}\colorbox{MyWhite}{$\begin{matrix}\vspace{0.1cm}&\vspace{0.1cm}\end{matrix}$}\end{minipage}&
  \begin{minipage}{0.50cm}\colorbox{MyWhite}{$\begin{matrix}\vspace{0.1cm}&\vspace{0.1cm}\end{matrix}$}\end{minipage}&
  \begin{minipage}{0.50cm}\colorbox{MyWhite}{$\begin{matrix}\vspace{0.1cm}&\vspace{0.1cm}\end{matrix}$}\end{minipage}&
  \begin{minipage}{0.50cm}\colorbox{MyWhite}{$\begin{matrix}\vspace{0.1cm}&\vspace{0.1cm}\end{matrix}$}\end{minipage}\\
  &&&&&&&&& \\[-0.15cm] \hline
  &&&&&&&&& \\[-0.15cm]
  \begin{minipage}{0.50cm}\colorbox{MyWhite}{$\begin{matrix}\vspace{0.1cm}&\vspace{0.1cm}\end{matrix}$}\end{minipage}&
  \begin{minipage}{0.50cm}\colorbox{MyWhite}{$\begin{matrix}\vspace{0.1cm}&\vspace{0.1cm}\end{matrix}$}\end{minipage}&
  \begin{minipage}{0.50cm}\colorbox{MyGreen}{\LARGE\hspace{-0.03cm}$\times$\hspace{-0.08cm}}\end{minipage}&
  \begin{minipage}{0.50cm}\colorbox{MyGreen}{\LARGE\hspace{-0.03cm}$\times$\hspace{-0.08cm}}\end{minipage}&
  \begin{minipage}{0.50cm}\colorbox{MyGreen}{\LARGE\hspace{-0.03cm}$\times$\hspace{-0.08cm}}\end{minipage}&
  \begin{minipage}{0.50cm}\colorbox{MyGreen}{\LARGE\hspace{-0.03cm}$\times$\hspace{-0.08cm}}\end{minipage}&
  \begin{minipage}{0.50cm}\colorbox{MyGreen}{\LARGE\hspace{-0.03cm}$\times$\hspace{-0.08cm}}\end{minipage}&
  \begin{minipage}{0.50cm}\colorbox{MyGreen}{\LARGE\hspace{-0.03cm}$\times$\hspace{-0.08cm}}\end{minipage}&
  \begin{minipage}{0.50cm}\colorbox{MyGreen}{\LARGE\hspace{-0.03cm}$\times$\hspace{-0.08cm}}\end{minipage}&
  \begin{minipage}{0.50cm}\colorbox{MyWhite}{$\begin{matrix}\vspace{0.1cm}&\vspace{0.1cm}\end{matrix}$}\end{minipage}\\
  &&&&&&&&& \\[-0.15cm] \hline
\end{tabular}
\begin{minipage}{0.50cm}
\hspace*{0.00cm}$\bra{x_{1}}\!\left\{\begin{matrix}\vspace{0.49cm}\end{matrix}\right.$\\       
\hspace*{0.00cm}$\bra{x_{2}}\!\left\{\begin{matrix}\vspace{0.49cm}\end{matrix}\right.$
\end{minipage}
\hspace{0.15cm}
\begin{tabular}{|p{0.60cm}p{0.60cm}|p{0.60cm}p{0.60cm}|p{0.60cm}p{0.60cm}p{0.60cm}p{0.60cm}p{0.60cm}|p{0.60cm}|} \hline
  &&&&&&&&& \\[-0.15cm]
  \begin{minipage}{0.50cm}\colorbox{MyOrangeBayes}{\LARGE\hspace{-0.03cm}$\otimes$\hspace{-0.08cm}}\end{minipage}&
  \begin{minipage}{0.50cm}\colorbox{MyOrangeBayes}{\LARGE\hspace{-0.03cm}$\otimes$\hspace{-0.08cm}}\end{minipage}&
  \begin{minipage}{0.50cm}\colorbox{MyOrangeBayes}{\LARGE\hspace{-0.03cm}$\otimes$\hspace{-0.08cm}}\end{minipage}&
  \begin{minipage}{0.50cm}\colorbox{MyOrangeBayes}{\LARGE\hspace{-0.03cm}$\otimes$\hspace{-0.08cm}}\end{minipage}&
  \begin{minipage}{0.50cm}\colorbox{MyOrangeBayes}{$\begin{matrix}\vspace{0.1cm}&\vspace{0.1cm}\end{matrix}$}\end{minipage}&
  \begin{minipage}{0.50cm}\colorbox{MyOrangeBayes}{$\begin{matrix}\vspace{0.1cm}&\vspace{0.1cm}\end{matrix}$}\end{minipage}&
  \begin{minipage}{0.50cm}\colorbox{MyOrangeBayes}{$\begin{matrix}\vspace{0.1cm}&\vspace{0.1cm}\end{matrix}$}\end{minipage}&
  \begin{minipage}{0.50cm}\colorbox{MyOrangeBayes}{$\begin{matrix}\vspace{0.1cm}&\vspace{0.1cm}\end{matrix}$}\end{minipage}&
  \begin{minipage}{0.50cm}\colorbox{MyOrangeBayes}{$\begin{matrix}\vspace{0.1cm}&\vspace{0.1cm}\end{matrix}$}\end{minipage}&
  \begin{minipage}{0.50cm}\colorbox{MyOrangeBayes}{$\begin{matrix}\vspace{0.1cm}&\vspace{0.1cm}\end{matrix}$}\end{minipage}\\
  &&&&&&&&& \\[-0.15cm] \hline
  &&&&&&&&& \\[-0.15cm]
  \begin{minipage}{0.50cm}\colorbox{MyOrangeBayes}{$\begin{matrix}\vspace{0.1cm}&\vspace{0.1cm}\end{matrix}$}\end{minipage}&
  \begin{minipage}{0.50cm}\colorbox{MyOrangeBayes}{$\begin{matrix}\vspace{0.1cm}&\vspace{0.1cm}\end{matrix}$}\end{minipage}&
  \begin{minipage}{0.50cm}\colorbox{MyOrangeBayes}{\LARGE\hspace{-0.03cm}$\otimes$\hspace{-0.08cm}}\end{minipage}&
  \begin{minipage}{0.50cm}\colorbox{MyOrangeBayes}{\LARGE\hspace{-0.03cm}$\otimes$\hspace{-0.08cm}}\end{minipage}&
  \begin{minipage}{0.50cm}\colorbox{MyOrangeBayes}{\LARGE\hspace{-0.03cm}$\otimes$\hspace{-0.08cm}}\end{minipage}&
  \begin{minipage}{0.50cm}\colorbox{MyOrangeBayes}{\LARGE\hspace{-0.03cm}$\otimes$\hspace{-0.08cm}}\end{minipage}&
  \begin{minipage}{0.50cm}\colorbox{MyOrangeBayes}{\LARGE\hspace{-0.03cm}$\otimes$\hspace{-0.08cm}}\end{minipage}&
  \begin{minipage}{0.50cm}\colorbox{MyOrangeBayes}{\LARGE\hspace{-0.03cm}$\otimes$\hspace{-0.08cm}}\end{minipage}&
  \begin{minipage}{0.50cm}\colorbox{MyOrangeBayes}{\LARGE\hspace{-0.03cm}$\otimes$\hspace{-0.08cm}}\end{minipage}&
  \begin{minipage}{0.50cm}\colorbox{MyOrangeBayes}{$\begin{matrix}\vspace{0.1cm}&\vspace{0.1cm}\end{matrix}$}\end{minipage}\\
  &&&&&&&&& \\[-0.15cm] \hline
\end{tabular}

\vspace{-3.88cm}

\hspace{8.65cm}{\Huge $\mathscr R$}

\vspace{2.20cm}

\hspace{0.79cm}{\Huge $\mathscr H$}\hspace{70mm}{\Huge $\mathscr M$}

\vspace{0.50cm}

\caption{Venn diagrams of the three co$\sim$events  $\mathscr H, \mathscr{R}$, and $\mathscr{M}=(\mathscr H\Delta\,\mathscr R)^c$ $\Big($
\colorbox{MyGreen}{\large$\times$} , \colorbox{MyAqua0}{\small$\bigcirc$} ,
\colorbox{MyOrangeBayes}{\large$\otimes$}\hspace{1pt}\colorbox{MyOrangeBayes}{\protect\phantom{\large$\otimes$}} $\Big)$ in the co$\sim$event-based
Bayesian scheme. \label{2x10Bayes}}
\end{table}

\begin{table}[!h]
\centering
\begin{tabular}{r|l|l|l|l|l|l|}
 &           &\hspace*{-0.40cm}&\footnotesize b.tumor: $\ket{\{x^{\circ}\}}$&\footnotesize b.tumor $\&$ cold: $\ket{\{x^{\circ},y^{\circ}\}}$&\footnotesize cold: $\ket{\{y^{\circ}\}}$&\footnotesize nothing: $\ket{\emptyset^{\circ}}$\\ \hline
 &           &\hspace*{-0.40cm}&\hfill $p(\{x^{\circ}\})$ &\hfill $p(\{x^{\circ},y^{\circ}\})$ &\hfill $p(\{y^{\circ}\})$&\hfill $p(\emptyset^{\circ})$ \\
 &           &\hspace*{-0.40cm}&\hfill$8/200$ &\hfill$2/200$ &\hfill$150/200$&\hfill$40/200$ \\ \hline
 &&\hspace*{-0.40cm}&&&&\\[-0.40cm] \hline
\footnotesize b.tumor: $\bra{x}$&\hfill $b_x$ &\hspace*{-0.40cm}&&&\hfill $\varphi_x(\{y^{\circ}\})$&\hfill $\varphi_x(\emptyset^{\circ})$ \\
                           & $1/2$ &\hspace*{-0.40cm}&\hfill $0$&\hfill $0$&\hfill 75/200&\hfill 20/200 \\ \hline
\footnotesize cold: $\bra{y}$&\hfill $b_y$ &\hspace*{-0.40cm}&\hfill $\varphi_y(\{x^{\circ}\})$&&&\hfill $\varphi_y(\emptyset^{\circ})$\\
                           &\hfill $1/2$ &\hspace*{-0.40cm}&\hfill 4/200&\hfill $0$&\hfill $0$&\hfill 20/200\\ \hline
\end{tabular}

\vspace{10pt}

\begin{tabular}{r|l|l|l|l|l|l|}
 &           &\hspace*{-0.40cm}&\footnotesize b.tumor: \ \ $\ket{\{x\}}$&\footnotesize b.tumor $\&$ cold: \ \ $\ket{\{x,y\}}$&\footnotesize cold: \ \ $\ket{\{y\}}$&\footnotesize nothing: \ \ $\ket{\emptyset}$\\ \hline
 &           &\hspace*{-0.40cm}&\hfill $p(\{x\})$ &\hfill $p(\{x,y\})$ &\hfill $p(\{y\})$&\hfill $p(\emptyset)$ \\
 &           &\hspace*{-0.40cm}&\hfill$8/200$ &\hfill$2/200$ &\hfill$150/200$&\hfill$40/200$ \\ \hline
 &&\hspace*{-0.40cm}&&&&\\[-0.40cm] \hline
\footnotesize b.tumor: $\bra{x}$&\hfill $b_x$ &\hspace*{-0.40cm}&&&& \\
                           &1/2&\hspace*{-0.40cm}&\hfill $0$&\hfill $0$&\hfill $0$&\hfill $0$ \\ \hline
\footnotesize cold: $\bra{y}$&\hfill $b_y$ &\hspace*{-0.40cm}&&&&\\
                           &1/2&\hspace*{-0.40cm}&\hfill $0$&\hfill $0$&\hfill $0$&\hfill $0$\\ \hline
 &&\hspace*{-0.40cm}&&&&\\[-0.40cm] \hline
\footnotesize b.tumor: $\bra{x}$&\hfill $b^{\mbox{\tiny post}}_x$ &\hspace*{-0.40cm}&&&& \\
                           &\tiny \!\!\!\!undefined\!\!\!\!&\hspace*{-0.40cm}&\hfill $0$&\hfill $0$&\hfill $0$&\hfill $0$ \\ \hline
\footnotesize cold: $\bra{y}$&\hfill $b^{\mbox{\tiny post}}_y$ &\hspace*{-0.40cm}&&&&\\
                           &\tiny \!\!\!\!undefined\!\!\!\!&\hspace*{-0.40cm}&\hfill $0$&\hfill $0$&\hfill $0$&\hfill $0$\\ \hline
\end{tabular}

\caption{\textbf{Example 2, at the top: A priori}. Venn diagram of the co$\sim$event $\mathscr R=$
``Reality''$=\protect\braket{x|\{x^{\circ}\}}+\protect\braket{\Omega|\{x^{\circ},y^{\circ}\}}+\protect\braket{y|\{y^{\circ}\}}\subseteq
\protect\braket{\Omega|\Omega}$ with the labelling $\protect\braket{\frak X_{\!\mathscr R}|\protect\rsS^{\frak X_{\!\mathscr
R}}}=\protect\braket{\{x,y\}|\{\{x^{\circ}\},\{x^{\circ},y^{\circ}\},\{y^{\circ}\},\emptyset\}}$, with real statistics of ket-events:
$p_{x^{\!\circ}}=190/200$, $p_{y^{\!\circ}}=48/200$, \emph{a priori} believabilities of bra-events: $b_x=1/2, b_y=1/2$, and with certainty
$\bfPhi(\mathscr R)=119/200$. \textbf{At the bottom: A posteriori}. Venn diagram of the co$\sim$event $\mathscr M=(\mathscr H\Delta\,\mathscr R)^c=$
``Match Hypotheses with Reality''$=\protect\braket{x|\{x\}}+\protect\braket{\Omega|\{x,y\}}+\protect\braket{y|\{y\}}\subseteq
\protect\braket{\Omega|\Omega}$ with the labelling $\protect\braket{\frak X_{\!\mathscr M}|\protect\rsS^{\frak X_{\!\mathscr
M}}}=\protect\braket{\{x,y\}|\{\{x\},\{x,y\},\{y\},\emptyset\}}$, with real statistics of ket-events: $p_x=0$, $p_y=0$, \emph{a priori} certainty
$\bfPhi(\mathscr M)=0$, undefined \emph{a posteriori} believabilities of bra-events: $b^{\mbox{\tiny post}}_x, b^{\mbox{\tiny post}}_y$, and
undefined \emph{a posteriori} certainty $\bfPhi^{\mbox{\tiny post}}(\mathscr M)$.\label{example2_TABFinal}}
\end{table}

\begin{table}[h!]

\hspace{0.5cm}
\begin{minipage}{80mm}
\tiny
\begin{equation}\nonumber
\begin{split}
\mbox{\normalsize Example 2: \hspace{2.5cm}} \frak X_{\!\mathscr{R}} &= \{x_{1},x_{2}\},\\
\ket{X_{1}}       &=
\ket{X_1^{\!\times}} \cap \ket{X_1^{\circ}},\\[-1pt]
\ket{X_{2}}       &=
\ket{X_2^{\!\times}} \cap \ket{X_2^{\circ}},\\[-1pt]
\ket{X_{3}}       &=
\ket{X_2^{\!\times}} \cap \ket{X_3^{\circ}},\\[-1pt]
\ket{X_{4}}       &=
\ket{X_4^{\!\times}} \cap \ket{X_4^{\circ}}.
\end{split}
\end{equation}
\end{minipage}

\vspace{-1.0cm}

\hspace{-7.20cm}
\begin{minipage}{80mm}
\normalsize $\left\{b_{x_{1}}^{\mbox{\tiny post}},b_{x_{2}}^{\mbox{\tiny post}}\right\}=$ ``undefined''
\end{minipage}

\vspace{0.5cm}

\vspace{-25mm}

\setlength{\tabcolsep}{1pt}        
\renewcommand{\arraystretch}{0.45} 
\centering
\hspace{8.65cm}$\overbrace{\hspace{6.75cm}}^{\mbox{\normalsize$\ket{\Omega}$}}$            

\hspace{8.70cm}$\overbrace{\hspace{1.3cm}}^{\ket{X^{\circ}_1}}
                \overbrace{\hspace{1.3cm}}^{\ket{X^{\circ}_2}}
                \overbrace{\hspace{3.3cm}}^{\ket{X^{\circ}_3}}
                \overbrace{\hspace{0.65cm}}^{\ket{X^{\circ}_4}}$

\hspace{6.84cm}\begin{minipage}{0.50cm}
\hspace*{0.30cm}$\bra{\Omega}\!\!\left\{\begin{matrix}\vspace{1.2cm}\end{matrix}\right.$ 
\end{minipage}
\hspace{0.15cm}
\begin{minipage}{0.50cm}
\hspace*{0.15cm}$\bra{x_{1}}\!\left\{\begin{matrix}\vspace{0.47cm}\end{matrix}\right.$\\       
\hspace*{0.15cm}$\bra{x_{2}}\!\left\{\begin{matrix}\vspace{0.47cm}\end{matrix}\right.$
\end{minipage}
\hspace{0.30cm}
\begin{tabular}{|p{0.60cm}p{0.60cm}|p{0.60cm}p{0.60cm}|p{0.60cm}p{0.60cm}p{0.60cm}p{0.60cm}p{0.60cm}|p{0.60cm}|} \hline
  &&&&&&&&& \\[-0.15cm]
  \begin{minipage}{0.50cm}\colorbox{MyWhite}{$\begin{matrix}\vspace{0.1cm}&\vspace{0.1cm}\end{matrix}$}\end{minipage}&
  \begin{minipage}{0.50cm}\colorbox{MyWhite}{$\begin{matrix}\vspace{0.1cm}&\vspace{0.1cm}\end{matrix}$}\end{minipage}&
  \begin{minipage}{0.50cm}\colorbox{MyWhite}{$\begin{matrix}\vspace{0.1cm}&\vspace{0.1cm}\end{matrix}$}\end{minipage}&
  \begin{minipage}{0.50cm}\colorbox{MyWhite}{$\begin{matrix}\vspace{0.1cm}&\vspace{0.1cm}\end{matrix}$}\end{minipage}&
  \begin{minipage}{0.50cm}\colorbox{MyAqua0}{\large\hspace{-0.03cm}$\bigcirc$\hspace{-0.08cm}}\end{minipage}&
  \begin{minipage}{0.50cm}\colorbox{MyAqua0}{\large\hspace{-0.03cm}$\bigcirc$\hspace{-0.08cm}}\end{minipage}&
  \begin{minipage}{0.50cm}\colorbox{MyAqua0}{\large\hspace{-0.03cm}$\bigcirc$\hspace{-0.08cm}}\end{minipage}&
  \begin{minipage}{0.50cm}\colorbox{MyAqua0}{\large\hspace{-0.03cm}$\bigcirc$\hspace{-0.08cm}}\end{minipage}&
  \begin{minipage}{0.50cm}\colorbox{MyAqua0}{\large\hspace{-0.03cm}$\bigcirc$\hspace{-0.08cm}}\end{minipage}&
  \begin{minipage}{0.50cm}\colorbox{MyAqua0}{\large\hspace{-0.03cm}$\bigcirc$\hspace{-0.08cm}}\end{minipage}\\
  &&&&&&&&& \\[-0.15cm] \hline
  &&&&&&&&& \\[-0.15cm]
  \begin{minipage}{0.50cm}\colorbox{MyAqua0}{\large\hspace{-0.03cm}$\bigcirc$\hspace{-0.08cm}}\end{minipage}&
  \begin{minipage}{0.50cm}\colorbox{MyAqua0}{\large\hspace{-0.03cm}$\bigcirc$\hspace{-0.08cm}}\end{minipage}&
  \begin{minipage}{0.50cm}\colorbox{MyWhite}{$\begin{matrix}\vspace{0.1cm}&\vspace{0.1cm}\end{matrix}$}\end{minipage}&
  \begin{minipage}{0.50cm}\colorbox{MyWhite}{$\begin{matrix}\vspace{0.1cm}&\vspace{0.1cm}\end{matrix}$}\end{minipage}&
  \begin{minipage}{0.50cm}\colorbox{MyWhite}{$\begin{matrix}\vspace{0.1cm}&\vspace{0.1cm}\end{matrix}$}\end{minipage}&
  \begin{minipage}{0.50cm}\colorbox{MyWhite}{$\begin{matrix}\vspace{0.1cm}&\vspace{0.1cm}\end{matrix}$}\end{minipage}&
  \begin{minipage}{0.50cm}\colorbox{MyWhite}{$\begin{matrix}\vspace{0.1cm}&\vspace{0.1cm}\end{matrix}$}\end{minipage}&
  \begin{minipage}{0.50cm}\colorbox{MyWhite}{$\begin{matrix}\vspace{0.1cm}&\vspace{0.1cm}\end{matrix}$}\end{minipage}&
  \begin{minipage}{0.50cm}\colorbox{MyWhite}{$\begin{matrix}\vspace{0.1cm}&\vspace{0.1cm}\end{matrix}$}\end{minipage}&
  \begin{minipage}{0.50cm}\colorbox{MyAqua0}{\large\hspace{-0.03cm}$\bigcirc$\hspace{-0.08cm}}\end{minipage}\\
  &&&&&&&&& \\[-0.15cm] \hline
\end{tabular}

\vspace{0.0cm}
\hspace{0.85cm}$\overbrace{\hspace{6.7cm}}^{\mbox{\normalsize$\ket{\Omega}$}}$  \hspace{0.78cm}  $\overbrace{\hspace{6.7cm}}^{\mbox{\normalsize$\ket{\Omega}$}}$  

\hspace{0.85cm}$\overbrace{\hspace{1.3cm}}^{\ket{X^{\times}_1}}
                \overbrace{\hspace{1.3cm}}^{\ket{X^{\times}_2}}
                \overbrace{\hspace{3.3cm}}^{\ket{X^{\times}_3}}
                \overbrace{\hspace{0.65cm}}^{\ket{X^{\times}_4}}$
\hspace{0.89cm}$\overbrace{\hspace{1.3cm}}^{\ket{X_1}}
                \overbrace{\hspace{1.3cm}}^{\ket{X_2}}
                \overbrace{\hspace{3.3cm}}^{\ket{X_3}}
                \overbrace{\hspace{0.65cm}}^{\ket{X_4}}$

\hspace{-0.82cm}\begin{minipage}{0.50cm} $\bra{\Omega}\!\!\left\{\begin{matrix}\vspace{1.2cm}\end{matrix}\right.$
\end{minipage}
\hspace{0.15cm}
\begin{minipage}{0.50cm}
\hspace*{0.00cm}$\bra{x_{1}}\!\left\{\begin{matrix}\vspace{0.49cm}\end{matrix}\right.$\\       
\hspace*{0.00cm}$\bra{x_{2}}\!\left\{\begin{matrix}\vspace{0.49cm}\end{matrix}\right.$
\end{minipage}
\hspace{0.15cm}
\begin{tabular}{|p{0.60cm}p{0.60cm}|p{0.60cm}p{0.60cm}|p{0.60cm}p{0.60cm}p{0.60cm}p{0.60cm}p{0.60cm}|p{0.60cm}|} \hline
  &&&&&&&&& \\[-0.15cm]
  \begin{minipage}{0.50cm}\colorbox{MyGreen}{\LARGE\hspace{-0.03cm}$\times$\hspace{-0.08cm}}\end{minipage}&
  \begin{minipage}{0.50cm}\colorbox{MyGreen}{\LARGE\hspace{-0.03cm}$\times$\hspace{-0.08cm}}\end{minipage}&
  \begin{minipage}{0.50cm}\colorbox{MyGreen}{\LARGE\hspace{-0.03cm}$\times$\hspace{-0.08cm}}\end{minipage}&
  \begin{minipage}{0.50cm}\colorbox{MyGreen}{\LARGE\hspace{-0.03cm}$\times$\hspace{-0.08cm}}\end{minipage}&
  \begin{minipage}{0.50cm}\colorbox{MyWhite}{$\begin{matrix}\vspace{0.1cm}&\vspace{0.1cm}\end{matrix}$}\end{minipage}&
  \begin{minipage}{0.50cm}\colorbox{MyWhite}{$\begin{matrix}\vspace{0.1cm}&\vspace{0.1cm}\end{matrix}$}\end{minipage}&
  \begin{minipage}{0.50cm}\colorbox{MyWhite}{$\begin{matrix}\vspace{0.1cm}&\vspace{0.1cm}\end{matrix}$}\end{minipage}&
  \begin{minipage}{0.50cm}\colorbox{MyWhite}{$\begin{matrix}\vspace{0.1cm}&\vspace{0.1cm}\end{matrix}$}\end{minipage}&
  \begin{minipage}{0.50cm}\colorbox{MyWhite}{$\begin{matrix}\vspace{0.1cm}&\vspace{0.1cm}\end{matrix}$}\end{minipage}&
  \begin{minipage}{0.50cm}\colorbox{MyWhite}{$\begin{matrix}\vspace{0.1cm}&\vspace{0.1cm}\end{matrix}$}\end{minipage}\\
  &&&&&&&&& \\[-0.15cm] \hline
  &&&&&&&&& \\[-0.15cm]
  \begin{minipage}{0.50cm}\colorbox{MyWhite}{$\begin{matrix}\vspace{0.1cm}&\vspace{0.1cm}\end{matrix}$}\end{minipage}&
  \begin{minipage}{0.50cm}\colorbox{MyWhite}{$\begin{matrix}\vspace{0.1cm}&\vspace{0.1cm}\end{matrix}$}\end{minipage}&
  \begin{minipage}{0.50cm}\colorbox{MyGreen}{\LARGE\hspace{-0.03cm}$\times$\hspace{-0.08cm}}\end{minipage}&
  \begin{minipage}{0.50cm}\colorbox{MyGreen}{\LARGE\hspace{-0.03cm}$\times$\hspace{-0.08cm}}\end{minipage}&
  \begin{minipage}{0.50cm}\colorbox{MyGreen}{\LARGE\hspace{-0.03cm}$\times$\hspace{-0.08cm}}\end{minipage}&
  \begin{minipage}{0.50cm}\colorbox{MyGreen}{\LARGE\hspace{-0.03cm}$\times$\hspace{-0.08cm}}\end{minipage}&
  \begin{minipage}{0.50cm}\colorbox{MyGreen}{\LARGE\hspace{-0.03cm}$\times$\hspace{-0.08cm}}\end{minipage}&
  \begin{minipage}{0.50cm}\colorbox{MyGreen}{\LARGE\hspace{-0.03cm}$\times$\hspace{-0.08cm}}\end{minipage}&
  \begin{minipage}{0.50cm}\colorbox{MyGreen}{\LARGE\hspace{-0.03cm}$\times$\hspace{-0.08cm}}\end{minipage}&
  \begin{minipage}{0.50cm}\colorbox{MyWhite}{$\begin{matrix}\vspace{0.1cm}&\vspace{0.1cm}\end{matrix}$}\end{minipage}\\
  &&&&&&&&& \\[-0.15cm] \hline
\end{tabular}
\begin{minipage}{0.50cm}
\hspace*{0.00cm}$\bra{x_{1}}\!\left\{\begin{matrix}\vspace{0.49cm}\end{matrix}\right.$\\       
\hspace*{0.00cm}$\bra{x_{2}}\!\left\{\begin{matrix}\vspace{0.49cm}\end{matrix}\right.$
\end{minipage}
\hspace{0.15cm}
\begin{tabular}{|p{0.60cm}p{0.60cm}|p{0.60cm}p{0.60cm}|p{0.60cm}p{0.60cm}p{0.60cm}p{0.60cm}p{0.60cm}|p{0.60cm}|} \hline
  &&&&&&&&& \\[-0.15cm]
  \begin{minipage}{0.50cm}\colorbox{MyWhite}{\LARGE\hspace{-0.05cm}$\times$\hspace{-0.05cm}}\end{minipage}&
  \begin{minipage}{0.50cm}\colorbox{MyWhite}{\LARGE\hspace{-0.05cm}$\times$\hspace{-0.05cm}}\end{minipage}&
  \begin{minipage}{0.50cm}\colorbox{MyWhite}{\LARGE\hspace{-0.05cm}$\times$\hspace{-0.05cm}}\end{minipage}&
  \begin{minipage}{0.50cm}\colorbox{MyWhite}{\LARGE\hspace{-0.05cm}$\times$\hspace{-0.05cm}}\end{minipage}&
  \begin{minipage}{0.50cm}\colorbox{MyWhite}{\large\hspace{-0.03cm}$\bigcirc$\hspace{-0.03cm}}\end{minipage}&
  \begin{minipage}{0.50cm}\colorbox{MyWhite}{\large\hspace{-0.03cm}$\bigcirc$\hspace{-0.03cm}}\end{minipage}&
  \begin{minipage}{0.50cm}\colorbox{MyWhite}{\large\hspace{-0.03cm}$\bigcirc$\hspace{-0.03cm}}\end{minipage}&
  \begin{minipage}{0.50cm}\colorbox{MyWhite}{\large\hspace{-0.03cm}$\bigcirc$\hspace{-0.03cm}}\end{minipage}&
  \begin{minipage}{0.50cm}\colorbox{MyWhite}{\large\hspace{-0.03cm}$\bigcirc$\hspace{-0.03cm}}\end{minipage}&
  \begin{minipage}{0.50cm}\colorbox{MyWhite}{\large\hspace{-0.03cm}$\bigcirc$\hspace{-0.03cm}}\end{minipage}\\
  &&&&&&&&& \\[-0.15cm] \hline
  &&&&&&&&& \\[-0.15cm]
  \begin{minipage}{0.50cm}\colorbox{MyWhite}{\large\hspace{-0.03cm}$\bigcirc$\hspace{-0.03cm}}\end{minipage}&
  \begin{minipage}{0.50cm}\colorbox{MyWhite}{\large\hspace{-0.03cm}$\bigcirc$\hspace{-0.03cm}}\end{minipage}&
  \begin{minipage}{0.50cm}\colorbox{MyWhite}{\LARGE\hspace{-0.05cm}$\times$\hspace{-0.05cm}}\end{minipage}&
  \begin{minipage}{0.50cm}\colorbox{MyWhite}{\LARGE\hspace{-0.05cm}$\times$\hspace{-0.05cm}}\end{minipage}&
  \begin{minipage}{0.50cm}\colorbox{MyWhite}{\LARGE\hspace{-0.05cm}$\times$\hspace{-0.05cm}}\end{minipage}&
  \begin{minipage}{0.50cm}\colorbox{MyWhite}{\LARGE\hspace{-0.05cm}$\times$\hspace{-0.05cm}}\end{minipage}&
  \begin{minipage}{0.50cm}\colorbox{MyWhite}{\LARGE\hspace{-0.05cm}$\times$\hspace{-0.05cm}}\end{minipage}&
  \begin{minipage}{0.50cm}\colorbox{MyWhite}{\LARGE\hspace{-0.05cm}$\times$\hspace{-0.05cm}}\end{minipage}&
  \begin{minipage}{0.50cm}\colorbox{MyWhite}{\LARGE\hspace{-0.05cm}$\times$\hspace{-0.05cm}}\end{minipage}&
  \begin{minipage}{0.50cm}\colorbox{MyWhite}{\large\hspace{-0.03cm}$\bigcirc$\hspace{-0.03cm}}\end{minipage}\\
  &&&&&&&&& \\[-0.15cm] \hline
\end{tabular}

\vspace{-3.88cm}

\hspace{8.65cm}{\Huge $\mathscr R$}

\vspace{2.20cm}

\hspace{0.79cm}{\Huge $\mathscr H$}\hspace{70mm}{\Huge $\mathscr M$}

\vspace{0.50cm}

\caption{Venn diagrams of the three co$\sim$events  $\mathscr H, \mathscr{R}$, and $\mathscr{M}=(\mathscr H\Delta\,\mathscr R)^c$ $\Big($
\colorbox{MyGreen}{\large$\times$} , \colorbox{MyAqua0}{\small$\bigcirc$} ,
\colorbox{MyOrangeBayes}{\large$\otimes$}\hspace{1pt}\colorbox{MyOrangeBayes}{\protect\phantom{\large$\otimes$}} $\Big)$ in the co$\sim$event-based
Bayesian scheme. \label{2x10Bayes}}
\end{table}

\clearpage
\begin{table}[!h]
\centering
\begin{tabular}{r|l|l|l|l|l|l|}
 &           &\hspace*{-0.40cm}&\footnotesize b.tumor: $\ket{\{x^{\circ}\}}$&\footnotesize b.tumor $\&$ cold: $\ket{\{x^{\circ},y^{\circ}\}}$&\footnotesize cold: $\ket{\{y^{\circ}\}}$&\footnotesize nothing: $\ket{\emptyset^{\circ}}$\\ \hline
 &           &\hspace*{-0.40cm}&\hfill $p(\{x^{\circ}\})$ &\hfill $p(\{x^{\circ},y^{\circ}\})$ &\hfill $p(\{y^{\circ}\})$&\hfill $p(\emptyset^{\circ})$ \\
 &           &\hspace*{-0.40cm}&\hfill$8/200$ &\hfill$2/200$ &\hfill$150/200$&\hfill$40/200$ \\ \hline
 &&\hspace*{-0.40cm}&&&&\\[-0.40cm] \hline
\footnotesize b.tumor: $\bra{x}$&\hfill $b_x$ &\hspace*{-0.40cm}&&&\hfill $\varphi_x(\{y^{\circ}\})$& \\
                           & $1/2$ &\hspace*{-0.40cm}&\hfill $0$&\hfill $0$&\hfill 75/200&\hfill 0 \\ \hline
\footnotesize cold: $\bra{y}$&\hfill $b_y$ &\hspace*{-0.40cm}&&\hfill $\varphi_y(\{x^{\circ},y^{\circ}\})$& \hfill$\varphi_y(\{y^{\circ}\})$&\\
                           &\hfill $1/2$ &\hspace*{-0.40cm}&\hfill 0&\hfill $1/200$&\hfill $75/200$&\hfill 0\\ \hline
\end{tabular}

\vspace{10pt}

\begin{tabular}{r|l|l|l|l|l|l|}
 &           &\hspace*{-0.40cm}&\footnotesize b.tumor: \ \ $\ket{\{x\}}$&\footnotesize b.tumor $\&$ cold: \ \ $\ket{\{x,y\}}$&\footnotesize cold: \ \ $\ket{\{y\}}$&\footnotesize nothing: \ \ $\ket{\emptyset}$\\ \hline
 &           &\hspace*{-0.40cm}&\hfill $p(\{x\})$ &\hfill $p(\{x,y\})$ &\hfill $p(\{y\})$&\hfill $p(\emptyset)$ \\
 &           &\hspace*{-0.40cm}&\hfill$8/200$ &\hfill$2/200$ &\hfill$150/200$&\hfill$40/200$ \\ \hline
 &&\hspace*{-0.40cm}&&&&\\[-0.40cm] \hline
\footnotesize b.tumor: $\bra{x}$&\hfill $b_x$ &\hspace*{-0.40cm}&&&&\hfill $\varphi_x(\emptyset)$ \\
                           &\hfill $1/2$ &\hspace*{-0.40cm}&\hfill $0$&\hfill $0$&\hfill $0$&\hfill $20/200$ \\ \hline
\footnotesize cold: $\bra{y}$&\hfill $b_y$ &\hspace*{-0.40cm}&\hfill $\varphi_y(\{x\})$&\hfill $\varphi_y(\{x,y\})$& \hfill$\varphi_y(\{y\})$&\hfill $\varphi_y(\emptyset)$\\
                           &\hfill $1/2$ &\hspace*{-0.40cm}&\hfill $4/200$&\hfill $1/200$&\hfill $75/200$&\hfill $20/200$\\ \hline
 &&\hspace*{-0.40cm}&&&&\\[-0.40cm] \hline
\footnotesize b.tumor: $\bra{x}$&\hfill $b^{\mbox{\tiny post}}_x$ &\hspace*{-0.40cm}&&&&\hfill $\varphi^{\mbox{\tiny post}}_x(\emptyset)$ \\
                           &\hfill $1/6$ &\hspace*{-0.40cm}&\hfill $0$&\hfill $0$&\hfill $0$&\hfill $40/1200$ \\ \hline
\footnotesize cold: $\bra{y}$&\hfill $b^{\mbox{\tiny post}}_y$ &\hspace*{-0.40cm}&\hfill $\varphi^{\mbox{\tiny post}}_y(\{x\})$&\hfill $\varphi^{\mbox{\tiny post}}_y(\{x,y\})$& \hfill$\varphi^{\mbox{\tiny post}}_y(\{y\})$&\hfill $\varphi^{\mbox{\tiny post}}_y(\emptyset)$\\
                           &\hfill $5/6$ &\hspace*{-0.40cm}&\hfill $40/1200$&\hfill $10/1200$&\hfill $750/1200$&\hfill $200/1200$\\ \hline
\end{tabular}

\caption{\textbf{Example 3, at the top: A priori}. Venn diagram of the co$\sim$event $\mathscr R=$
``Reality''$=\protect\braket{x|\{x^{\circ}\}}+\protect\braket{\Omega|\{x^{\circ},y^{\circ}\}}+\protect\braket{y|\{y^{\circ}\}}\subseteq
\protect\braket{\Omega|\Omega}$ with the labelling $\protect\braket{\frak X_{\!\mathscr R}|\protect\rsS^{\frak X_{\!\mathscr
R}}}=\protect\braket{\{x,y\}|\{\{x^{\circ}\},\{x^{\circ},y^{\circ}\},\{y^{\circ}\},\emptyset\}}$, with real statistics of ket-events:
$p_{x^{\!\circ}}=150/200$, $p_{y^{\!\circ}}=152/200$, \emph{a priori} believabilities of bra-events: $b_x=1/2, b_y=1/2$, and with certainty
$\bfPhi(\mathscr R)=151/200$. \textbf{At the bottom: A posteriori}. Venn diagram of the co$\sim$event $\mathscr M=(\mathscr H\Delta\,\mathscr R)^c=$
``Match Hypotheses with Reality''$=\protect\braket{x|\{x\}}+\protect\braket{\Omega|\{x,y\}}+\protect\braket{y|\{y\}}\subseteq
\protect\braket{\Omega|\Omega}$ with the labelling $\protect\braket{\frak X_{\!\mathscr M}|\protect\rsS^{\frak X_{\!\mathscr
M}}}=\protect\braket{\{x,y\}|\{\{x\},\{x,y\},\{y\},\emptyset\}}$, with real statistics of ket-events: $p_x=40/200$, $p_y=1$, \emph{a posteriori}
believabilities of bra-events: $b^{\mbox{\tiny post}}_x=1/6, b^{\mbox{\tiny post}}_y=5/6$, and with \emph{a priori} certainty $\bfPhi(\mathscr
M)~=~120/200~=~0.6$, and \emph{a posteriori} certainty $\bfPhi^{\mbox{\tiny post}}(\mathscr M)=1040/1200=13/15\approx
0.867$.\label{example3_TABFinal}}
\end{table}

\begin{table}[!h]
\centering
\begin{tabular}{r|l|l|l|l|l|l|l|}
 &           &\hspace*{-0.40cm}&
 \scriptsize b.tumor \& cold: $\ket{\{x^{\circ},y^{\circ}\}}$&
 \scriptsize b.tumor: $\ket{\{x^{\circ}\}}$&
 \scriptsize b.tumor \& cold: $\ket{\{x^{\circ},y^{\circ}\}}$&
 \scriptsize cold: \hspace*{7pt}$\ket{\{y^{\circ}\}}$\hspace*{7pt}&
 \scriptsize nothing: \hspace*{5pt}$\ket{\emptyset^{\circ}}$\hspace*{5pt}\\ \hline 
 &           &\hspace*{-0.40cm}&
 \hfill $p(\{x^{\circ},y^{\circ}\})$ &
 \hfill $p(\{y^{\circ}\})$ &
 \hfill $p(\{x^{\circ},y^{\circ}\})$ &
 \hfill $p(\{y^{\circ}\})$&
 \hfill $p(\emptyset^{\circ})$ \\ \hline 
 &           &\hspace*{-0.40cm}&
 \hfill$8/200$ &
 \hfill$2/200$ &
 \hfill$120/200$ &
 \hfill$30/200$&
 \hfill$40/200$ \\ \hline
 &&\hspace*{-0.40cm}&&&&&\\[-0.40cm] \hline 
\scriptsize b.tumor: $\bra{x}$&\hfill $b_x$ &\hspace*{-0.40cm}&
\hfill $\varphi_x(\{x^{\circ},y^{\circ}\})$&
&
\hfill $\varphi_x(\{x^{\circ},y^{\circ}\})$&
\hfill $\varphi_x(\{y^{\circ}\})$& \\ 
&1/2&\hspace*{-0.40cm}&
\hfill $4/200$&
\hfill $0$&
\hfill $60/200$&
\hfill 15/200&
\hfill 0 \\ \hline 
\scriptsize cold: $\bra{y}$&\hfill $b_y$ &\hspace*{-0.40cm}&
\hfill $\varphi_y(\{x^{\circ},y^{\circ}\})$&
\hfill $\varphi'_y(\{y^{\circ}\})$&&
\hfill$\varphi''_y(\{y^{\circ}\})$& \\ 
&1/2&\hspace*{-0.40cm}&
\hfill $4/200$&
\hfill $1/200$&
\hfill $0$&
\hfill $15/200$&
\hfill 0\\ \hline 
\end{tabular}

\vspace{10pt}

\begin{tabular}{r|l|l|l|l|l|l|l|}
 &           &\hspace*{-0.40cm}&
 \scriptsize \hspace*{22pt}$\ket{X_1}=\ket{\{x\}}$\hspace*{22pt}&
 \hfill\scriptsize \hspace*{2pt}$\ket{X_2}'=\ket{\{y\}}'$\hspace*{2pt}&
 \hfill\scriptsize  \hspace*{26pt}$\ket{X_3}=\ket{\emptyset}$\hspace*{26pt}&
 \hfill\scriptsize  $\ket{X_2}''=\ket{\{y\}}''$&
 \hfill\scriptsize   $\ket{X_4}=\ket{\{x,y\}}$\\ \hline 
 &           &\hspace*{-0.40cm}&
 \hfill $p(\{x\})$ &
 \hfill $p'(\{y\})$ &
 \hfill $p(\emptyset)$&
 \hfill $p''(\{y\})$&
 \hfill $p(\{x,y\})$ \\ \hline 
 &           &\hspace*{-0.40cm}&
 \hfill$8/200$ &
 \hfill$2/200$ &
 \hfill$120/200$ &
 \hfill$30/200$&
 \hfill$40/200$ \\ \hline
 &&\hspace*{-0.40cm}&&&&&\\[-0.40cm] \hline 
\scriptsize b.tumor: $\bra{x}$&\hfill $b_x$ &\hspace*{-0.40cm}&
\hfill $\varphi_x(\{x\})$&
&
\hfill &
\hfill &
$\varphi_x(\{x,y\})$ \\ 
&1/2&\hspace*{-0.40cm}&
\hfill $4/200$&
\hfill $0$&
\hfill $0$&
\hfill $0$&
\hfill $20/200$\\ \hline 
\scriptsize cold: $\bra{y}$&\hfill $b_y$ &\hspace*{-0.40cm}&
\hfill &
\hfill $\varphi'_y(\{y\})$&
&
\hfill $\varphi''_y(\{y\})$&
$\varphi_y(\{x,y\})$ \\ 
&1/2&\hspace*{-0.40cm}&
\hfill $0$&
\hfill $1/200$&
\hfill $0$&
\hfill $15/200$&
\hfill $20/200$\\ \hline 
 &&\hspace*{-0.40cm}&&&&&\\[-0.40cm] \hline 
\scriptsize b.tumor: $\bra{x}$&\hfill $b^{\mbox{\tiny post}}_x$ &\hspace*{-0.40cm}&
\hfill $\varphi^{\mbox{\tiny post}}_x(\{x\})$&
&
\hfill &
\hfill &
$\varphi^{\mbox{\tiny post}}_x(\{x,y\})$ \\ 
&2/5&\hspace*{-0.40cm}&
\hfill $16/1000$&
\hfill $0$&
\hfill $0$&
\hfill $0$&
\hfill $80/1000$\\ \hline 
\scriptsize cold: $\bra{y}$&\hfill $b^{\mbox{\tiny post}}_y$ &\hspace*{-0.40cm}&
\hfill &
\hfill ${\varphi'}^{\mbox{\tiny post}}_y(\{y\})$&
&
\hfill ${\varphi''}^{\mbox{\tiny post}}_y(\{y\})$&
$\varphi^{\mbox{\tiny post}}_y(\{x,y\})$ \\ 
&3/5&\hspace*{-0.40cm}&
\hfill $0$&
\hfill $6/1000$&
\hfill $0$&
\hfill $90/1000$&
\hfill $120/1000$\\ \hline 
\end{tabular}

\caption{\textbf{Example 4, at the top: A priori}. Venn diagram of the co$\sim$event $\mathscr R=$
``Reality''$=\protect\braket{x|\{x^{\circ}\}}+\protect\braket{\Omega|\{x^{\circ},y^{\circ}\}}+\protect\braket{y|\{y^{\circ}\}}\subseteq
\protect\braket{\Omega|\Omega}$ with the labelling $\protect\braket{\frak X_{\!\mathscr R}|\protect\rsS^{\frak X_{\!\mathscr
R}}}=\protect\braket{\{x,y\}|\{\{x^{\circ}\},\{x^{\circ},y^{\circ}\},\{y^{\circ}\},\emptyset\}}$, with real statistics of ket-events:
$p_{x^{\!\circ}}=158/200$, $p_{y^{\!\circ}}=40/200$, \emph{a priori} believabilities of bra-events: $b_x=1/2, b_y=1/2$, and with certainty
$\bfPhi(\mathscr R)=99/200$. \textbf{At the bottom: A posteriori}. Venn diagram of the co$\sim$event $\mathscr M=(\mathscr H\Delta\,\mathscr R)^c=$
``Match Hypotheses with Reality''$=\protect\braket{x|\{x\}}+\protect\braket{\Omega|\{x,y\}}+\protect\braket{y|\{y\}}\subseteq
\protect\braket{\Omega|\Omega}$ with the labelling $\protect\braket{\frak X_{\!\mathscr M}|\protect\rsS^{\frak X_{\!\mathscr
M}}}=\protect\braket{\{x,y\}|\{\{x\},\{x,y\},\{y\},\emptyset\}}$, with real statistics of ket-events: $p_x=48/200$, $p_y=72/200$, \emph{a posteriori}
believabilities of bra-events: $b^{\mbox{\tiny post}}_x=2/5, b^{\mbox{\tiny post}}_y=3/5$, and with \emph{a priori} certainty $\bfPhi(\mathscr
M)=0.300$, and \emph{a posteriori} certainty $\bfPhi^{\mbox{\tiny post}}(\mathscr M)=0.312$.\label{example4_TABFinal}}
\end{table}

\clearpage
\begin{table}[h!]

\vspace{-0.0cm}

\hspace{0.5cm}
\begin{minipage}{80mm}
\tiny
\begin{equation}\nonumber
\begin{split}
\mbox{\normalsize Example 3: \hspace{2.5cm}} \frak X_{\!\mathscr{R}} &= \{x_{1},x_{2}\},\\
\ket{X_{1}}       &=
\ket{X_1^{\!\times}} \cap \ket{X_1^{\circ}},\\[-1pt]
\ket{X_{2}}       &=
\ket{X_2^{\!\times}} \cap \ket{X_2^{\circ}},\\[-1pt]
\ket{X_{3}}       &=
\ket{X_2^{\!\times}} \cap \ket{X_3^{\circ}},\\[-1pt]
\ket{X_{4}}       &=
\ket{X_4^{\!\times}} \cap \ket{X_4^{\circ}}.
\end{split}
\end{equation}
\end{minipage}

\vspace{-1.0cm}

\hspace{-7.20cm}
\begin{minipage}{80mm}
\normalsize $\left\{b_{x_{1}}^{\mbox{\tiny post}},b_{x_{2}}^{\mbox{\tiny post}}\right\}=\{1/6, \ 5/6\}$
\end{minipage}

\vspace{0.5cm}

\vspace{-25mm}

\setlength{\tabcolsep}{1pt}        
\renewcommand{\arraystretch}{0.45} 
\centering
\hspace{8.65cm}$\overbrace{\hspace{6.75cm}}^{\mbox{\normalsize$\ket{\Omega}$}}$            

\hspace{8.70cm}$\overbrace{\hspace{1.3cm}}^{\ket{X^{\circ}_1}}
                \overbrace{\hspace{1.3cm}}^{\ket{X^{\circ}_2}}
                \overbrace{\hspace{3.3cm}}^{\ket{X^{\circ}_3}}
                \overbrace{\hspace{0.65cm}}^{\ket{X^{\circ}_4}}$

\hspace{6.84cm}\begin{minipage}{0.50cm}
\hspace*{0.30cm}$\bra{\Omega}\!\!\left\{\begin{matrix}\vspace{1.2cm}\end{matrix}\right.$ 
\end{minipage}
\hspace{0.15cm}
\begin{minipage}{0.50cm}
\hspace*{0.15cm}$\bra{x_{1}}\!\left\{\begin{matrix}\vspace{0.47cm}\end{matrix}\right.$\\       
\hspace*{0.15cm}$\bra{x_{2}}\!\left\{\begin{matrix}\vspace{0.47cm}\end{matrix}\right.$
\end{minipage}
\hspace{0.30cm}
\begin{tabular}{|p{0.60cm}p{0.60cm}|p{0.60cm}p{0.60cm}|p{0.60cm}p{0.60cm}p{0.60cm}p{0.60cm}p{0.60cm}|p{0.60cm}|} \hline
  &&&&&&&&& \\[-0.15cm]
  \begin{minipage}{0.50cm}\colorbox{MyWhite}{$\begin{matrix}\vspace{0.1cm}&\vspace{0.1cm}\end{matrix}$}\end{minipage}&
  \begin{minipage}{0.50cm}\colorbox{MyWhite}{$\begin{matrix}\vspace{0.1cm}&\vspace{0.1cm}\end{matrix}$}\end{minipage}&
  \begin{minipage}{0.50cm}\colorbox{MyWhite}{$\begin{matrix}\vspace{0.1cm}&\vspace{0.1cm}\end{matrix}$}\end{minipage}&
  \begin{minipage}{0.50cm}\colorbox{MyWhite}{$\begin{matrix}\vspace{0.1cm}&\vspace{0.1cm}\end{matrix}$}\end{minipage}&
  \begin{minipage}{0.50cm}\colorbox{MyAqua0}{\large\hspace{-0.03cm}$\bigcirc$\hspace{-0.08cm}}\end{minipage}&
  \begin{minipage}{0.50cm}\colorbox{MyAqua0}{\large\hspace{-0.03cm}$\bigcirc$\hspace{-0.08cm}}\end{minipage}&
  \begin{minipage}{0.50cm}\colorbox{MyAqua0}{\large\hspace{-0.03cm}$\bigcirc$\hspace{-0.08cm}}\end{minipage}&
  \begin{minipage}{0.50cm}\colorbox{MyAqua0}{\large\hspace{-0.03cm}$\bigcirc$\hspace{-0.08cm}}\end{minipage}&
  \begin{minipage}{0.50cm}\colorbox{MyAqua0}{\large\hspace{-0.03cm}$\bigcirc$\hspace{-0.08cm}}\end{minipage}&
  \begin{minipage}{0.50cm}\colorbox{MyWhite}{$\begin{matrix}\vspace{0.1cm}&\vspace{0.1cm}\end{matrix}$}\end{minipage}\\
  &&&&&&&&& \\[-0.15cm] \hline
  &&&&&&&&& \\[-0.15cm]
  \begin{minipage}{0.50cm}\colorbox{MyWhite}{$\begin{matrix}\vspace{0.1cm}&\vspace{0.1cm}\end{matrix}$}\end{minipage}&
  \begin{minipage}{0.50cm}\colorbox{MyWhite}{$\begin{matrix}\vspace{0.1cm}&\vspace{0.1cm}\end{matrix}$}\end{minipage}&
  \begin{minipage}{0.50cm}\colorbox{MyAqua0}{\large\hspace{-0.03cm}$\bigcirc$\hspace{-0.08cm}}\end{minipage}&
  \begin{minipage}{0.50cm}\colorbox{MyAqua0}{\large\hspace{-0.03cm}$\bigcirc$\hspace{-0.08cm}}\end{minipage}&
  \begin{minipage}{0.50cm}\colorbox{MyAqua0}{\large\hspace{-0.03cm}$\bigcirc$\hspace{-0.08cm}}\end{minipage}&
  \begin{minipage}{0.50cm}\colorbox{MyAqua0}{\large\hspace{-0.03cm}$\bigcirc$\hspace{-0.08cm}}\end{minipage}&
  \begin{minipage}{0.50cm}\colorbox{MyAqua0}{\large\hspace{-0.03cm}$\bigcirc$\hspace{-0.08cm}}\end{minipage}&
  \begin{minipage}{0.50cm}\colorbox{MyAqua0}{\large\hspace{-0.03cm}$\bigcirc$\hspace{-0.08cm}}\end{minipage}&
  \begin{minipage}{0.50cm}\colorbox{MyAqua0}{\large\hspace{-0.03cm}$\bigcirc$\hspace{-0.08cm}}\end{minipage}&
  \begin{minipage}{0.50cm}\colorbox{MyWhite}{$\begin{matrix}\vspace{0.1cm}&\vspace{0.1cm}\end{matrix}$}\end{minipage}\\
  &&&&&&&&& \\[-0.15cm] \hline
\end{tabular}

\vspace{0.0cm}
\hspace{0.85cm}$\overbrace{\hspace{6.7cm}}^{\mbox{\normalsize$\ket{\Omega}$}}$  \hspace{0.78cm}  $\overbrace{\hspace{6.7cm}}^{\mbox{\normalsize$\ket{\Omega}$}}$  

\hspace{0.85cm}$\overbrace{\hspace{1.3cm}}^{\ket{X^{\times}_1}}
                \overbrace{\hspace{1.3cm}}^{\ket{X^{\times}_2}}
                \overbrace{\hspace{3.3cm}}^{\ket{X^{\times}_3}}
                \overbrace{\hspace{0.65cm}}^{\ket{X^{\times}_4}}$
\hspace{0.89cm}$\overbrace{\hspace{1.3cm}}^{\ket{X_1}}
                \overbrace{\hspace{1.3cm}}^{\ket{X_2}}
                \overbrace{\hspace{3.3cm}}^{\ket{X_3}}
                \overbrace{\hspace{0.65cm}}^{\ket{X_4}}$

\hspace{-0.82cm}\begin{minipage}{0.50cm} $\bra{\Omega}\!\!\left\{\begin{matrix}\vspace{1.2cm}\end{matrix}\right.$
\end{minipage}
\hspace{0.15cm}
\begin{minipage}{0.50cm}
\hspace*{0.00cm}$\bra{x_{1}}\!\left\{\begin{matrix}\vspace{0.49cm}\end{matrix}\right.$\\       
\hspace*{0.00cm}$\bra{x_{2}}\!\left\{\begin{matrix}\vspace{0.49cm}\end{matrix}\right.$
\end{minipage}
\hspace{0.15cm}
\begin{tabular}{|p{0.60cm}p{0.60cm}|p{0.60cm}p{0.60cm}|p{0.60cm}p{0.60cm}p{0.60cm}p{0.60cm}p{0.60cm}|p{0.60cm}|} \hline
  &&&&&&&&& \\[-0.15cm]
  \begin{minipage}{0.50cm}\colorbox{MyGreen}{\LARGE\hspace{-0.03cm}$\times$\hspace{-0.08cm}}\end{minipage}&
  \begin{minipage}{0.50cm}\colorbox{MyGreen}{\LARGE\hspace{-0.03cm}$\times$\hspace{-0.08cm}}\end{minipage}&
  \begin{minipage}{0.50cm}\colorbox{MyGreen}{\LARGE\hspace{-0.03cm}$\times$\hspace{-0.08cm}}\end{minipage}&
  \begin{minipage}{0.50cm}\colorbox{MyGreen}{\LARGE\hspace{-0.03cm}$\times$\hspace{-0.08cm}}\end{minipage}&
  \begin{minipage}{0.50cm}\colorbox{MyWhite}{$\begin{matrix}\vspace{0.1cm}&\vspace{0.1cm}\end{matrix}$}\end{minipage}&
  \begin{minipage}{0.50cm}\colorbox{MyWhite}{$\begin{matrix}\vspace{0.1cm}&\vspace{0.1cm}\end{matrix}$}\end{minipage}&
  \begin{minipage}{0.50cm}\colorbox{MyWhite}{$\begin{matrix}\vspace{0.1cm}&\vspace{0.1cm}\end{matrix}$}\end{minipage}&
  \begin{minipage}{0.50cm}\colorbox{MyWhite}{$\begin{matrix}\vspace{0.1cm}&\vspace{0.1cm}\end{matrix}$}\end{minipage}&
  \begin{minipage}{0.50cm}\colorbox{MyWhite}{$\begin{matrix}\vspace{0.1cm}&\vspace{0.1cm}\end{matrix}$}\end{minipage}&
  \begin{minipage}{0.50cm}\colorbox{MyWhite}{$\begin{matrix}\vspace{0.1cm}&\vspace{0.1cm}\end{matrix}$}\end{minipage}\\
  &&&&&&&&& \\[-0.15cm] \hline
  &&&&&&&&& \\[-0.15cm]
  \begin{minipage}{0.50cm}\colorbox{MyWhite}{$\begin{matrix}\vspace{0.1cm}&\vspace{0.1cm}\end{matrix}$}\end{minipage}&
  \begin{minipage}{0.50cm}\colorbox{MyWhite}{$\begin{matrix}\vspace{0.1cm}&\vspace{0.1cm}\end{matrix}$}\end{minipage}&
  \begin{minipage}{0.50cm}\colorbox{MyGreen}{\LARGE\hspace{-0.03cm}$\times$\hspace{-0.08cm}}\end{minipage}&
  \begin{minipage}{0.50cm}\colorbox{MyGreen}{\LARGE\hspace{-0.03cm}$\times$\hspace{-0.08cm}}\end{minipage}&
  \begin{minipage}{0.50cm}\colorbox{MyGreen}{\LARGE\hspace{-0.03cm}$\times$\hspace{-0.08cm}}\end{minipage}&
  \begin{minipage}{0.50cm}\colorbox{MyGreen}{\LARGE\hspace{-0.03cm}$\times$\hspace{-0.08cm}}\end{minipage}&
  \begin{minipage}{0.50cm}\colorbox{MyGreen}{\LARGE\hspace{-0.03cm}$\times$\hspace{-0.08cm}}\end{minipage}&
  \begin{minipage}{0.50cm}\colorbox{MyGreen}{\LARGE\hspace{-0.03cm}$\times$\hspace{-0.08cm}}\end{minipage}&
  \begin{minipage}{0.50cm}\colorbox{MyGreen}{\LARGE\hspace{-0.03cm}$\times$\hspace{-0.08cm}}\end{minipage}&
  \begin{minipage}{0.50cm}\colorbox{MyWhite}{$\begin{matrix}\vspace{0.1cm}&\vspace{0.1cm}\end{matrix}$}\end{minipage}\\
  &&&&&&&&& \\[-0.15cm] \hline
\end{tabular}
\begin{minipage}{0.50cm}
\hspace*{0.00cm}$\bra{x_{1}}\!\left\{\begin{matrix}\vspace{0.49cm}\end{matrix}\right.$\\       
\hspace*{0.00cm}$\bra{x_{2}}\!\left\{\begin{matrix}\vspace{0.49cm}\end{matrix}\right.$
\end{minipage}
\hspace{0.15cm}
\begin{tabular}{|p{0.60cm}p{0.60cm}|p{0.60cm}p{0.60cm}|p{0.60cm}p{0.60cm}p{0.60cm}p{0.60cm}p{0.60cm}|p{0.60cm}|} \hline
  &&&&&&&&& \\[-0.15cm]
  \begin{minipage}{0.50cm}\colorbox{MyWhite}{\LARGE\hspace{-0.03cm}$\times$\hspace{-0.03cm}}\end{minipage}&
  \begin{minipage}{0.50cm}\colorbox{MyWhite}{\LARGE\hspace{-0.03cm}$\times$\hspace{-0.03cm}}\end{minipage}&
  \begin{minipage}{0.50cm}\colorbox{MyWhite}{\LARGE\hspace{-0.03cm}$\times$\hspace{-0.03cm}}\end{minipage}&
  \begin{minipage}{0.50cm}\colorbox{MyWhite}{\LARGE\hspace{-0.03cm}$\times$\hspace{-0.03cm}}\end{minipage}&
  \begin{minipage}{0.50cm}\colorbox{MyWhite}{\large\hspace{-0.03cm}$\bigcirc$\hspace{-0.03cm}}\end{minipage}&
  \begin{minipage}{0.50cm}\colorbox{MyWhite}{\large\hspace{-0.03cm}$\bigcirc$\hspace{-0.03cm}}\end{minipage}&
  \begin{minipage}{0.50cm}\colorbox{MyWhite}{\large\hspace{-0.03cm}$\bigcirc$\hspace{-0.03cm}}\end{minipage}&
  \begin{minipage}{0.50cm}\colorbox{MyWhite}{\large\hspace{-0.03cm}$\bigcirc$\hspace{-0.03cm}}\end{minipage}&
  \begin{minipage}{0.50cm}\colorbox{MyWhite}{\large\hspace{-0.03cm}$\bigcirc$\hspace{-0.03cm}}\end{minipage}&
  \begin{minipage}{0.50cm}\colorbox{MyOrangeBayes}{$\begin{matrix}\vspace{0.1cm}&\vspace{0.1cm}\end{matrix}$}\end{minipage}\\
  &&&&&&&&& \\[-0.15cm] \hline
  &&&&&&&&& \\[-0.15cm]
  \begin{minipage}{0.50cm}\colorbox{MyOrangeBayes}{$\begin{matrix}\vspace{0.1cm}&\vspace{0.1cm}\end{matrix}$}\end{minipage}&  \begin{minipage}{0.50cm}\colorbox{MyOrangeBayes}{$\begin{matrix}\vspace{0.1cm}&\vspace{0.1cm}\end{matrix}$}\end{minipage}&  \begin{minipage}{0.50cm}\colorbox{MyOrangeBayes}{\LARGE\hspace{-0.03cm}$\otimes$\hspace{-0.08cm}}\end{minipage}&
  \begin{minipage}{0.50cm}\colorbox{MyOrangeBayes}{\LARGE\hspace{-0.03cm}$\otimes$\hspace{-0.08cm}}\end{minipage}&
  \begin{minipage}{0.50cm}\colorbox{MyOrangeBayes}{\LARGE\hspace{-0.03cm}$\otimes$\hspace{-0.08cm}}\end{minipage}&
  \begin{minipage}{0.50cm}\colorbox{MyOrangeBayes}{\LARGE\hspace{-0.03cm}$\otimes$\hspace{-0.08cm}}\end{minipage}&
  \begin{minipage}{0.50cm}\colorbox{MyOrangeBayes}{\LARGE\hspace{-0.03cm}$\otimes$\hspace{-0.08cm}}\end{minipage}&
  \begin{minipage}{0.50cm}\colorbox{MyOrangeBayes}{\LARGE\hspace{-0.03cm}$\otimes$\hspace{-0.08cm}}\end{minipage}&
  \begin{minipage}{0.50cm}\colorbox{MyOrangeBayes}{\LARGE\hspace{-0.03cm}$\otimes$\hspace{-0.08cm}}\end{minipage}&
  \begin{minipage}{0.50cm}\colorbox{MyOrangeBayes}{$\begin{matrix}\vspace{0.1cm}&\vspace{0.1cm}\end{matrix}$}\end{minipage}\\
  &&&&&&&&& \\[-0.15cm] \hline
\end{tabular}

\vspace{-3.88cm}

\hspace{8.65cm}{\Huge $\mathscr R$}

\vspace{2.20cm}

\hspace{0.79cm}{\Huge $\mathscr H$}\hspace{70mm}{\Huge $\mathscr M$}

\vspace{0.60cm}

\hspace{-8.5cm}
\begin{minipage}{80mm}
\tiny
\begin{equation}\nonumber
\begin{split}
\mbox{\normalsize Example 4: \hspace{2.5cm}} \frak X_\mathscr{R} &= \{x_{1},x_{2}\},\\
\ket{X_{1}}       &=
\ket{X_1^{\!\times}} \cap \ket{X_1^{\circ}},\\[-1pt]
\ket{X_{2}}       &=
\ket{X_2^{\!\times}} \cap \ket{X_1^{\circ}},\\[-1pt]
\ket{X_{3}}       &=
\ket{X_3^{\!\times}} \cap \ket{X_3^{\circ}},\\[-1pt]
\ket{X_{4}}       &=
\ket{X_3^{\!\times}} \cap \ket{X_1^{\circ}},\\[-1pt]
\ket{X_{5}}       &=
\ket{X_4^{\!\times}} \cap \ket{X_4^{\circ}}.
\end{split}
\end{equation}
\end{minipage}

\vspace{-1.5cm}

\hspace{-7.20cm}
\begin{minipage}{80mm}
\normalsize $\left\{b_{x_{1}}^{\mbox{\tiny post}},b_{x_{2}}^{\mbox{\tiny post}}\right\}=\{2/5, \ 3/5\}$
\end{minipage}

\vspace{0.75cm}

\vspace{-25mm}

\setlength{\tabcolsep}{1pt}        
\renewcommand{\arraystretch}{0.45} 
\centering
\hspace{8.65cm}$\overbrace{\hspace{6.75cm}}^{\mbox{\normalsize$\ket{\Omega}$}}$            

\hspace{8.70cm}$\overbrace{\hspace{1.3cm}}^{\ket{X^{\circ}_1}}
                \overbrace{\hspace{1.3cm}}^{\ket{X^{\circ}_2}}
                \overbrace{\hspace{2.0cm}}^{\ket{X^{\circ}_3}}
                \overbrace{\hspace{1.3cm}}^{\ket{X^{\circ}_1}}
                \overbrace{\hspace{0.65cm}}^{\ket{X^{\circ}_4}}$

\hspace{6.84cm}\begin{minipage}{0.50cm}
\hspace*{0.30cm}$\bra{\Omega}\!\!\left\{\begin{matrix}\vspace{1.2cm}\end{matrix}\right.$ 
\end{minipage}
\hspace{0.15cm}
\begin{minipage}{0.50cm}
\hspace*{0.15cm}$\bra{x_{1}}\!\left\{\begin{matrix}\vspace{0.47cm}\end{matrix}\right.$\\       
\hspace*{0.15cm}$\bra{x_{2}}\!\left\{\begin{matrix}\vspace{0.47cm}\end{matrix}\right.$
\end{minipage}
\hspace{0.30cm}
\begin{tabular}{|p{0.60cm}p{0.60cm}|p{0.60cm}p{0.60cm}|p{0.60cm}p{0.60cm}p{0.60cm}|p{0.60cm}p{0.60cm}|p{0.60cm}|} \hline
  &&&&&&&&& \\[-0.15cm]
  \begin{minipage}{0.50cm}\colorbox{MyAqua0}{\large\hspace{-0.03cm}$\bigcirc$\hspace{-0.08cm}}\end{minipage}&
  \begin{minipage}{0.50cm}\colorbox{MyAqua0}{\large\hspace{-0.03cm}$\bigcirc$\hspace{-0.08cm}}\end{minipage}&
  \begin{minipage}{0.50cm}\colorbox{MyWhite}{$\begin{matrix}\vspace{0.1cm}&\vspace{0.1cm}\end{matrix}$}\end{minipage}&
  \begin{minipage}{0.50cm}\colorbox{MyWhite}{$\begin{matrix}\vspace{0.1cm}&\vspace{0.1cm}\end{matrix}$}\end{minipage}&
  \begin{minipage}{0.50cm}\colorbox{MyAqua0}{\large\hspace{-0.03cm}$\bigcirc$\hspace{-0.08cm}}\end{minipage}&
  \begin{minipage}{0.50cm}\colorbox{MyAqua0}{\large\hspace{-0.03cm}$\bigcirc$\hspace{-0.08cm}}\end{minipage}&
  \begin{minipage}{0.50cm}\colorbox{MyAqua0}{\large\hspace{-0.03cm}$\bigcirc$\hspace{-0.08cm}}\end{minipage}&
  \begin{minipage}{0.50cm}\colorbox{MyAqua0}{\large\hspace{-0.03cm}$\bigcirc$\hspace{-0.08cm}}\end{minipage}&
  \begin{minipage}{0.50cm}\colorbox{MyAqua0}{\large\hspace{-0.03cm}$\bigcirc$\hspace{-0.08cm}}\end{minipage}&
  \begin{minipage}{0.50cm}\colorbox{MyWhite}{$\begin{matrix}\vspace{0.1cm}&\vspace{0.1cm}\end{matrix}$}\end{minipage}\\
  &&&&&&&&& \\[-0.15cm] \hline
  &&&&&&&&& \\[-0.15cm]
  \begin{minipage}{0.50cm}\colorbox{MyAqua0}{\large\hspace{-0.03cm}$\bigcirc$\hspace{-0.08cm}}\end{minipage}&
  \begin{minipage}{0.50cm}\colorbox{MyAqua0}{\large\hspace{-0.03cm}$\bigcirc$\hspace{-0.08cm}}\end{minipage}&
  \begin{minipage}{0.50cm}\colorbox{MyAqua0}{\large\hspace{-0.03cm}$\bigcirc$\hspace{-0.08cm}}\end{minipage}&
  \begin{minipage}{0.50cm}\colorbox{MyAqua0}{\large\hspace{-0.03cm}$\bigcirc$\hspace{-0.08cm}}\end{minipage}&
  \begin{minipage}{0.50cm}\colorbox{MyWhite}{$\begin{matrix}\vspace{0.1cm}&\vspace{0.1cm}\end{matrix}$}\end{minipage}&
  \begin{minipage}{0.50cm}\colorbox{MyWhite}{$\begin{matrix}\vspace{0.1cm}&\vspace{0.1cm}\end{matrix}$}\end{minipage}&
  \begin{minipage}{0.50cm}\colorbox{MyWhite}{$\begin{matrix}\vspace{0.1cm}&\vspace{0.1cm}\end{matrix}$}\end{minipage}&
  \begin{minipage}{0.50cm}\colorbox{MyAqua0}{\large\hspace{-0.03cm}$\bigcirc$\hspace{-0.08cm}}\end{minipage}&
  \begin{minipage}{0.50cm}\colorbox{MyAqua0}{\large\hspace{-0.03cm}$\bigcirc$\hspace{-0.08cm}}\end{minipage}&
  \begin{minipage}{0.50cm}\colorbox{MyWhite}{$\begin{matrix}\vspace{0.1cm}&\vspace{0.1cm}\end{matrix}$}\end{minipage}\\
  &&&&&&&&& \\[-0.15cm] \hline
\end{tabular}

\vspace{0.0cm}
\hspace{0.85cm}$\overbrace{\hspace{6.7cm}}^{\mbox{\normalsize$\ket{\Omega}$}}$  \hspace{0.78cm}  $\overbrace{\hspace{6.7cm}}^{\mbox{\normalsize$\ket{\Omega}$}}$  

\hspace{0.85cm}$\overbrace{\hspace{1.3cm}}^{\ket{X^{\times}_1}}
                \overbrace{\hspace{1.3cm}}^{\ket{X^{\times}_2}}
                \overbrace{\hspace{3.3cm}}^{\ket{X^{\times}_3}}
                \overbrace{\hspace{0.65cm}}^{\ket{X^{\times}_4}}$
\hspace{0.89cm}$\overbrace{\hspace{1.3cm}}^{\ket{X_1}}
                \overbrace{\hspace{1.3cm}}^{\ket{X_2}}
                \overbrace{\hspace{2.0cm}}^{\ket{X_3}}
                \overbrace{\hspace{1.3cm}}^{\ket{X_4}}
                \overbrace{\hspace{0.65cm}}^{\ket{X_5}}$

\hspace{-0.82cm}\begin{minipage}{0.50cm} $\bra{\Omega}\!\!\left\{\begin{matrix}\vspace{1.2cm}\end{matrix}\right.$
\end{minipage}
\hspace{0.15cm}
\begin{minipage}{0.50cm}
\hspace*{0.00cm}$\bra{x_{1}}\!\left\{\begin{matrix}\vspace{0.49cm}\end{matrix}\right.$\\       
\hspace*{0.00cm}$\bra{x_{2}}\!\left\{\begin{matrix}\vspace{0.49cm}\end{matrix}\right.$
\end{minipage}
\hspace{0.15cm}
\begin{tabular}{|p{0.60cm}p{0.60cm}|p{0.60cm}p{0.60cm}|p{0.60cm}p{0.60cm}p{0.60cm}p{0.60cm}p{0.60cm}|p{0.60cm}|} \hline
  &&&&&&&&& \\[-0.15cm]
  \begin{minipage}{0.50cm}\colorbox{MyGreen}{\LARGE\hspace{-0.03cm}$\times$\hspace{-0.08cm}}\end{minipage}&
  \begin{minipage}{0.50cm}\colorbox{MyGreen}{\LARGE\hspace{-0.03cm}$\times$\hspace{-0.08cm}}\end{minipage}&
  \begin{minipage}{0.50cm}\colorbox{MyGreen}{\LARGE\hspace{-0.03cm}$\times$\hspace{-0.08cm}}\end{minipage}&
  \begin{minipage}{0.50cm}\colorbox{MyGreen}{\LARGE\hspace{-0.03cm}$\times$\hspace{-0.08cm}}\end{minipage}&
  \begin{minipage}{0.50cm}\colorbox{MyWhite}{$\begin{matrix}\vspace{0.1cm}&\vspace{0.1cm}\end{matrix}$}\end{minipage}&
  \begin{minipage}{0.50cm}\colorbox{MyWhite}{$\begin{matrix}\vspace{0.1cm}&\vspace{0.1cm}\end{matrix}$}\end{minipage}&
  \begin{minipage}{0.50cm}\colorbox{MyWhite}{$\begin{matrix}\vspace{0.1cm}&\vspace{0.1cm}\end{matrix}$}\end{minipage}&
  \begin{minipage}{0.50cm}\colorbox{MyWhite}{$\begin{matrix}\vspace{0.1cm}&\vspace{0.1cm}\end{matrix}$}\end{minipage}&
  \begin{minipage}{0.50cm}\colorbox{MyWhite}{$\begin{matrix}\vspace{0.1cm}&\vspace{0.1cm}\end{matrix}$}\end{minipage}&
  \begin{minipage}{0.50cm}\colorbox{MyWhite}{$\begin{matrix}\vspace{0.1cm}&\vspace{0.1cm}\end{matrix}$}\end{minipage}\\
  &&&&&&&&& \\[-0.15cm] \hline
  &&&&&&&&& \\[-0.15cm]
  \begin{minipage}{0.50cm}\colorbox{MyWhite}{$\begin{matrix}\vspace{0.1cm}&\vspace{0.1cm}\end{matrix}$}\end{minipage}&
  \begin{minipage}{0.50cm}\colorbox{MyWhite}{$\begin{matrix}\vspace{0.1cm}&\vspace{0.1cm}\end{matrix}$}\end{minipage}&
  \begin{minipage}{0.50cm}\colorbox{MyGreen}{\LARGE\hspace{-0.03cm}$\times$\hspace{-0.08cm}}\end{minipage}&
  \begin{minipage}{0.50cm}\colorbox{MyGreen}{\LARGE\hspace{-0.03cm}$\times$\hspace{-0.08cm}}\end{minipage}&
  \begin{minipage}{0.50cm}\colorbox{MyGreen}{\LARGE\hspace{-0.03cm}$\times$\hspace{-0.08cm}}\end{minipage}&
  \begin{minipage}{0.50cm}\colorbox{MyGreen}{\LARGE\hspace{-0.03cm}$\times$\hspace{-0.08cm}}\end{minipage}&
  \begin{minipage}{0.50cm}\colorbox{MyGreen}{\LARGE\hspace{-0.03cm}$\times$\hspace{-0.08cm}}\end{minipage}&
  \begin{minipage}{0.50cm}\colorbox{MyGreen}{\LARGE\hspace{-0.03cm}$\times$\hspace{-0.08cm}}\end{minipage}&
  \begin{minipage}{0.50cm}\colorbox{MyGreen}{\LARGE\hspace{-0.03cm}$\times$\hspace{-0.08cm}}\end{minipage}&
  \begin{minipage}{0.50cm}\colorbox{MyWhite}{$\begin{matrix}\vspace{0.1cm}&\vspace{0.1cm}\end{matrix}$}\end{minipage}\\
  &&&&&&&&& \\[-0.15cm] \hline
\end{tabular}
\begin{minipage}{0.50cm}
\hspace*{0.00cm}$\bra{x_{1}}\!\left\{\begin{matrix}\vspace{0.49cm}\end{matrix}\right.$\\       
\hspace*{0.00cm}$\bra{x_{2}}\!\left\{\begin{matrix}\vspace{0.49cm}\end{matrix}\right.$
\end{minipage}
\hspace{0.15cm}
\begin{tabular}{|p{0.60cm}p{0.60cm}|p{0.60cm}p{0.60cm}|p{0.60cm}p{0.60cm}p{0.60cm}|p{0.60cm}p{0.60cm}|p{0.60cm}|} \hline
  &&&&&&&&& \\[-0.15cm]
  \begin{minipage}{0.50cm}\colorbox{MyOrangeBayes}{\LARGE\hspace{-0.03cm}$\otimes$\hspace{-0.08cm}}\end{minipage}&
  \begin{minipage}{0.50cm}\colorbox{MyOrangeBayes}{\LARGE\hspace{-0.03cm}$\otimes$\hspace{-0.08cm}}\end{minipage}&
  \begin{minipage}{0.50cm}\colorbox{MyWhite}{\LARGE\hspace{-0.05cm}$\times$\hspace{-0.05cm}}\end{minipage}&
  \begin{minipage}{0.50cm}\colorbox{MyWhite}{\LARGE\hspace{-0.05cm}$\times$\hspace{-0.05cm}}\end{minipage}&
  \begin{minipage}{0.50cm}\colorbox{MyWhite}{\large\hspace{-0.03cm}$\bigcirc$\hspace{-0.03cm}}\end{minipage}&
  \begin{minipage}{0.50cm}\colorbox{MyWhite}{\large\hspace{-0.03cm}$\bigcirc$\hspace{-0.03cm}}\end{minipage}&
  \begin{minipage}{0.50cm}\colorbox{MyWhite}{\large\hspace{-0.03cm}$\bigcirc$\hspace{-0.03cm}}\end{minipage}&
  \begin{minipage}{0.50cm}\colorbox{MyWhite}{\large\hspace{-0.03cm}$\bigcirc$\hspace{-0.03cm}}\end{minipage}&
  \begin{minipage}{0.50cm}\colorbox{MyWhite}{\large\hspace{-0.03cm}$\bigcirc$\hspace{-0.03cm}}\end{minipage}&
  \begin{minipage}{0.50cm}\colorbox{MyOrangeBayes}{$\begin{matrix}\vspace{0.1cm}&\vspace{0.1cm}\end{matrix}$}\end{minipage}\\
  &&&&&&&&& \\[-0.15cm] \hline
  &&&&&&&&& \\[-0.15cm]
  \begin{minipage}{0.50cm}\colorbox{MyWhite}{\large\hspace{-0.03cm}$\bigcirc$\hspace{-0.03cm}}\end{minipage}&
  \begin{minipage}{0.50cm}\colorbox{MyWhite}{\large\hspace{-0.03cm}$\bigcirc$\hspace{-0.03cm}}\end{minipage}&
  \begin{minipage}{0.50cm}\colorbox{MyOrangeBayes}{\LARGE\hspace{-0.03cm}$\otimes$\hspace{-0.08cm}}\end{minipage}&
  \begin{minipage}{0.50cm}\colorbox{MyOrangeBayes}{\LARGE\hspace{-0.03cm}$\otimes$\hspace{-0.08cm}}\end{minipage}&
  \begin{minipage}{0.50cm}\colorbox{MyWhite}{\LARGE\hspace{-0.05cm}$\times$\hspace{-0.05cm}}\end{minipage}&
  \begin{minipage}{0.50cm}\colorbox{MyWhite}{\LARGE\hspace{-0.05cm}$\times$\hspace{-0.05cm}}\end{minipage}&
  \begin{minipage}{0.50cm}\colorbox{MyWhite}{\LARGE\hspace{-0.05cm}$\times$\hspace{-0.05cm}}\end{minipage}&
  \begin{minipage}{0.50cm}\colorbox{MyOrangeBayes}{\LARGE\hspace{-0.03cm}$\otimes$\hspace{-0.08cm}}\end{minipage}&
  \begin{minipage}{0.50cm}\colorbox{MyOrangeBayes}{\LARGE\hspace{-0.03cm}$\otimes$\hspace{-0.08cm}}\end{minipage}&
  \begin{minipage}{0.50cm}\colorbox{MyOrangeBayes}{$\begin{matrix}\vspace{0.1cm}&\vspace{0.1cm}\end{matrix}$}\end{minipage}\\
  &&&&&&&&& \\[-0.15cm] \hline
\end{tabular}

\vspace{-3.88cm}

\hspace{8.65cm}{\Huge $\mathscr R$}

\vspace{2.20cm}

\hspace{0.79cm}{\Huge $\mathscr H$}\hspace{70mm}{\Huge $\mathscr M$}

\vspace{0.50cm}

\caption{Venn diagrams of the three co$\sim$events  $\mathscr H, \mathscr{R}$, and $\mathscr{M}=(\mathscr H\Delta\,\mathscr R)^c$ $\Big($
\colorbox{MyGreen}{\large$\times$} , \colorbox{MyAqua0}{\small$\bigcirc$} ,
\colorbox{MyOrangeBayes}{\large$\otimes$}\hspace{1pt}\colorbox{MyOrangeBayes}{\protect\phantom{\large$\otimes$}} $\Big)$ in the co$\sim$event-based
Bayesian scheme. \label{2x10Bayes}}
\end{table}

In the third example (Tables \ref{example1_priorTABxFinal}, \ref{example3_TABFinal}) the \textbf{prior} certainty is 0.6 and \emph{with \textbf{a
posteriori} certainty 0.867} the posteriori believability distribution of your hypotheses has the form:
\begin{eqnarray}
\label{example3_posterior_x}
b^{\mbox{\tiny post}}_x&=&b_x\frac{p_x}{\bfPhi(\mathscr M)}=1/6,\\[-3pt]
\label{example3_posterior_y}
b^{\mbox{\tiny post}}_y&=&b_y\frac{p_y}{\bfPhi(\mathscr M)}=5/6,\\[-3pt]
\label{example3_posterior_phi}
\bfPhi^{\mbox{\tiny post}}(\mathscr M)&=&\sum_{z\in\{x,y\}}\sum_{Z\in\{\{x\},\{x,y\},\{y\},\emptyset\}}\varphi^{\mbox{\tiny post}}_z(Z)=0.867.
\end{eqnarray}

In the fourth example (Tables \ref{example1_priorTABxFinal}, \ref{example4_TABFinal}) the \textbf{prior} certainty is 0.6 and \emph{with \textbf{a
posteriori} certainty 0.867} the posteriori believability distribution of your hypotheses has the form:
\begin{eqnarray}
\label{example4_posterior_x}
b^{\mbox{\tiny post}}_x&=&b_x\frac{p_x}{\bfPhi(\mathscr M)}=2/5,\\[-3pt]
\label{example4_posterior_y}
b^{\mbox{\tiny post}}_y&=&b_y\frac{p_y}{\bfPhi(\mathscr M)}=3/5,\\[-3pt]
\label{example4_posterior_phi}
\bfPhi^{\mbox{\tiny post}}(\mathscr M)&=&\sum_{z\in\{x,y\}}\sum_{Z\in\{\{x\},\{x,y\},\{y\},\emptyset\}}\varphi^{\mbox{\tiny post}}_z(Z)=0.312.
\end{eqnarray}

\section{Once more example: ``10 tasters trying wine from 10 bottles''}

Imagine 10 tasters $\{x_1,\ldots,x_{10}\}$ trying wine from 10 bottles whose winemakers are known. First, each taster, after recognizing the wine
maker of the wine, puts forward hypothesis whether this wine is good or not. As a result of such an experience-random experiment the co$\sim$event
\begin{equation}\label{HHHwine}
\mathscr H= \mbox{ ``\emph{Hypotheses}'' }=\braket{x_1|x^{\!\times}_1}+\ldots+\braket{x_{10}|x^{\!\times}_{10}}\subseteq\braket{\Omega|\Omega}
\end{equation}
occurs. Here $\ket{x_j^{\!\times}}\subseteq\ket{\Omega}$ is the $j$-th subset of 10 bottles such that the $j$-th taster assumes the wine from this
subset of bottles would be good. Second, each taster tasting wine from each bottle. As a result of such an experience-random experiment the
co$\sim$event
\begin{equation}\label{RRRwine}
\mathscr R= \mbox{ ``\emph{Reality}'' } =\braket{x_1|x^{\!\circ}_1}+\ldots+\braket{x_{10}|x^{\!\circ}_{10}}\subseteq\braket{\Omega|\Omega}
\end{equation}
occurs. Here $\ket{x_j^{\!\circ}}\subseteq\ket{\Omega}$ is the $j$-th subset of 10 bottles such that the $j$-th taster considers the wine from this
subset of bottles is good. After comparing the co$\sim$events $\mathscr H$ and $\mathscr R$ the co$\sim$event
\begin{equation}\label{MMM}
\mathscr M=(\mathscr H\Delta\,\mathscr R)^c= \mbox{ ``\emph{Match Hypotheses with Reality}'' } =
\braket{x_1|(x^{\!\times}_1\Delta\,x^{\!\circ}_1)^c}+\ldots+\braket{x_{M}|(x^{\!\times}_1\Delta\,x^{\!\circ}_M)^c}
\end{equation}
occurs.

\begin{table}[h!]

\vspace{0.0cm}

\hspace{0.5cm}
\begin{minipage}{80mm}
\begin{equation}\nonumber
\begin{split}
&\mbox{\normalsize Example 5:}\\
&\frak X_\mathscr{R} = \{x_{1},x_{2},x_{3},x_{4},x_{5},x_{6},x_{7},x_{8},x_{9},x_{10}\},
\end{split}
\end{equation}
\begin{equation}\nonumber
\begin{split}
\ket{X_{1}}       &=
\ket{X_1^{\!\times}},\\[-1pt]
\ket{X_{2}}       &=
\ket{X_2^{\!\times}} \cap \ket{X_1^{\circ}},\\[-1pt]
\ket{X_{3}}       &=
\ket{X_2^{\!\times}} \cap \ket{X_2^{\circ}},\\[-1pt]
\ket{X_{4}}       &=
\ket{X_3^{\!\times}} \cap \ket{X_2^{\circ}},\\[-1pt]
\ket{X_{5}}       &=
\ket{X_3^{\!\times}} \cap \ket{X_3^{\circ}},\\[-1pt]
\ket{X_{6}}       &=
\ket{X_4^{\!\times}} \cap \ket{X_3^{\circ}},\\[-1pt]
\ket{X_{7}}       &=
\ket{X_4^{\!\times}} \cap \ket{X_4^{\circ}},\\[-1pt]
\ket{X_{8}}       &=
\ket{X_5^{\!\times}} \cap \ket{X_4^{\circ}},\\[-1pt]
\ket{X_{9}}       &=
\ket{X_5^{\!\times}} \cap \ket{X_5^{\circ}},\\[-1pt]
\ket{X_{10}}       &=
\ket{X_{6}^{\!\times}}.
\end{split}
\end{equation}
\end{minipage}

\vspace{-70mm}

\setlength{\tabcolsep}{1pt}        
\renewcommand{\arraystretch}{0.45} 
\centering
\hspace{8.65cm}$\overbrace{\hspace{6.83cm}}^{\mbox{\normalsize$\ket{\Omega}$}}$            

\hspace{8.70cm}$\overbrace{\hspace{1.3cm}}^{\ket{X^{\circ}_1}}             
                \overbrace{\hspace{1.3cm}}^{\ket{X^{\circ}_2}}
                \overbrace{\hspace{1.3cm}}^{\ket{X^{\circ}_3}}
                \overbrace{\hspace{1.3cm}}^{\ket{X^{\circ}_4}}
                \overbrace{\hspace{1.3cm}}^{\ket{X^{\circ}_5}}$      

\hspace{6.84cm}\begin{minipage}{0.50cm}
\hspace*{0.12cm}$\bra{\Omega}\!\!\left\{\begin{matrix}\vspace{6.74cm}\end{matrix}\right.$ 
\end{minipage}
\hspace{0.15cm}
\begin{minipage}{0.50cm}
\hspace*{0.15cm}$\bra{x_{1}}\!\left\{\begin{matrix}\vspace{0.47cm}\end{matrix}\right.$\\       
\hspace*{0.15cm}$\bra{x_{2}}\!\left\{\begin{matrix}\vspace{0.47cm}\end{matrix}\right.$\\       
\hspace*{0.15cm}$\bra{x_{3}}\!\left\{\begin{matrix}\vspace{0.47cm}\end{matrix}\right.$\\       
\hspace*{0.15cm}$\bra{x_{4}}\!\left\{\begin{matrix}\vspace{0.47cm}\end{matrix}\right.$\\       
\hspace*{0.15cm}$\bra{x_{5}}\!\left\{\begin{matrix}\vspace{0.47cm}\end{matrix}\right.$\\       
\hspace*{0.15cm}$\bra{x_{6}}\!\left\{\begin{matrix}\vspace{0.47cm}\end{matrix}\right.$\\       
\hspace*{0.15cm}$\bra{x_{7}}\!\left\{\begin{matrix}\vspace{0.47cm}\end{matrix}\right.$\\       
\hspace*{0.15cm}$\bra{x_{8}}\!\left\{\begin{matrix}\vspace{0.47cm}\end{matrix}\right.$\\       
\hspace*{0.15cm}$\bra{x_{9}}\!\left\{\begin{matrix}\vspace{0.47cm}\end{matrix}\right.$\\       
\hspace*{0.00cm}$\bra{x_{10}}\!\left\{\begin{matrix}\vspace{0.47cm}\end{matrix}\right.$
\end{minipage}
\hspace{0.30cm}
\begin{tabular}{|p{0.60cm}p{0.60cm}|p{0.60cm}p{0.60cm}|p{0.60cm}p{0.60cm}|p{0.60cm}p{0.60cm}|p{0.60cm}p{0.60cm}|} \hline
  &&&&&&&&& \\[-0.15cm]
  \begin{minipage}{0.50cm}\colorbox{MyAqua0}{\large\hspace{-0.03cm}$\bigcirc$\hspace{-0.08cm}}\end{minipage}&
  \begin{minipage}{0.50cm}\colorbox{MyAqua0}{\large\hspace{-0.03cm}$\bigcirc$\hspace{-0.08cm}}\end{minipage}&
  \begin{minipage}{0.50cm}\colorbox{MyAqua0}{\large\hspace{-0.03cm}$\bigcirc$\hspace{-0.08cm}}\end{minipage}&
  \begin{minipage}{0.50cm}\colorbox{MyAqua0}{\large\hspace{-0.03cm}$\bigcirc$\hspace{-0.08cm}}\end{minipage}&
  \begin{minipage}{0.50cm}\colorbox{MyWhite}{$\begin{matrix}\vspace{0.1cm}&\vspace{0.1cm}\end{matrix}$}\end{minipage}&
  \begin{minipage}{0.50cm}\colorbox{MyWhite}{$\begin{matrix}\vspace{0.1cm}&\vspace{0.1cm}\end{matrix}$}\end{minipage}&
  \begin{minipage}{0.50cm}\colorbox{MyAqua0}{\large\hspace{-0.03cm}$\bigcirc$\hspace{-0.08cm}}\end{minipage}&
  \begin{minipage}{0.50cm}\colorbox{MyAqua0}{\large\hspace{-0.03cm}$\bigcirc$\hspace{-0.08cm}}\end{minipage}&
  \begin{minipage}{0.50cm}\colorbox{MyAqua0}{\large\hspace{-0.03cm}$\bigcirc$\hspace{-0.08cm}}\end{minipage}&
  \begin{minipage}{0.50cm}\colorbox{MyAqua0}{\large\hspace{-0.03cm}$\bigcirc$\hspace{-0.08cm}}\end{minipage}\\
  &&&&&&&&& \\[-0.15cm] \hline
  &&&&&&&&& \\[-0.15cm]
  \begin{minipage}{0.50cm}\colorbox{MyWhite}{$\begin{matrix}\vspace{0.1cm}&\vspace{0.1cm}\end{matrix}$}\end{minipage}&
  \begin{minipage}{0.50cm}\colorbox{MyWhite}{$\begin{matrix}\vspace{0.1cm}&\vspace{0.1cm}\end{matrix}$}\end{minipage}&
  \begin{minipage}{0.50cm}\colorbox{MyAqua0}{\large\hspace{-0.03cm}$\bigcirc$\hspace{-0.08cm}}\end{minipage}&
  \begin{minipage}{0.50cm}\colorbox{MyAqua0}{\large\hspace{-0.03cm}$\bigcirc$\hspace{-0.08cm}}\end{minipage}&
  \begin{minipage}{0.50cm}\colorbox{MyAqua0}{\large\hspace{-0.03cm}$\bigcirc$\hspace{-0.08cm}}\end{minipage}&
  \begin{minipage}{0.50cm}\colorbox{MyAqua0}{\large\hspace{-0.03cm}$\bigcirc$\hspace{-0.08cm}}\end{minipage}&
  \begin{minipage}{0.50cm}\colorbox{MyAqua0}{\large\hspace{-0.03cm}$\bigcirc$\hspace{-0.08cm}}\end{minipage}&
  \begin{minipage}{0.50cm}\colorbox{MyAqua0}{\large\hspace{-0.03cm}$\bigcirc$\hspace{-0.08cm}}\end{minipage}&
  \begin{minipage}{0.50cm}\colorbox{MyWhite}{$\begin{matrix}\vspace{0.1cm}&\vspace{0.1cm}\end{matrix}$}\end{minipage}&
  \begin{minipage}{0.50cm}\colorbox{MyWhite}{$\begin{matrix}\vspace{0.1cm}&\vspace{0.1cm}\end{matrix}$}\end{minipage}\\
  &&&&&&&&& \\[-0.15cm] \hline
  &&&&&&&&& \\[-0.15cm]
  \begin{minipage}{0.50cm}\colorbox{MyAqua0}{\large\hspace{-0.03cm}$\bigcirc$\hspace{-0.08cm}}\end{minipage}&
  \begin{minipage}{0.50cm}\colorbox{MyAqua0}{\large\hspace{-0.03cm}$\bigcirc$\hspace{-0.08cm}}\end{minipage}&
  \begin{minipage}{0.50cm}\colorbox{MyAqua0}{\large\hspace{-0.03cm}$\bigcirc$\hspace{-0.08cm}}\end{minipage}&
  \begin{minipage}{0.50cm}\colorbox{MyAqua0}{\large\hspace{-0.03cm}$\bigcirc$\hspace{-0.08cm}}\end{minipage}&
  \begin{minipage}{0.50cm}\colorbox{MyAqua0}{\large\hspace{-0.03cm}$\bigcirc$\hspace{-0.08cm}}\end{minipage}&
  \begin{minipage}{0.50cm}\colorbox{MyAqua0}{\large\hspace{-0.03cm}$\bigcirc$\hspace{-0.08cm}}\end{minipage}&
  \begin{minipage}{0.50cm}\colorbox{MyWhite}{$\begin{matrix}\vspace{0.1cm}&\vspace{0.1cm}\end{matrix}$}\end{minipage}&
  \begin{minipage}{0.50cm}\colorbox{MyWhite}{$\begin{matrix}\vspace{0.1cm}&\vspace{0.1cm}\end{matrix}$}\end{minipage}&
  \begin{minipage}{0.50cm}\colorbox{MyWhite}{$\begin{matrix}\vspace{0.1cm}&\vspace{0.1cm}\end{matrix}$}\end{minipage}&
  \begin{minipage}{0.50cm}\colorbox{MyWhite}{$\begin{matrix}\vspace{0.1cm}&\vspace{0.1cm}\end{matrix}$}\end{minipage}\\
  &&&&&&&&& \\[-0.15cm] \hline
  &&&&&&&&& \\[-0.15cm]
  \begin{minipage}{0.50cm}\colorbox{MyWhite}{$\begin{matrix}\vspace{0.1cm}&\vspace{0.1cm}\end{matrix}$}\end{minipage}&
  \begin{minipage}{0.50cm}\colorbox{MyWhite}{$\begin{matrix}\vspace{0.1cm}&\vspace{0.1cm}\end{matrix}$}\end{minipage}&
  \begin{minipage}{0.50cm}\colorbox{MyWhite}{$\begin{matrix}\vspace{0.1cm}&\vspace{0.1cm}\end{matrix}$}\end{minipage}&
  \begin{minipage}{0.50cm}\colorbox{MyWhite}{$\begin{matrix}\vspace{0.1cm}&\vspace{0.1cm}\end{matrix}$}\end{minipage}&
  \begin{minipage}{0.50cm}\colorbox{MyAqua0}{\large\hspace{-0.03cm}$\bigcirc$\hspace{-0.08cm}}\end{minipage}&
  \begin{minipage}{0.50cm}\colorbox{MyAqua0}{\large\hspace{-0.03cm}$\bigcirc$\hspace{-0.08cm}}\end{minipage}&
  \begin{minipage}{0.50cm}\colorbox{MyAqua0}{\large\hspace{-0.03cm}$\bigcirc$\hspace{-0.08cm}}\end{minipage}&
  \begin{minipage}{0.50cm}\colorbox{MyAqua0}{\large\hspace{-0.03cm}$\bigcirc$\hspace{-0.08cm}}\end{minipage}&
  \begin{minipage}{0.50cm}\colorbox{MyAqua0}{\large\hspace{-0.03cm}$\bigcirc$\hspace{-0.08cm}}\end{minipage}&
  \begin{minipage}{0.50cm}\colorbox{MyAqua0}{\large\hspace{-0.03cm}$\bigcirc$\hspace{-0.08cm}}\end{minipage}\\
  &&&&&&&&& \\[-0.15cm] \hline
  &&&&&&&&& \\[-0.15cm]
  \begin{minipage}{0.50cm}\colorbox{MyAqua0}{\large\hspace{-0.03cm}$\bigcirc$\hspace{-0.08cm}}\end{minipage}&
  \begin{minipage}{0.50cm}\colorbox{MyAqua0}{\large\hspace{-0.03cm}$\bigcirc$\hspace{-0.08cm}}\end{minipage}&
  \begin{minipage}{0.50cm}\colorbox{MyWhite}{$\begin{matrix}\vspace{0.1cm}&\vspace{0.1cm}\end{matrix}$}\end{minipage}&
  \begin{minipage}{0.50cm}\colorbox{MyWhite}{$\begin{matrix}\vspace{0.1cm}&\vspace{0.1cm}\end{matrix}$}\end{minipage}&
  \begin{minipage}{0.50cm}\colorbox{MyAqua0}{\large\hspace{-0.03cm}$\bigcirc$\hspace{-0.08cm}}\end{minipage}&
  \begin{minipage}{0.50cm}\colorbox{MyAqua0}{\large\hspace{-0.03cm}$\bigcirc$\hspace{-0.08cm}}\end{minipage}&
  \begin{minipage}{0.50cm}\colorbox{MyWhite}{$\begin{matrix}\vspace{0.1cm}&\vspace{0.1cm}\end{matrix}$}\end{minipage}&
  \begin{minipage}{0.50cm}\colorbox{MyWhite}{$\begin{matrix}\vspace{0.1cm}&\vspace{0.1cm}\end{matrix}$}\end{minipage}&
  \begin{minipage}{0.50cm}\colorbox{MyAqua0}{\large\hspace{-0.03cm}$\bigcirc$\hspace{-0.08cm}}\end{minipage}&
  \begin{minipage}{0.50cm}\colorbox{MyAqua0}{\large\hspace{-0.03cm}$\bigcirc$\hspace{-0.08cm}}\end{minipage}\\
  &&&&&&&&& \\[-0.15cm] \hline
  &&&&&&&&& \\[-0.15cm]
  \begin{minipage}{0.50cm}\colorbox{MyWhite}{$\begin{matrix}\vspace{0.1cm}&\vspace{0.1cm}\end{matrix}$}\end{minipage}&
  \begin{minipage}{0.50cm}\colorbox{MyWhite}{$\begin{matrix}\vspace{0.1cm}&\vspace{0.1cm}\end{matrix}$}\end{minipage}&
  \begin{minipage}{0.50cm}\colorbox{MyAqua0}{\large\hspace{-0.03cm}$\bigcirc$\hspace{-0.08cm}}\end{minipage}&
  \begin{minipage}{0.50cm}\colorbox{MyAqua0}{\large\hspace{-0.03cm}$\bigcirc$\hspace{-0.08cm}}\end{minipage}&
  \begin{minipage}{0.50cm}\colorbox{MyAqua0}{\large\hspace{-0.03cm}$\bigcirc$\hspace{-0.08cm}}\end{minipage}&
  \begin{minipage}{0.50cm}\colorbox{MyAqua0}{\large\hspace{-0.03cm}$\bigcirc$\hspace{-0.08cm}}\end{minipage}&
  \begin{minipage}{0.50cm}\colorbox{MyWhite}{$\begin{matrix}\vspace{0.1cm}&\vspace{0.1cm}\end{matrix}$}\end{minipage}&
  \begin{minipage}{0.50cm}\colorbox{MyWhite}{$\begin{matrix}\vspace{0.1cm}&\vspace{0.1cm}\end{matrix}$}\end{minipage}&
  \begin{minipage}{0.50cm}\colorbox{MyWhite}{$\begin{matrix}\vspace{0.1cm}&\vspace{0.1cm}\end{matrix}$}\end{minipage}&
  \begin{minipage}{0.50cm}\colorbox{MyWhite}{$\begin{matrix}\vspace{0.1cm}&\vspace{0.1cm}\end{matrix}$}\end{minipage}\\
  &&&&&&&&& \\[-0.15cm] \hline
  &&&&&&&&& \\[-0.15cm]
  \begin{minipage}{0.50cm}\colorbox{MyWhite}{$\begin{matrix}\vspace{0.1cm}&\vspace{0.1cm}\end{matrix}$}\end{minipage}&
  \begin{minipage}{0.50cm}\colorbox{MyWhite}{$\begin{matrix}\vspace{0.1cm}&\vspace{0.1cm}\end{matrix}$}\end{minipage}&
  \begin{minipage}{0.50cm}\colorbox{MyWhite}{$\begin{matrix}\vspace{0.1cm}&\vspace{0.1cm}\end{matrix}$}\end{minipage}&
  \begin{minipage}{0.50cm}\colorbox{MyWhite}{$\begin{matrix}\vspace{0.1cm}&\vspace{0.1cm}\end{matrix}$}\end{minipage}&
  \begin{minipage}{0.50cm}\colorbox{MyAqua0}{\large\hspace{-0.03cm}$\bigcirc$\hspace{-0.08cm}}\end{minipage}&
  \begin{minipage}{0.50cm}\colorbox{MyAqua0}{\large\hspace{-0.03cm}$\bigcirc$\hspace{-0.08cm}}\end{minipage}&
  \begin{minipage}{0.50cm}\colorbox{MyAqua0}{\large\hspace{-0.03cm}$\bigcirc$\hspace{-0.08cm}}\end{minipage}&
  \begin{minipage}{0.50cm}\colorbox{MyAqua0}{\large\hspace{-0.03cm}$\bigcirc$\hspace{-0.08cm}}\end{minipage}&
  \begin{minipage}{0.50cm}\colorbox{MyWhite}{$\begin{matrix}\vspace{0.1cm}&\vspace{0.1cm}\end{matrix}$}\end{minipage}&
  \begin{minipage}{0.50cm}\colorbox{MyWhite}{$\begin{matrix}\vspace{0.1cm}&\vspace{0.1cm}\end{matrix}$}\end{minipage}\\
  &&&&&&&&& \\[-0.15cm] \hline
  &&&&&&&&& \\[-0.15cm]
  \begin{minipage}{0.50cm}\colorbox{MyAqua0}{\large\hspace{-0.03cm}$\bigcirc$\hspace{-0.08cm}}\end{minipage}&
  \begin{minipage}{0.50cm}\colorbox{MyAqua0}{\large\hspace{-0.03cm}$\bigcirc$\hspace{-0.08cm}}\end{minipage}&
  \begin{minipage}{0.50cm}\colorbox{MyWhite}{$\begin{matrix}\vspace{0.1cm}&\vspace{0.1cm}\end{matrix}$}\end{minipage}&
  \begin{minipage}{0.50cm}\colorbox{MyWhite}{$\begin{matrix}\vspace{0.1cm}&\vspace{0.1cm}\end{matrix}$}\end{minipage}&
  \begin{minipage}{0.50cm}\colorbox{MyWhite}{$\begin{matrix}\vspace{0.1cm}&\vspace{0.1cm}\end{matrix}$}\end{minipage}&
  \begin{minipage}{0.50cm}\colorbox{MyWhite}{$\begin{matrix}\vspace{0.1cm}&\vspace{0.1cm}\end{matrix}$}\end{minipage}&
  \begin{minipage}{0.50cm}\colorbox{MyAqua0}{\large\hspace{-0.03cm}$\bigcirc$\hspace{-0.08cm}}\end{minipage}&
  \begin{minipage}{0.50cm}\colorbox{MyAqua0}{\large\hspace{-0.03cm}$\bigcirc$\hspace{-0.08cm}}\end{minipage}&
  \begin{minipage}{0.50cm}\colorbox{MyAqua0}{\large\hspace{-0.03cm}$\bigcirc$\hspace{-0.08cm}}\end{minipage}&
  \begin{minipage}{0.50cm}\colorbox{MyAqua0}{\large\hspace{-0.03cm}$\bigcirc$\hspace{-0.08cm}}\end{minipage}\\
  &&&&&&&&& \\[-0.15cm] \hline
  &&&&&&&&& \\[-0.15cm]
  \begin{minipage}{0.50cm}\colorbox{MyAqua0}{\large\hspace{-0.03cm}$\bigcirc$\hspace{-0.08cm}}\end{minipage}&
  \begin{minipage}{0.50cm}\colorbox{MyAqua0}{\large\hspace{-0.03cm}$\bigcirc$\hspace{-0.08cm}}\end{minipage}&
  \begin{minipage}{0.50cm}\colorbox{MyWhite}{$\begin{matrix}\vspace{0.1cm}&\vspace{0.1cm}\end{matrix}$}\end{minipage}&
  \begin{minipage}{0.50cm}\colorbox{MyWhite}{$\begin{matrix}\vspace{0.1cm}&\vspace{0.1cm}\end{matrix}$}\end{minipage}&
  \begin{minipage}{0.50cm}\colorbox{MyAqua0}{\large\hspace{-0.03cm}$\bigcirc$\hspace{-0.08cm}}\end{minipage}&
  \begin{minipage}{0.50cm}\colorbox{MyAqua0}{\large\hspace{-0.03cm}$\bigcirc$\hspace{-0.08cm}}\end{minipage}&
  \begin{minipage}{0.50cm}\colorbox{MyAqua0}{\large\hspace{-0.03cm}$\bigcirc$\hspace{-0.08cm}}\end{minipage}&
  \begin{minipage}{0.50cm}\colorbox{MyAqua0}{\large\hspace{-0.03cm}$\bigcirc$\hspace{-0.08cm}}\end{minipage}&
  \begin{minipage}{0.50cm}\colorbox{MyWhite}{$\begin{matrix}\vspace{0.1cm}&\vspace{0.1cm}\end{matrix}$}\end{minipage}&
  \begin{minipage}{0.50cm}\colorbox{MyWhite}{$\begin{matrix}\vspace{0.1cm}&\vspace{0.1cm}\end{matrix}$}\end{minipage}\\
  &&&&&&&&& \\[-0.15cm] \hline
  &&&&&&&&& \\[-0.15cm]
  \begin{minipage}{0.50cm}\colorbox{MyAqua0}{\large\hspace{-0.03cm}$\bigcirc$\hspace{-0.08cm}}\end{minipage}&
  \begin{minipage}{0.50cm}\colorbox{MyAqua0}{\large\hspace{-0.03cm}$\bigcirc$\hspace{-0.08cm}}\end{minipage}&
  \begin{minipage}{0.50cm}\colorbox{MyAqua0}{\large\hspace{-0.03cm}$\bigcirc$\hspace{-0.08cm}}\end{minipage}&
  \begin{minipage}{0.50cm}\colorbox{MyAqua0}{\large\hspace{-0.03cm}$\bigcirc$\hspace{-0.08cm}}\end{minipage}&
  \begin{minipage}{0.50cm}\colorbox{MyAqua0}{\large\hspace{-0.03cm}$\bigcirc$\hspace{-0.08cm}}\end{minipage}&
  \begin{minipage}{0.50cm}\colorbox{MyAqua0}{\large\hspace{-0.03cm}$\bigcirc$\hspace{-0.08cm}}\end{minipage}&
  \begin{minipage}{0.50cm}\colorbox{MyWhite}{$\begin{matrix}\vspace{0.1cm}&\vspace{0.1cm}\end{matrix}$}\end{minipage}&
  \begin{minipage}{0.50cm}\colorbox{MyWhite}{$\begin{matrix}\vspace{0.1cm}&\vspace{0.1cm}\end{matrix}$}\end{minipage}&
  \begin{minipage}{0.50cm}\colorbox{MyAqua0}{\large\hspace{-0.03cm}$\bigcirc$\hspace{-0.08cm}}\end{minipage}&
  \begin{minipage}{0.50cm}\colorbox{MyAqua0}{\large\hspace{-0.03cm}$\bigcirc$\hspace{-0.08cm}}\end{minipage}\\
  &&&&&&&&& \\[-0.15cm] \hline
\end{tabular}

\vspace{0.0cm}
\hspace{0.85cm}$\overbrace{\hspace{6.84cm}}^{\mbox{\normalsize$\ket{\Omega}$}}$  \hspace{0.78cm}  $\overbrace{\hspace{6.84cm}}^{\mbox{\normalsize$\ket{\Omega}$}}$  

\hspace{0.85cm}$\overbrace{\hspace{0.62cm}}^{\ket{X^{\times}_1}}
                \overbrace{\hspace{1.29cm}}^{\ket{X^{\times}_2}} 
                \overbrace{\hspace{1.29cm}}^{\ket{X^{\times}_3}} 
                \overbrace{\hspace{1.29cm}}^{\ket{X^{\times}_4}} 
                \overbrace{\hspace{1.29cm}}^{\ket{X^{\times}_5}} 
                \overbrace{\hspace{0.62cm}}^{\ket{X^{\times}_6}}$            
\hspace{0.89cm}$
                \overbrace{\hspace{0.55cm}}^{\ket{X_{\!1}}}  
                \overbrace{\hspace{0.55cm}}^{\ket{X_{\!2}}}  
                \overbrace{\hspace{0.55cm}}^{\ket{X_{\!3}}}  
                \overbrace{\hspace{0.55cm}}^{\ket{X_{\!4}}}  
                \overbrace{\hspace{0.55cm}}^{\ket{X_{\!5}}}  
                \overbrace{\hspace{0.55cm}}^{\ket{X_{\!6}}}  
                \overbrace{\hspace{0.55cm}}^{\ket{X_{\!7}}}  
                \overbrace{\hspace{0.55cm}}^{\ket{X_{\!8}}}  
                \overbrace{\hspace{0.55cm}}^{\ket{X_{\!9}}}  
                \overbrace{\hspace{0.55cm}}^{\ket{X_{\!10}}}  
                $    

\hspace{-0.82cm}\begin{minipage}{0.50cm} $\bra{\Omega}\!\!\left\{\begin{matrix}\vspace{6.75cm}\end{matrix}\right.$
\end{minipage}
\hspace{0.15cm}
\begin{minipage}{0.50cm}
\hspace*{0.00cm}$\bra{x_{1}}\!\left\{\begin{matrix}\vspace{0.49cm}\end{matrix}\right.$\\       
\hspace*{0.00cm}$\bra{x_{2}}\!\left\{\begin{matrix}\vspace{0.49cm}\end{matrix}\right.$\\       
\hspace*{0.00cm}$\bra{x_{3}}\!\left\{\begin{matrix}\vspace{0.49cm}\end{matrix}\right.$\\       
\hspace*{0.00cm}$\bra{x_{4}}\!\left\{\begin{matrix}\vspace{0.49cm}\end{matrix}\right.$\\       
\hspace*{0.00cm}$\bra{x_{5}}\!\left\{\begin{matrix}\vspace{0.49cm}\end{matrix}\right.$\\       
\hspace*{0.00cm}$\bra{x_{6}}\!\left\{\begin{matrix}\vspace{0.49cm}\end{matrix}\right.$\\       
\hspace*{0.00cm}$\bra{x_{7}}\!\left\{\begin{matrix}\vspace{0.49cm}\end{matrix}\right.$\\       
\hspace*{0.00cm}$\bra{x_{8}}\!\left\{\begin{matrix}\vspace{0.49cm}\end{matrix}\right.$\\       
\hspace*{0.00cm}$\bra{x_{9}}\!\left\{\begin{matrix}\vspace{0.49cm}\end{matrix}\right.$\\       
\hspace*{-0.15cm}$\bra{x_{10}}\!\left\{\begin{matrix}\vspace{0.49cm}\end{matrix}\right.$
\end{minipage}
\hspace{0.15cm}
\begin{tabular}{|p{0.60cm}|p{0.60cm}p{0.60cm}|p{0.60cm}p{0.60cm}|p{0.60cm}p{0.60cm}|p{0.60cm}p{0.60cm}|p{0.60cm}|} \hline
  &&&&&&&&& \\[-0.15cm]
  \begin{minipage}{0.50cm}\colorbox{MyWhite}{$\begin{matrix}\vspace{0.1cm}&\vspace{0.1cm}\end{matrix}$}\end{minipage}&
  \begin{minipage}{0.50cm}\colorbox{MyGreen}{\LARGE\hspace{-0.03cm}$\times$\hspace{-0.08cm}}\end{minipage}&
  \begin{minipage}{0.50cm}\colorbox{MyGreen}{\LARGE\hspace{-0.03cm}$\times$\hspace{-0.08cm}}\end{minipage}&
  \begin{minipage}{0.50cm}\colorbox{MyWhite}{$\begin{matrix}\vspace{0.1cm}&\vspace{0.1cm}\end{matrix}$}\end{minipage}&
  \begin{minipage}{0.50cm}\colorbox{MyWhite}{$\begin{matrix}\vspace{0.1cm}&\vspace{0.1cm}\end{matrix}$}\end{minipage}&
  \begin{minipage}{0.50cm}\colorbox{MyGreen}{\LARGE\hspace{-0.03cm}$\times$\hspace{-0.08cm}}\end{minipage}&
  \begin{minipage}{0.50cm}\colorbox{MyGreen}{\LARGE\hspace{-0.03cm}$\times$\hspace{-0.08cm}}\end{minipage}&
  \begin{minipage}{0.50cm}\colorbox{MyGreen}{\LARGE\hspace{-0.03cm}$\times$\hspace{-0.08cm}}\end{minipage}&
  \begin{minipage}{0.50cm}\colorbox{MyGreen}{\LARGE\hspace{-0.03cm}$\times$\hspace{-0.08cm}}\end{minipage}&
  \begin{minipage}{0.50cm}\colorbox{MyWhite}{$\begin{matrix}\vspace{0.1cm}&\vspace{0.1cm}\end{matrix}$}\end{minipage}\\
  &&&&&&&&& \\[-0.15cm] \hline
  &&&&&&&&& \\[-0.15cm]
  \begin{minipage}{0.50cm}\colorbox{MyGreen}{\LARGE\hspace{-0.03cm}$\times$\hspace{-0.08cm}}\end{minipage}&
  \begin{minipage}{0.50cm}\colorbox{MyGreen}{\LARGE\hspace{-0.03cm}$\times$\hspace{-0.08cm}}\end{minipage}&
  \begin{minipage}{0.50cm}\colorbox{MyGreen}{\LARGE\hspace{-0.03cm}$\times$\hspace{-0.08cm}}\end{minipage}&
  \begin{minipage}{0.50cm}\colorbox{MyGreen}{\LARGE\hspace{-0.03cm}$\times$\hspace{-0.08cm}}\end{minipage}&
  \begin{minipage}{0.50cm}\colorbox{MyGreen}{\LARGE\hspace{-0.03cm}$\times$\hspace{-0.08cm}}\end{minipage}&
  \begin{minipage}{0.50cm}\colorbox{MyWhite}{$\begin{matrix}\vspace{0.1cm}&\vspace{0.1cm}\end{matrix}$}\end{minipage}&
  \begin{minipage}{0.50cm}\colorbox{MyWhite}{$\begin{matrix}\vspace{0.1cm}&\vspace{0.1cm}\end{matrix}$}\end{minipage}&
  \begin{minipage}{0.50cm}\colorbox{MyWhite}{$\begin{matrix}\vspace{0.1cm}&\vspace{0.1cm}\end{matrix}$}\end{minipage}&
  \begin{minipage}{0.50cm}\colorbox{MyWhite}{$\begin{matrix}\vspace{0.1cm}&\vspace{0.1cm}\end{matrix}$}\end{minipage}&
  \begin{minipage}{0.50cm}\colorbox{MyWhite}{$\begin{matrix}\vspace{0.1cm}&\vspace{0.1cm}\end{matrix}$}\end{minipage}\\
  &&&&&&&&& \\[-0.15cm] \hline
  &&&&&&&&& \\[-0.15cm]
  \begin{minipage}{0.50cm}\colorbox{MyWhite}{$\begin{matrix}\vspace{0.1cm}&\vspace{0.1cm}\end{matrix}$}\end{minipage}&
  \begin{minipage}{0.50cm}\colorbox{MyWhite}{$\begin{matrix}\vspace{0.1cm}&\vspace{0.1cm}\end{matrix}$}\end{minipage}&
  \begin{minipage}{0.50cm}\colorbox{MyWhite}{$\begin{matrix}\vspace{0.1cm}&\vspace{0.1cm}\end{matrix}$}\end{minipage}&
  \begin{minipage}{0.50cm}\colorbox{MyGreen}{\LARGE\hspace{-0.03cm}$\times$\hspace{-0.08cm}}\end{minipage}&
  \begin{minipage}{0.50cm}\colorbox{MyGreen}{\LARGE\hspace{-0.03cm}$\times$\hspace{-0.08cm}}\end{minipage}&
  \begin{minipage}{0.50cm}\colorbox{MyGreen}{\LARGE\hspace{-0.03cm}$\times$\hspace{-0.08cm}}\end{minipage}&
  \begin{minipage}{0.50cm}\colorbox{MyGreen}{\LARGE\hspace{-0.03cm}$\times$\hspace{-0.08cm}}\end{minipage}&
  \begin{minipage}{0.50cm}\colorbox{MyGreen}{\LARGE\hspace{-0.03cm}$\times$\hspace{-0.08cm}}\end{minipage}&
  \begin{minipage}{0.50cm}\colorbox{MyGreen}{\LARGE\hspace{-0.03cm}$\times$\hspace{-0.08cm}}\end{minipage}&
  \begin{minipage}{0.50cm}\colorbox{MyWhite}{$\begin{matrix}\vspace{0.1cm}&\vspace{0.1cm}\end{matrix}$}\end{minipage}\\
  &&&&&&&&& \\[-0.15cm] \hline
  &&&&&&&&& \\[-0.15cm]
  \begin{minipage}{0.50cm}\colorbox{MyGreen}{\LARGE\hspace{-0.03cm}$\times$\hspace{-0.08cm}}\end{minipage}&
  \begin{minipage}{0.50cm}\colorbox{MyGreen}{\LARGE\hspace{-0.03cm}$\times$\hspace{-0.08cm}}\end{minipage}&
  \begin{minipage}{0.50cm}\colorbox{MyGreen}{\LARGE\hspace{-0.03cm}$\times$\hspace{-0.08cm}}\end{minipage}&
  \begin{minipage}{0.50cm}\colorbox{MyWhite}{$\begin{matrix}\vspace{0.1cm}&\vspace{0.1cm}\end{matrix}$}\end{minipage}&
  \begin{minipage}{0.50cm}\colorbox{MyWhite}{$\begin{matrix}\vspace{0.1cm}&\vspace{0.1cm}\end{matrix}$}\end{minipage}&
  \begin{minipage}{0.50cm}\colorbox{MyWhite}{$\begin{matrix}\vspace{0.1cm}&\vspace{0.1cm}\end{matrix}$}\end{minipage}&
  \begin{minipage}{0.50cm}\colorbox{MyWhite}{$\begin{matrix}\vspace{0.1cm}&\vspace{0.1cm}\end{matrix}$}\end{minipage}&
  \begin{minipage}{0.50cm}\colorbox{MyGreen}{\LARGE\hspace{-0.03cm}$\times$\hspace{-0.08cm}}\end{minipage}&
  \begin{minipage}{0.50cm}\colorbox{MyGreen}{\LARGE\hspace{-0.03cm}$\times$\hspace{-0.08cm}}\end{minipage}&
  \begin{minipage}{0.50cm}\colorbox{MyGreen}{\LARGE\hspace{-0.03cm}$\times$\hspace{-0.08cm}}\end{minipage}\\
  &&&&&&&&& \\[-0.15cm] \hline
  &&&&&&&&& \\[-0.15cm]
  \begin{minipage}{0.50cm}\colorbox{MyWhite}{$\begin{matrix}\vspace{0.1cm}&\vspace{0.1cm}\end{matrix}$}\end{minipage}&
  \begin{minipage}{0.50cm}\colorbox{MyWhite}{$\begin{matrix}\vspace{0.1cm}&\vspace{0.1cm}\end{matrix}$}\end{minipage}&
  \begin{minipage}{0.50cm}\colorbox{MyWhite}{$\begin{matrix}\vspace{0.1cm}&\vspace{0.1cm}\end{matrix}$}\end{minipage}&
  \begin{minipage}{0.50cm}\colorbox{MyGreen}{\LARGE\hspace{-0.03cm}$\times$\hspace{-0.08cm}}\end{minipage}&
  \begin{minipage}{0.50cm}\colorbox{MyGreen}{\LARGE\hspace{-0.03cm}$\times$\hspace{-0.08cm}}\end{minipage}&
  \begin{minipage}{0.50cm}\colorbox{MyGreen}{\LARGE\hspace{-0.03cm}$\times$\hspace{-0.08cm}}\end{minipage}&
  \begin{minipage}{0.50cm}\colorbox{MyGreen}{\LARGE\hspace{-0.03cm}$\times$\hspace{-0.08cm}}\end{minipage}&
  \begin{minipage}{0.50cm}\colorbox{MyWhite}{$\begin{matrix}\vspace{0.1cm}&\vspace{0.1cm}\end{matrix}$}\end{minipage}&
  \begin{minipage}{0.50cm}\colorbox{MyWhite}{$\begin{matrix}\vspace{0.1cm}&\vspace{0.1cm}\end{matrix}$}\end{minipage}&
  \begin{minipage}{0.50cm}\colorbox{MyWhite}{$\begin{matrix}\vspace{0.1cm}&\vspace{0.1cm}\end{matrix}$}\end{minipage}\\
  &&&&&&&&& \\[-0.15cm] \hline
  &&&&&&&&& \\[-0.15cm]
  \begin{minipage}{0.50cm}\colorbox{MyWhite}{$\begin{matrix}\vspace{0.1cm}&\vspace{0.1cm}\end{matrix}$}\end{minipage}&
  \begin{minipage}{0.50cm}\colorbox{MyWhite}{$\begin{matrix}\vspace{0.1cm}&\vspace{0.1cm}\end{matrix}$}\end{minipage}&
  \begin{minipage}{0.50cm}\colorbox{MyWhite}{$\begin{matrix}\vspace{0.1cm}&\vspace{0.1cm}\end{matrix}$}\end{minipage}&
  \begin{minipage}{0.50cm}\colorbox{MyWhite}{$\begin{matrix}\vspace{0.1cm}&\vspace{0.1cm}\end{matrix}$}\end{minipage}&
  \begin{minipage}{0.50cm}\colorbox{MyWhite}{$\begin{matrix}\vspace{0.1cm}&\vspace{0.1cm}\end{matrix}$}\end{minipage}&
  \begin{minipage}{0.50cm}\colorbox{MyGreen}{\LARGE\hspace{-0.03cm}$\times$\hspace{-0.08cm}}\end{minipage}&
  \begin{minipage}{0.50cm}\colorbox{MyGreen}{\LARGE\hspace{-0.03cm}$\times$\hspace{-0.08cm}}\end{minipage}&
  \begin{minipage}{0.50cm}\colorbox{MyGreen}{\LARGE\hspace{-0.03cm}$\times$\hspace{-0.08cm}}\end{minipage}&
  \begin{minipage}{0.50cm}\colorbox{MyGreen}{\LARGE\hspace{-0.03cm}$\times$\hspace{-0.08cm}}\end{minipage}&
  \begin{minipage}{0.50cm}\colorbox{MyGreen}{\LARGE\hspace{-0.03cm}$\times$\hspace{-0.08cm}}\end{minipage}\\
  &&&&&&&&& \\[-0.15cm] \hline
  &&&&&&&&& \\[-0.15cm]
  \begin{minipage}{0.50cm}\colorbox{MyWhite}{$\begin{matrix}\vspace{0.1cm}&\vspace{0.1cm}\end{matrix}$}\end{minipage}&
  \begin{minipage}{0.50cm}\colorbox{MyGreen}{\LARGE\hspace{-0.03cm}$\times$\hspace{-0.08cm}}\end{minipage}&
  \begin{minipage}{0.50cm}\colorbox{MyGreen}{\LARGE\hspace{-0.03cm}$\times$\hspace{-0.08cm}}\end{minipage}&
  \begin{minipage}{0.50cm}\colorbox{MyGreen}{\LARGE\hspace{-0.03cm}$\times$\hspace{-0.08cm}}\end{minipage}&
  \begin{minipage}{0.50cm}\colorbox{MyGreen}{\LARGE\hspace{-0.03cm}$\times$\hspace{-0.08cm}}\end{minipage}&
  \begin{minipage}{0.50cm}\colorbox{MyGreen}{\LARGE\hspace{-0.03cm}$\times$\hspace{-0.08cm}}\end{minipage}&
  \begin{minipage}{0.50cm}\colorbox{MyGreen}{\LARGE\hspace{-0.03cm}$\times$\hspace{-0.08cm}}\end{minipage}&
  \begin{minipage}{0.50cm}\colorbox{MyWhite}{$\begin{matrix}\vspace{0.1cm}&\vspace{0.1cm}\end{matrix}$}\end{minipage}&
  \begin{minipage}{0.50cm}\colorbox{MyWhite}{$\begin{matrix}\vspace{0.1cm}&\vspace{0.1cm}\end{matrix}$}\end{minipage}&
  \begin{minipage}{0.50cm}\colorbox{MyWhite}{$\begin{matrix}\vspace{0.1cm}&\vspace{0.1cm}\end{matrix}$}\end{minipage}\\
  &&&&&&&&& \\[-0.15cm] \hline
  &&&&&&&&& \\[-0.15cm]
  \begin{minipage}{0.50cm}\colorbox{MyGreen}{\LARGE\hspace{-0.03cm}$\times$\hspace{-0.08cm}}\end{minipage}&
  \begin{minipage}{0.50cm}\colorbox{MyWhite}{$\begin{matrix}\vspace{0.1cm}&\vspace{0.1cm}\end{matrix}$}\end{minipage}&
  \begin{minipage}{0.50cm}\colorbox{MyWhite}{$\begin{matrix}\vspace{0.1cm}&\vspace{0.1cm}\end{matrix}$}\end{minipage}&
  \begin{minipage}{0.50cm}\colorbox{MyGreen}{\LARGE\hspace{-0.03cm}$\times$\hspace{-0.08cm}}\end{minipage}&
  \begin{minipage}{0.50cm}\colorbox{MyGreen}{\LARGE\hspace{-0.03cm}$\times$\hspace{-0.08cm}}\end{minipage}&
  \begin{minipage}{0.50cm}\colorbox{MyWhite}{$\begin{matrix}\vspace{0.1cm}&\vspace{0.1cm}\end{matrix}$}\end{minipage}&
  \begin{minipage}{0.50cm}\colorbox{MyWhite}{$\begin{matrix}\vspace{0.1cm}&\vspace{0.1cm}\end{matrix}$}\end{minipage}&
  \begin{minipage}{0.50cm}\colorbox{MyGreen}{\LARGE\hspace{-0.03cm}$\times$\hspace{-0.08cm}}\end{minipage}&
  \begin{minipage}{0.50cm}\colorbox{MyGreen}{\LARGE\hspace{-0.03cm}$\times$\hspace{-0.08cm}}\end{minipage}&
  \begin{minipage}{0.50cm}\colorbox{MyWhite}{$\begin{matrix}\vspace{0.1cm}&\vspace{0.1cm}\end{matrix}$}\end{minipage}\\
  &&&&&&&&& \\[-0.15cm] \hline
  &&&&&&&&& \\[-0.15cm]
  \begin{minipage}{0.50cm}\colorbox{MyWhite}{$\begin{matrix}\vspace{0.1cm}&\vspace{0.1cm}\end{matrix}$}\end{minipage}&
  \begin{minipage}{0.50cm}\colorbox{MyGreen}{\LARGE\hspace{-0.03cm}$\times$\hspace{-0.08cm}}\end{minipage}&
  \begin{minipage}{0.50cm}\colorbox{MyGreen}{\LARGE\hspace{-0.03cm}$\times$\hspace{-0.08cm}}\end{minipage}&
  \begin{minipage}{0.50cm}\colorbox{MyWhite}{$\begin{matrix}\vspace{0.1cm}&\vspace{0.1cm}\end{matrix}$}\end{minipage}&
  \begin{minipage}{0.50cm}\colorbox{MyWhite}{$\begin{matrix}\vspace{0.1cm}&\vspace{0.1cm}\end{matrix}$}\end{minipage}&
  \begin{minipage}{0.50cm}\colorbox{MyGreen}{\LARGE\hspace{-0.03cm}$\times$\hspace{-0.08cm}}\end{minipage}&
  \begin{minipage}{0.50cm}\colorbox{MyGreen}{\LARGE\hspace{-0.03cm}$\times$\hspace{-0.08cm}}\end{minipage}&
  \begin{minipage}{0.50cm}\colorbox{MyWhite}{$\begin{matrix}\vspace{0.1cm}&\vspace{0.1cm}\end{matrix}$}\end{minipage}&
  \begin{minipage}{0.50cm}\colorbox{MyWhite}{$\begin{matrix}\vspace{0.1cm}&\vspace{0.1cm}\end{matrix}$}\end{minipage}&
  \begin{minipage}{0.50cm}\colorbox{MyGreen}{\LARGE\hspace{-0.03cm}$\times$\hspace{-0.08cm}}\end{minipage}\\
  &&&&&&&&& \\[-0.15cm] \hline
  &&&&&&&&& \\[-0.15cm]
  \begin{minipage}{0.50cm}\colorbox{MyWhite}{$\begin{matrix}\vspace{0.1cm}&\vspace{0.1cm}\end{matrix}$}\end{minipage}&
  \begin{minipage}{0.50cm}\colorbox{MyGreen}{\LARGE\hspace{-0.03cm}$\times$\hspace{-0.08cm}}\end{minipage}&
  \begin{minipage}{0.50cm}\colorbox{MyGreen}{\LARGE\hspace{-0.03cm}$\times$\hspace{-0.08cm}}\end{minipage}&
  \begin{minipage}{0.50cm}\colorbox{MyGreen}{\LARGE\hspace{-0.03cm}$\times$\hspace{-0.08cm}}\end{minipage}&
  \begin{minipage}{0.50cm}\colorbox{MyGreen}{\LARGE\hspace{-0.03cm}$\times$\hspace{-0.08cm}}\end{minipage}&
  \begin{minipage}{0.50cm}\colorbox{MyWhite}{$\begin{matrix}\vspace{0.1cm}&\vspace{0.1cm}\end{matrix}$}\end{minipage}&
  \begin{minipage}{0.50cm}\colorbox{MyWhite}{$\begin{matrix}\vspace{0.1cm}&\vspace{0.1cm}\end{matrix}$}\end{minipage}&
  \begin{minipage}{0.50cm}\colorbox{MyGreen}{\LARGE\hspace{-0.03cm}$\times$\hspace{-0.08cm}}\end{minipage}&
  \begin{minipage}{0.50cm}\colorbox{MyGreen}{\LARGE\hspace{-0.03cm}$\times$\hspace{-0.08cm}}\end{minipage}&
  \begin{minipage}{0.50cm}\colorbox{MyWhite}{$\begin{matrix}\vspace{0.1cm}&\vspace{0.1cm}\end{matrix}$}\end{minipage}\\
  &&&&&&&&& \\[-0.15cm] \hline
\end{tabular}
\begin{minipage}{0.50cm}
\hspace*{0.00cm}$\bra{x_{1}}\!\left\{\begin{matrix}\vspace{0.49cm}\end{matrix}\right.$\\       
\hspace*{0.00cm}$\bra{x_{2}}\!\left\{\begin{matrix}\vspace{0.49cm}\end{matrix}\right.$\\       
\hspace*{0.00cm}$\bra{x_{3}}\!\left\{\begin{matrix}\vspace{0.49cm}\end{matrix}\right.$\\       
\hspace*{0.00cm}$\bra{x_{4}}\!\left\{\begin{matrix}\vspace{0.49cm}\end{matrix}\right.$\\       
\hspace*{0.00cm}$\bra{x_{5}}\!\left\{\begin{matrix}\vspace{0.49cm}\end{matrix}\right.$\\       
\hspace*{0.00cm}$\bra{x_{6}}\!\left\{\begin{matrix}\vspace{0.49cm}\end{matrix}\right.$\\       
\hspace*{0.00cm}$\bra{x_{7}}\!\left\{\begin{matrix}\vspace{0.49cm}\end{matrix}\right.$\\       
\hspace*{0.00cm}$\bra{x_{8}}\!\left\{\begin{matrix}\vspace{0.49cm}\end{matrix}\right.$\\       
\hspace*{0.00cm}$\bra{x_{9}}\!\left\{\begin{matrix}\vspace{0.49cm}\end{matrix}\right.$\\       
\hspace*{-0.08cm}$\bra{x_{10}}\!\!\left\{\begin{matrix}\vspace{0.49cm}\end{matrix}\right.$
\end{minipage}
\hspace{0.15cm}
\begin{tabular}{|p{0.60cm}|p{0.60cm}|p{0.60cm}|p{0.60cm}|p{0.60cm}|p{0.60cm}|p{0.60cm}|p{0.60cm}|p{0.60cm}|p{0.60cm}|} \hline
  &&&&&&&&& \\[-0.15cm]
  \begin{minipage}{0.50cm}\colorbox{MyWhite}{\large\hspace{-0.03cm}$\bigcirc$\hspace{-0.03cm}}\end{minipage}&
  \begin{minipage}{0.50cm}\colorbox{MyOrangeBayes}{\LARGE\hspace{-0.03cm}$\otimes$\hspace{-0.08cm}}\end{minipage}&
  \begin{minipage}{0.50cm}\colorbox{MyOrangeBayes}{\LARGE\hspace{-0.03cm}$\otimes$\hspace{-0.08cm}}\end{minipage}&
  \begin{minipage}{0.50cm}\colorbox{MyWhite}{\large\hspace{-0.03cm}$\bigcirc$\hspace{-0.03cm}}\end{minipage}&
  \begin{minipage}{0.50cm}\colorbox{MyOrangeBayes}{$\begin{matrix}\vspace{0.1cm}&\vspace{0.1cm}\end{matrix}$}\end{minipage}&
  \begin{minipage}{0.50cm}\colorbox{MyWhite}{\LARGE\hspace{-0.05cm}$\times$\hspace{-0.05cm}}\end{minipage}&
  \begin{minipage}{0.50cm}\colorbox{MyOrangeBayes}{\LARGE\hspace{-0.03cm}$\otimes$\hspace{-0.08cm}}\end{minipage}&
  \begin{minipage}{0.50cm}\colorbox{MyOrangeBayes}{\LARGE\hspace{-0.03cm}$\otimes$\hspace{-0.08cm}}\end{minipage}&
  \begin{minipage}{0.50cm}\colorbox{MyOrangeBayes}{\LARGE\hspace{-0.03cm}$\otimes$\hspace{-0.08cm}}\end{minipage}&
  \begin{minipage}{0.50cm}\colorbox{MyWhite}{\large\hspace{-0.03cm}$\bigcirc$\hspace{-0.03cm}}\end{minipage}\\
  &&&&&&&&& \\[-0.15cm] \hline
  &&&&&&&&& \\[-0.15cm]
  \begin{minipage}{0.50cm}\colorbox{MyWhite}{\LARGE\hspace{-0.05cm}$\times$\hspace{-0.05cm}}\end{minipage}&
  \begin{minipage}{0.50cm}\colorbox{MyWhite}{\LARGE\hspace{-0.05cm}$\times$\hspace{-0.05cm}}\end{minipage}&
  \begin{minipage}{0.50cm}\colorbox{MyOrangeBayes}{\LARGE\hspace{-0.03cm}$\otimes$\hspace{-0.08cm}}\end{minipage}&
  \begin{minipage}{0.50cm}\colorbox{MyOrangeBayes}{\LARGE\hspace{-0.03cm}$\otimes$\hspace{-0.08cm}}\end{minipage}&
  \begin{minipage}{0.50cm}\colorbox{MyOrangeBayes}{\LARGE\hspace{-0.03cm}$\otimes$\hspace{-0.08cm}}\end{minipage}&
  \begin{minipage}{0.50cm}\colorbox{MyWhite}{\large\hspace{-0.03cm}$\bigcirc$\hspace{-0.03cm}}\end{minipage}&
  \begin{minipage}{0.50cm}\colorbox{MyWhite}{\large\hspace{-0.03cm}$\bigcirc$\hspace{-0.03cm}}\end{minipage}&
  \begin{minipage}{0.50cm}\colorbox{MyWhite}{\large\hspace{-0.03cm}$\bigcirc$\hspace{-0.03cm}}\end{minipage}&
  \begin{minipage}{0.50cm}\colorbox{MyOrangeBayes}{$\begin{matrix}\vspace{0.1cm}&\vspace{0.1cm}\end{matrix}$}\end{minipage}&
  \begin{minipage}{0.50cm}\colorbox{MyOrangeBayes}{$\begin{matrix}\vspace{0.1cm}&\vspace{0.1cm}\end{matrix}$}\end{minipage}\\
  &&&&&&&&& \\[-0.15cm] \hline
  &&&&&&&&& \\[-0.15cm]
  \begin{minipage}{0.50cm}\colorbox{MyWhite}{\large\hspace{-0.03cm}$\bigcirc$\hspace{-0.03cm}}\end{minipage}&
  \begin{minipage}{0.50cm}\colorbox{MyWhite}{\large\hspace{-0.03cm}$\bigcirc$\hspace{-0.03cm}}\end{minipage}&
  \begin{minipage}{0.50cm}\colorbox{MyWhite}{\large\hspace{-0.03cm}$\bigcirc$\hspace{-0.03cm}}\end{minipage}&
  \begin{minipage}{0.50cm}\colorbox{MyOrangeBayes}{\LARGE\hspace{-0.03cm}$\otimes$\hspace{-0.08cm}}\end{minipage}&
  \begin{minipage}{0.50cm}\colorbox{MyOrangeBayes}{\LARGE\hspace{-0.03cm}$\otimes$\hspace{-0.08cm}}\end{minipage}&
  \begin{minipage}{0.50cm}\colorbox{MyOrangeBayes}{\LARGE\hspace{-0.03cm}$\otimes$\hspace{-0.08cm}}\end{minipage}&
  \begin{minipage}{0.50cm}\colorbox{MyWhite}{\LARGE\hspace{-0.05cm}$\times$\hspace{-0.05cm}}\end{minipage}&
  \begin{minipage}{0.50cm}\colorbox{MyWhite}{\LARGE\hspace{-0.05cm}$\times$\hspace{-0.05cm}}\end{minipage}&
  \begin{minipage}{0.50cm}\colorbox{MyWhite}{\LARGE\hspace{-0.05cm}$\times$\hspace{-0.05cm}}\end{minipage}&
  \begin{minipage}{0.50cm}\colorbox{MyOrangeBayes}{$\begin{matrix}\vspace{0.1cm}&\vspace{0.1cm}\end{matrix}$}\end{minipage}\\
  &&&&&&&&& \\[-0.15cm] \hline
  &&&&&&&&& \\[-0.15cm]
  \begin{minipage}{0.50cm}\colorbox{MyWhite}{\LARGE\hspace{-0.05cm}$\times$\hspace{-0.05cm}}\end{minipage}&
  \begin{minipage}{0.50cm}\colorbox{MyWhite}{\LARGE\hspace{-0.05cm}$\times$\hspace{-0.05cm}}\end{minipage}&
  \begin{minipage}{0.50cm}\colorbox{MyWhite}{\LARGE\hspace{-0.05cm}$\times$\hspace{-0.05cm}}\end{minipage}&
  \begin{minipage}{0.50cm}\colorbox{MyOrangeBayes}{$\begin{matrix}\vspace{0.1cm}&\vspace{0.1cm}\end{matrix}$}\end{minipage}&
  \begin{minipage}{0.50cm}\colorbox{MyWhite}{\large\hspace{-0.03cm}$\bigcirc$\hspace{-0.03cm}}\end{minipage}&
  \begin{minipage}{0.50cm}\colorbox{MyWhite}{\large\hspace{-0.03cm}$\bigcirc$\hspace{-0.03cm}}\end{minipage}&
  \begin{minipage}{0.50cm}\colorbox{MyWhite}{\large\hspace{-0.03cm}$\bigcirc$\hspace{-0.03cm}}\end{minipage}&
  \begin{minipage}{0.50cm}\colorbox{MyOrangeBayes}{\LARGE\hspace{-0.03cm}$\otimes$\hspace{-0.08cm}}\end{minipage}&
  \begin{minipage}{0.50cm}\colorbox{MyOrangeBayes}{\LARGE\hspace{-0.03cm}$\otimes$\hspace{-0.08cm}}\end{minipage}&
  \begin{minipage}{0.50cm}\colorbox{MyOrangeBayes}{\LARGE\hspace{-0.03cm}$\otimes$\hspace{-0.08cm}}\end{minipage}\\
  &&&&&&&&& \\[-0.15cm] \hline
  &&&&&&&&& \\[-0.15cm]
  \begin{minipage}{0.50cm}\colorbox{MyWhite}{\large\hspace{-0.03cm}$\bigcirc$\hspace{-0.03cm}}\end{minipage}&
  \begin{minipage}{0.50cm}\colorbox{MyWhite}{\large\hspace{-0.03cm}$\bigcirc$\hspace{-0.03cm}}\end{minipage}&
  \begin{minipage}{0.50cm}\colorbox{MyOrangeBayes}{$\begin{matrix}\vspace{0.1cm}&\vspace{0.1cm}\end{matrix}$}\end{minipage}&
  \begin{minipage}{0.50cm}\colorbox{MyWhite}{\LARGE\hspace{-0.05cm}$\times$\hspace{-0.05cm}}\end{minipage}&
  \begin{minipage}{0.50cm}\colorbox{MyOrangeBayes}{\LARGE\hspace{-0.03cm}$\otimes$\hspace{-0.08cm}}\end{minipage}&
  \begin{minipage}{0.50cm}\colorbox{MyOrangeBayes}{\LARGE\hspace{-0.03cm}$\otimes$\hspace{-0.08cm}}\end{minipage}&
  \begin{minipage}{0.50cm}\colorbox{MyWhite}{\LARGE\hspace{-0.05cm}$\times$\hspace{-0.05cm}}\end{minipage}&
  \begin{minipage}{0.50cm}\colorbox{MyOrangeBayes}{$\begin{matrix}\vspace{0.1cm}&\vspace{0.1cm}\end{matrix}$}\end{minipage}&
  \begin{minipage}{0.50cm}\colorbox{MyWhite}{\large\hspace{-0.03cm}$\bigcirc$\hspace{-0.03cm}}\end{minipage}&
  \begin{minipage}{0.50cm}\colorbox{MyWhite}{\large\hspace{-0.03cm}$\bigcirc$\hspace{-0.03cm}}\end{minipage}\\
  &&&&&&&&& \\[-0.15cm] \hline
  &&&&&&&&& \\[-0.15cm]
  \begin{minipage}{0.50cm}\colorbox{MyOrangeBayes}{$\begin{matrix}\vspace{0.1cm}&\vspace{0.1cm}\end{matrix}$}\end{minipage}&
  \begin{minipage}{0.50cm}\colorbox{MyOrangeBayes}{$\begin{matrix}\vspace{0.1cm}&\vspace{0.1cm}\end{matrix}$}\end{minipage}&
  \begin{minipage}{0.50cm}\colorbox{MyWhite}{\large\hspace{-0.03cm}$\bigcirc$\hspace{-0.03cm}}\end{minipage}&
  \begin{minipage}{0.50cm}\colorbox{MyWhite}{\large\hspace{-0.03cm}$\bigcirc$\hspace{-0.03cm}}\end{minipage}&
  \begin{minipage}{0.50cm}\colorbox{MyWhite}{\large\hspace{-0.03cm}$\bigcirc$\hspace{-0.03cm}}\end{minipage}&
  \begin{minipage}{0.50cm}\colorbox{MyOrangeBayes}{\LARGE\hspace{-0.03cm}$\otimes$\hspace{-0.08cm}}\end{minipage}&
  \begin{minipage}{0.50cm}\colorbox{MyWhite}{\LARGE\hspace{-0.05cm}$\times$\hspace{-0.05cm}}\end{minipage}&
  \begin{minipage}{0.50cm}\colorbox{MyWhite}{\LARGE\hspace{-0.05cm}$\times$\hspace{-0.05cm}}\end{minipage}&
  \begin{minipage}{0.50cm}\colorbox{MyWhite}{\LARGE\hspace{-0.05cm}$\times$\hspace{-0.05cm}}\end{minipage}&
  \begin{minipage}{0.50cm}\colorbox{MyWhite}{\LARGE\hspace{-0.05cm}$\times$\hspace{-0.05cm}}\end{minipage}\\
  &&&&&&&&& \\[-0.15cm] \hline
  &&&&&&&&& \\[-0.15cm]
  \begin{minipage}{0.50cm}\colorbox{MyOrangeBayes}{$\begin{matrix}\vspace{0.1cm}&\vspace{0.1cm}\end{matrix}$}\end{minipage}&
  \begin{minipage}{0.50cm}\colorbox{MyWhite}{\LARGE\hspace{-0.05cm}$\times$\hspace{-0.05cm}}\end{minipage}&
  \begin{minipage}{0.50cm}\colorbox{MyWhite}{\LARGE\hspace{-0.05cm}$\times$\hspace{-0.05cm}}\end{minipage}&
  \begin{minipage}{0.50cm}\colorbox{MyWhite}{\LARGE\hspace{-0.05cm}$\times$\hspace{-0.05cm}}\end{minipage}&
  \begin{minipage}{0.50cm}\colorbox{MyOrangeBayes}{\LARGE\hspace{-0.03cm}$\otimes$\hspace{-0.08cm}}\end{minipage}&
  \begin{minipage}{0.50cm}\colorbox{MyOrangeBayes}{\LARGE\hspace{-0.03cm}$\otimes$\hspace{-0.08cm}}\end{minipage}&
  \begin{minipage}{0.50cm}\colorbox{MyOrangeBayes}{\LARGE\hspace{-0.03cm}$\otimes$\hspace{-0.08cm}}\end{minipage}&
  \begin{minipage}{0.50cm}\colorbox{MyWhite}{\large\hspace{-0.03cm}$\bigcirc$\hspace{-0.03cm}}\end{minipage}&
  \begin{minipage}{0.50cm}\colorbox{MyOrangeBayes}{$\begin{matrix}\vspace{0.1cm}&\vspace{0.1cm}\end{matrix}$}\end{minipage}&
  \begin{minipage}{0.50cm}\colorbox{MyOrangeBayes}{$\begin{matrix}\vspace{0.1cm}&\vspace{0.1cm}\end{matrix}$}\end{minipage}\\
  &&&&&&&&& \\[-0.15cm] \hline
  &&&&&&&&& \\[-0.15cm]
  \begin{minipage}{0.50cm}\colorbox{MyOrangeBayes}{\LARGE\hspace{-0.03cm}$\otimes$\hspace{-0.08cm}}\end{minipage}&
  \begin{minipage}{0.50cm}\colorbox{MyWhite}{\large\hspace{-0.03cm}$\bigcirc$\hspace{-0.03cm}}\end{minipage}&
  \begin{minipage}{0.50cm}\colorbox{MyOrangeBayes}{$\begin{matrix}\vspace{0.1cm}&\vspace{0.1cm}\end{matrix}$}\end{minipage}&
  \begin{minipage}{0.50cm}\colorbox{MyWhite}{\LARGE\hspace{-0.05cm}$\times$\hspace{-0.05cm}}\end{minipage}&
  \begin{minipage}{0.50cm}\colorbox{MyWhite}{\LARGE\hspace{-0.05cm}$\times$\hspace{-0.05cm}}\end{minipage}&
  \begin{minipage}{0.50cm}\colorbox{MyOrangeBayes}{$\begin{matrix}\vspace{0.1cm}&\vspace{0.1cm}\end{matrix}$}\end{minipage}&
  \begin{minipage}{0.50cm}\colorbox{MyWhite}{\large\hspace{-0.03cm}$\bigcirc$\hspace{-0.03cm}}\end{minipage}&
  \begin{minipage}{0.50cm}\colorbox{MyOrangeBayes}{\LARGE\hspace{-0.03cm}$\otimes$\hspace{-0.08cm}}\end{minipage}&
  \begin{minipage}{0.50cm}\colorbox{MyOrangeBayes}{\LARGE\hspace{-0.03cm}$\otimes$\hspace{-0.08cm}}\end{minipage}&
  \begin{minipage}{0.50cm}\colorbox{MyWhite}{\large\hspace{-0.03cm}$\bigcirc$\hspace{-0.03cm}}\end{minipage}\\
  &&&&&&&&& \\[-0.15cm] \hline
  &&&&&&&&& \\[-0.15cm]
  \begin{minipage}{0.50cm}\colorbox{MyWhite}{\large\hspace{-0.03cm}$\bigcirc$\hspace{-0.03cm}}\end{minipage}&
  \begin{minipage}{0.50cm}\colorbox{MyOrangeBayes}{\LARGE\hspace{-0.03cm}$\otimes$\hspace{-0.08cm}}\end{minipage}&
  \begin{minipage}{0.50cm}\colorbox{MyWhite}{\LARGE\hspace{-0.05cm}$\times$\hspace{-0.05cm}}\end{minipage}&
  \begin{minipage}{0.50cm}\colorbox{MyOrangeBayes}{$\begin{matrix}\vspace{0.1cm}&\vspace{0.1cm}\end{matrix}$}\end{minipage}&
  \begin{minipage}{0.50cm}\colorbox{MyWhite}{\large\hspace{-0.03cm}$\bigcirc$\hspace{-0.03cm}}\end{minipage}&
  \begin{minipage}{0.50cm}\colorbox{MyOrangeBayes}{\LARGE\hspace{-0.03cm}$\otimes$\hspace{-0.08cm}}\end{minipage}&
  \begin{minipage}{0.50cm}\colorbox{MyOrangeBayes}{\LARGE\hspace{-0.03cm}$\otimes$\hspace{-0.08cm}}\end{minipage}&
  \begin{minipage}{0.50cm}\colorbox{MyWhite}{\large\hspace{-0.03cm}$\bigcirc$\hspace{-0.03cm}}\end{minipage}&
  \begin{minipage}{0.50cm}\colorbox{MyOrangeBayes}{$\begin{matrix}\vspace{0.1cm}&\vspace{0.1cm}\end{matrix}$}\end{minipage}&
  \begin{minipage}{0.50cm}\colorbox{MyWhite}{\LARGE\hspace{-0.05cm}$\times$\hspace{-0.05cm}}\end{minipage}\\
  &&&&&&&&& \\[-0.15cm] \hline
  &&&&&&&&& \\[-0.15cm]
  \begin{minipage}{0.50cm}\colorbox{MyWhite}{\large\hspace{-0.03cm}$\bigcirc$\hspace{-0.03cm}}\end{minipage}&
  \begin{minipage}{0.50cm}\colorbox{MyOrangeBayes}{\LARGE\hspace{-0.03cm}$\otimes$\hspace{-0.08cm}}\end{minipage}&
  \begin{minipage}{0.50cm}\colorbox{MyOrangeBayes}{\LARGE\hspace{-0.03cm}$\otimes$\hspace{-0.08cm}}\end{minipage}&
  \begin{minipage}{0.50cm}\colorbox{MyOrangeBayes}{\LARGE\hspace{-0.03cm}$\otimes$\hspace{-0.08cm}}\end{minipage}&
  \begin{minipage}{0.50cm}\colorbox{MyOrangeBayes}{\LARGE\hspace{-0.03cm}$\otimes$\hspace{-0.08cm}}\end{minipage}&
  \begin{minipage}{0.50cm}\colorbox{MyWhite}{\large\hspace{-0.03cm}$\bigcirc$\hspace{-0.03cm}}\end{minipage}&
  \begin{minipage}{0.50cm}\colorbox{MyOrangeBayes}{$\begin{matrix}\vspace{0.1cm}&\vspace{0.1cm}\end{matrix}$}\end{minipage}&
  \begin{minipage}{0.50cm}\colorbox{MyWhite}{\LARGE\hspace{-0.05cm}$\times$\hspace{-0.05cm}}\end{minipage}&
  \begin{minipage}{0.50cm}\colorbox{MyOrangeBayes}{\LARGE\hspace{-0.03cm}$\otimes$\hspace{-0.08cm}}\end{minipage}&
  \begin{minipage}{0.50cm}\colorbox{MyWhite}{\large\hspace{-0.03cm}$\bigcirc$\hspace{-0.03cm}}\end{minipage}\\
  &&&&&&&&& \\[-0.15cm] \hline
\end{tabular}

\vspace{-12.33cm}

\hspace{8.65cm}{\Huge $\mathscr R$}

\vspace{7.80cm}

\hspace{0.79cm}{\Huge $\mathscr H$}\hspace{70mm}{\Huge $\mathscr M$}

\vspace{3.10cm}

\caption{Venn diagrams of the three co$\sim$events  $\mathscr H, \mathscr{R}$, and $\mathscr{M}=(\mathscr H\Delta\,\mathscr R)^c$ $\Big($
\colorbox{MyGreen}{\large$\times$} , \colorbox{MyAqua0}{\small$\bigcirc$} ,
\colorbox{MyOrangeBayes}{\large$\otimes$}\hspace{1pt}\colorbox{MyOrangeBayes}{\protect\phantom{\large$\otimes$}} $\Big)$ in the co$\sim$event-based
Bayesian scheme. \label{10x10Bayes}}
\end{table}

\clearpage Table \ref{10x10Bayes} shows the results of the tasting. The believability and probability characteristics of the co$\sim$event $\mathscr
M$ have the following forms
\begin{eqnarray}\nonumber
\breve{b}_{\frak X_{\!\mathscr M}}=&\{b_{x}\colon x\in\frak X_{\!\mathscr M}\}=\{b_1,\ldots,b_{10}\}&
=\{0.02,0.04,0.06,0.09,0.11,0.13,0.15,0.17,0.19,0.04\},\\
\nonumber
\breve{p}_{\frak X_{\!\mathscr M}}=&\{p_{x}\colon x\in\frak X_{\!\mathscr M}\}=\{p_1,\ldots,p_{10}\}&
=\{0.60,0.50,0.40,0.40,0.40,0.30,0.60,0.50,0.50,0.60\},\\
\nonumber
\mathbf p_{\frak X_{\!\mathscr M}}=&\{p(X\!/\!\!/\frak X_{\!\mathscr M})\colon\!X\in\rS^{\frak X_{\!\mathscr M}}\}=\{p(1),\ldots,p(10)\}&
=\{0.10,0.10,0.10,0.10,0.10,0.10,0.10,0.10,0.10,0.10\}
\end{eqnarray}
where \cite{Vorobyev2016famems2,Vorobyev2016famems3} for $i=1,\ldots,10$, and  $j=1,\ldots,10$
\begin{eqnarray}
\nonumber
b_i&=&b_{x_i}=\mathbf B(\bra{x_i}),\\
\nonumber
p_i&=&p_{x_i}=\mathbf P(\ket{x_i}),\\
\nonumber
p(j)&=&p(X_j\!/\!\!/\frak X_{\!\mathscr M})\!=\!\mathbf P(\ket{\textsf{ter}(X_j)}).
\end{eqnarray}

The fact that the wine from 10 bottles has good quality in the opinion of 10 tasters has the \emph{prior certainty}
\begin{equation}\label{certainty}
\bfPhi(\mathscr M)=
\sum_{\braket{x_i|X_j}\in\mathscr M} \bfPhi(\braket{x_i|X_j})=
\sum_{j=0}^{10}
\sum_{i=0}^{10} b_i p(j) \mathbf 1_{\!X_{\!j}}\!(x_i)
=0.470,
\end{equation}

With the \emph{posterior certainty} 0.489 the \emph{posterior believability distribution} of the co$\sim$event $\mathscr M$ calculated by the
co$\sim$event-based Bayes' theorem have the following values\footnote{Here the new co$\sim$event-based semantics is that a believability distribution
and a probability distribution of each co$\sim$event has the corresponding value of the certainty measure of this co$\sim$event.}:
\begin{eqnarray}
\label{b-aposteriori}
\breve{b}^{\mathrm{post}}_{\frak X_{\!\mathscr M}}&=&
\{0.027,0.045,0.054,0.072,0.090,0.081,0.190,0.181,0.204,0.054\},\\
\label{Phi-aposteriori}
\bfPhi^{\mathrm{post}}(\mathscr M)&=&
\sum_{j=0}^{10}
\sum_{i=0}^{10} b^{\mathrm{post}}_i p(j) \mathbf 1_{\!X_{\!j}}\!(x_i)
= 0.489.
\end{eqnarray}

\section{A recurrent sequence of co$\sim$event-based Bayes' formulas}

After computing the a posteriori characteristics of the co$\sim$event $\mathscr M$ (see, for example, (\ref{b-aposteriori}) and
(\ref{Phi-aposteriori})), it becomes possible to use the co$\sim$event-based Bayes' theorem once again, applying it not to the a priori, but to the a
posteriori characteristics obtained in the first step. This recurrent procedure can be applied repeatedly to trace changes in the posterior
characteristics of the event $\mathscr M$.

\texttt{\indent Theorem \!\refstepcounter{ctrTh}\arabic{ctrTh}\,\label{Th-recurrent}\itshape\footnotesize (the limit believability distribution in a
recurrent co$\sim$event-based Bayes' formula)\!.} \emph{Let the subset $X_{\max}=\{x\colon x\in\frak X_{\!\mathscr M}, p_x=\max_{y\in \frak
X_{\!\mathscr M}}p_y\}\subseteq\frak X_{\!\mathscr M}$ consists of all such $x\in\frak X_{\!\mathscr M}$ for which the probabilities $p_x$ take a
maximum value. Then the limit believability distribution in a recurrent co$\sim$event-based Bayes' formula
\begin{eqnarray}
\label{b-n}
b^{(n+1)}_{i}&=&b^{(n)}_{i}\frac{p_{i}}{\bfPhi^{(n)}(\mathscr M)}, \ \ \ i=1,\ldots,N,\\
\label{Phi-n}
\bfPhi^{(n+1)}(\mathscr M)&=&
\sum_{j=0}^{M}
\sum_{i=0}^{N} b^{(n+1)}_i p(j) \mathbf 1_{\!X_{\!j}}\!(x_i)
\end{eqnarray}
where\\[-30pt]
\begin{eqnarray}
\label{b-1}
b^{(1)}_{i}&=&b^{\mathrm{post}}_{i}, \ \ \ i=1,\ldots,N,\\
\label{Phi-1}
\bfPhi^{(1)}(\mathscr M)&=&\bfPhi^{\mathrm{post}}(\mathscr M)
\end{eqnarray}
are the characteristics of the event $\mathscr M$, computed in the first step, has the following form:
\begin{eqnarray}
\label{b-limit}
\lim_{n\to\infty} b^{(n)}_{x}&=&
\begin{cases}
\frac{\displaystyle b_{x}}{\displaystyle \sum_{x\in X_{\max}} b_{x}}, & x\in X_{\max},\\
0,                                        & \mbox{otherwise}.
\end{cases}
\end{eqnarray}
} \textsf{Proof} follows from the Banach fixed-point theorem \cite{Banach1922}.

\clearpage
\begin{figure}[ht!]
\scriptsize
\centering
\includegraphics[width=3.30in]{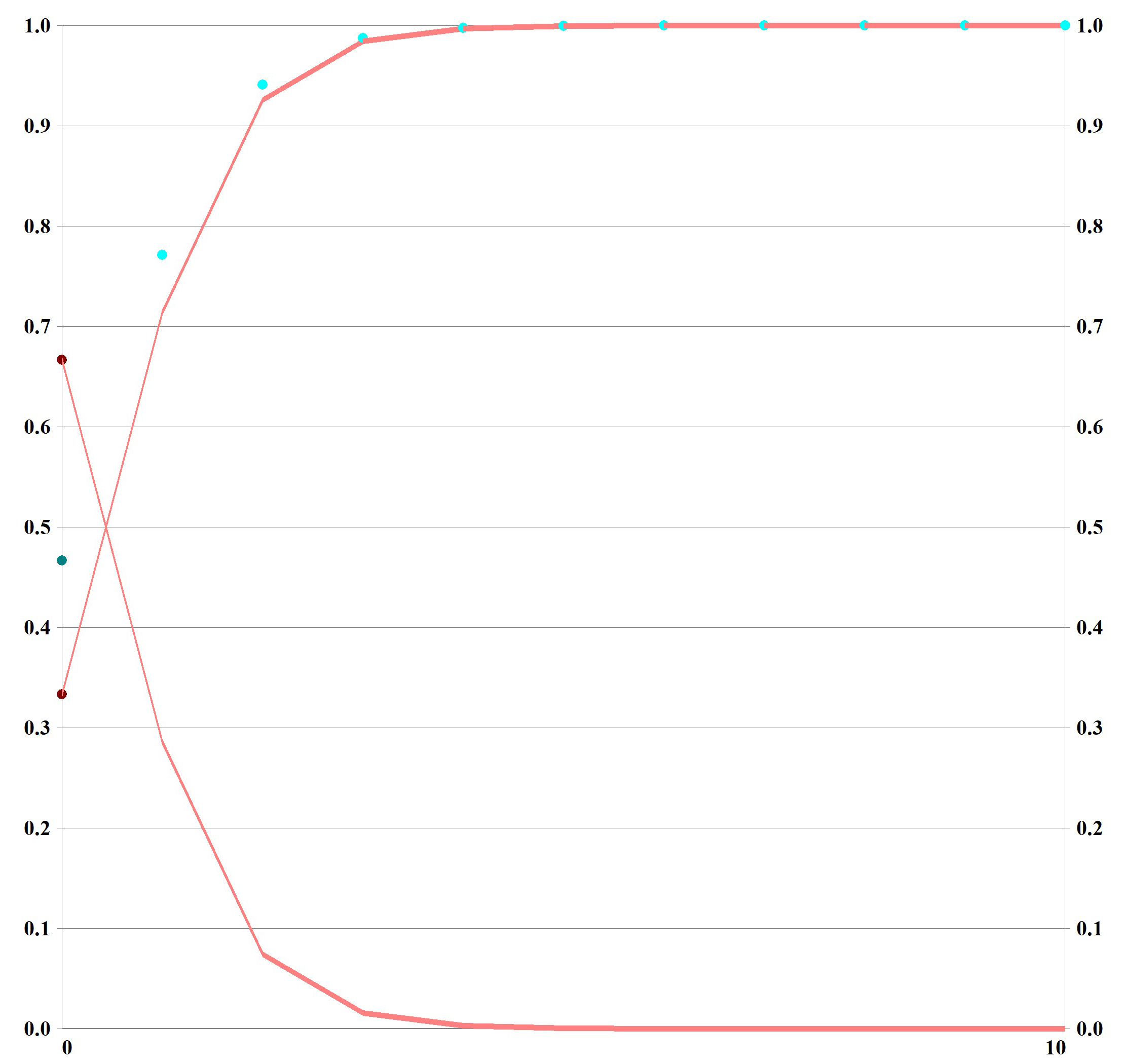}\hspace{10pt}
\includegraphics[width=3.30in]{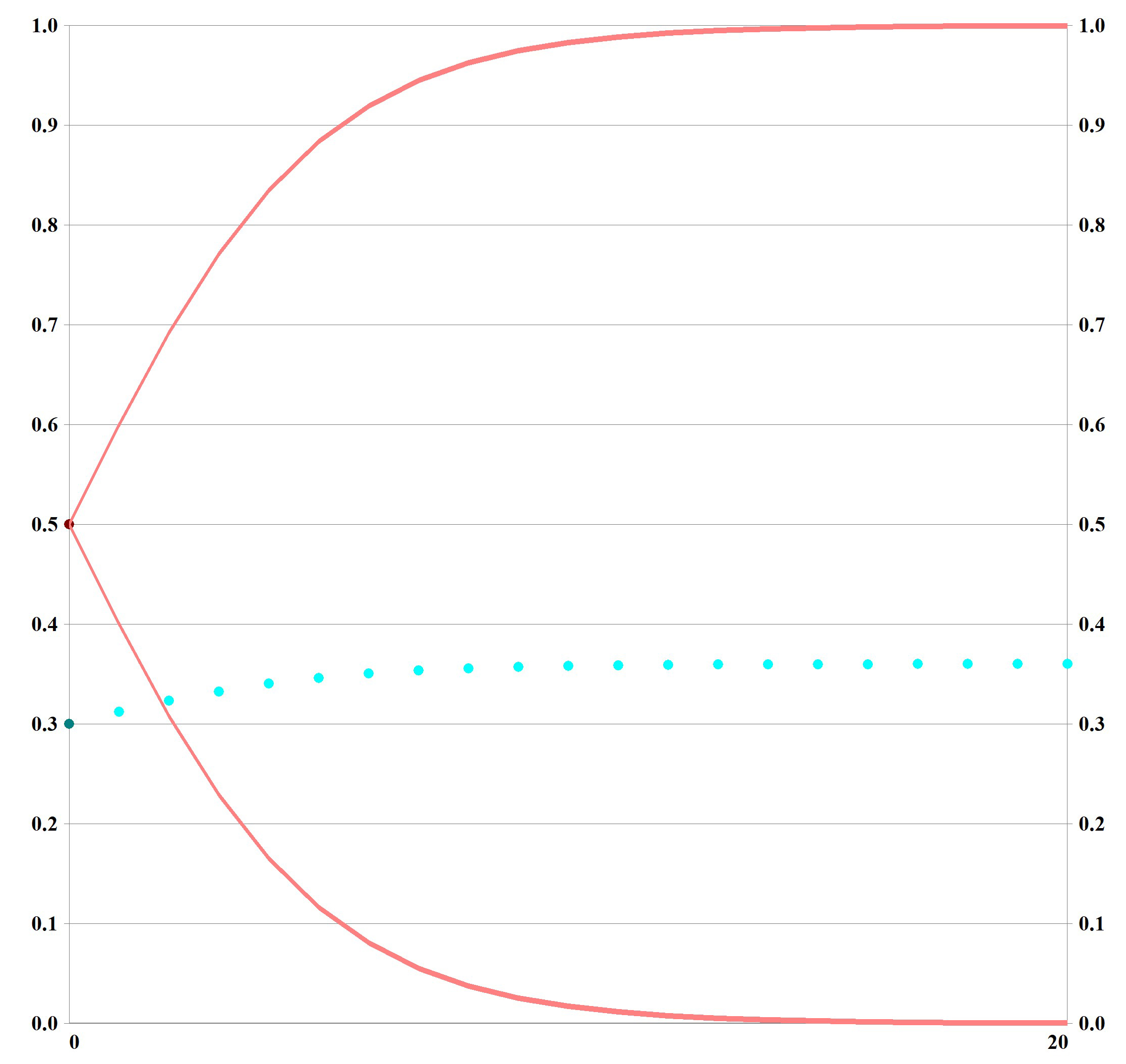}

\caption{The recurrent sequence of calculations by the co$\sim$event-based Bayes' formula: \emph{a posteriori} believability distributions
(\colorbox{MyOrangeBayes}{orange}) and \emph{a posteriori} certainty of $\mathscr M$ (\colorbox{MyAqua}{aqua}). Example 3 (left) and Example 4
(right).\label{Ex3Ex4}}
\end{figure}

\begin{table}[!h]\small\centering\begin{tabular}{ll|l|l|l|l|l|l|l|l|l|l|}&$p(X/\!\!/\frak X_{\!\mathscr M})$
&0.020&0.020&0.005&0.005&0.200&0.200&0.200&0.075&0.075&0.200\\ $p_{x/\!\!/\frak X_{\!\mathscr M}}$&$b_{x/\!\!/\frak X_{\!\mathscr M}}$  &  &  &  &  &  &  &  &  &  &\\[1pt] \hline
&&&&&&&&&&&\\[-10pt]
0.200&0.667&\colorbox{MyWhite}{0.013}&\colorbox{MyWhite}{0.013}&\colorbox{MyWhite}{0.003}&\colorbox{MyWhite}{0.003}&\colorbox{MyWhite}{0.133}&\colorbox{MyWhite}{0.133}&\colorbox{MyWhite}{0.133}&\colorbox{MyWhite}{0.050}&\colorbox{MyWhite}{0.050}&\colorbox{MyOrangeBayes}{0.133}\\[1pt] \hline
&&&&&&&&&&&\\[-10pt]
1.000&0.333&\colorbox{MyOrangeBayes}{0.007}&\colorbox{MyOrangeBayes}{0.007}&\colorbox{MyOrangeBayes}{0.002}&\colorbox{MyOrangeBayes}{0.002}&\colorbox{MyOrangeBayes}{0.067}&\colorbox{MyOrangeBayes}{0.067}&\colorbox{MyOrangeBayes}{0.067}&\colorbox{MyOrangeBayes}{0.025}&\colorbox{MyOrangeBayes}{0.025}&\colorbox{MyOrangeBayes}{0.067}\\[1pt] \hline
&&&&&&&&&&&\\[-10pt]
\end{tabular}

\vspace{10pt}

\begin{tabular}{ll|l|l|l|l|l|l|l|l|l|l|} &$p(X/\!\!/\frak X_{\!\mathscr M})$
&0.020&0.020&0.005&0.005&0.200&0.200&0.200&0.075&0.075&0.200\\ $p_{x/\!\!/\frak X_{\!\mathscr M}}$&$b^{\mbox{\tiny final}}_{x/\!\!/\frak X_{\!\mathscr M}}$  &  &  &  &  &  &  &  &  &  &\\[1pt] \hline
&&&&&&&&&&&\\[-10pt]
0.200&0\hspace{18pt}&\colorbox{MyWhite}{0\hspace{18pt}}&\colorbox{MyWhite}{0\hspace{18pt}}&\colorbox{MyWhite}{0\hspace{18pt}}&\colorbox{MyWhite}{0\hspace{18pt}}&\colorbox{MyWhite}{0\hspace{18pt}}&\colorbox{MyWhite}{0\hspace{18pt}}&\colorbox{MyWhite}{0\hspace{18pt}}&\colorbox{MyWhite}{0\hspace{18pt}}&\colorbox{MyWhite}{0\hspace{18pt}}&\colorbox{MyOrangeBayes}{0\hspace{18pt}}\\[1pt] \hline
&&&&&&&&&&&\\[-10pt]
1.000&1.000&\colorbox{MyOrangeBayes}{0.020}&\colorbox{MyOrangeBayes}{0.020}&\colorbox{MyOrangeBayes}{0.005}&\colorbox{MyOrangeBayes}{0.005}&\colorbox{MyOrangeBayes}{0.200}&\colorbox{MyOrangeBayes}{0.200}&\colorbox{MyOrangeBayes}{0.200}&\colorbox{MyOrangeBayes}{0.075}&\colorbox{MyOrangeBayes}{0.075}&\colorbox{MyOrangeBayes}{0.200}\\[1pt] \hline
&&&&&&&&&&&\\[-10pt]
\end{tabular}
\caption{The believability $\left\{b_{x/\!\!/\frak X_{\!\mathscr M}}\colon x\in\frak X_{\!\mathscr M}\right\}$, probability $\left\{p(X/\!\!/\frak
X_{\!\mathscr M})\colon X\in\protect\rsS^{\frak X_{\!\mathscr M}}\right\}$, and  certainty $\left\{\varphi_{x}(X/\!\!/\frak X_{\!\mathscr M})\colon
x\in\frak X_{\!\mathscr M}, X\in\protect\rsS^{\frak X_{\!\mathscr M}}\right\}$  distributions of the co$\sim$event \colorbox{MyOrangeBayes}{$\mathscr
M$} for Example 3. $\bfPhi(\mathscr M)= 0.467$, and $\bfPhi^{\mbox{\tiny final}}(\mathscr M)= 1.000$. Initial  distributions (in the top table) and
final distributions (in the bottom table) for the 10-th iterative calculation by the co$\sim$event-based Bayes theorem.\label{tabEx3}} \end{table}

\begin{table}[!h]\small\centering\begin{tabular}{ll|l|l|l|l|l|l|l|l|l|l|}&$p(X/\!\!/\frak X_{\!\mathscr M})$
&0.020&0.020&0.005&0.005&0.200&0.200&0.200&0.075&0.075&0.200\\ $p_{x/\!\!/\frak X_{\!\mathscr M}}$&$b_{x/\!\!/\frak X_{\!\mathscr M}}$  &  &  &  &  &  &  &  &  &  &\\[1pt] \hline
&&&&&&&&&&&\\[-10pt]
0.240&0.500&\colorbox{MyOrangeBayes}{0.010}&\colorbox{MyOrangeBayes}{0.010}&\colorbox{MyWhite}{0.003}&\colorbox{MyWhite}{0.003}&\colorbox{MyWhite}{0.100}&\colorbox{MyWhite}{0.100}&\colorbox{MyWhite}{0.100}&\colorbox{MyWhite}{0.038}&\colorbox{MyWhite}{0.038}&\colorbox{MyOrangeBayes}{0.100}\\[1pt] \hline
&&&&&&&&&&&\\[-10pt]
0.360&0.500&\colorbox{MyWhite}{0.010}&\colorbox{MyWhite}{0.010}&\colorbox{MyOrangeBayes}{0.003}&\colorbox{MyOrangeBayes}{0.003}&\colorbox{MyWhite}{0.100}&\colorbox{MyWhite}{0.100}&\colorbox{MyWhite}{0.100}&\colorbox{MyOrangeBayes}{0.038}&\colorbox{MyOrangeBayes}{0.038}&\colorbox{MyOrangeBayes}{0.100}\\[1pt] \hline
&&&&&&&&&&&\\[-10pt]
\end{tabular}

\vspace{10pt}

\begin{tabular}{ll|l|l|l|l|l|l|l|l|l|l|} &$p(X/\!\!/\frak X_{\!\mathscr M})$
&0.020&0.020&0.005&0.005&0.200&0.200&0.200&0.075&0.075&0.200\\ $p_{x/\!\!/\frak X_{\!\mathscr M}}$&$b^{\mbox{\tiny final}}_{x/\!\!/\frak X_{\!\mathscr M}}$  &  &  &  &  &  &  &  &  &  &\\[1pt] \hline
&&&&&&&&&&&\\[-10pt]
0.240&0\hspace{18pt}&\colorbox{MyOrangeBayes}{0\hspace{18pt}}&\colorbox{MyOrangeBayes}{0\hspace{18pt}}&\colorbox{MyWhite}{0\hspace{18pt}}&\colorbox{MyWhite}{0\hspace{18pt}}&\colorbox{MyWhite}{0\hspace{18pt}}&\colorbox{MyWhite}{0\hspace{18pt}}&\colorbox{MyWhite}{0\hspace{18pt}}&\colorbox{MyWhite}{0\hspace{18pt}}&\colorbox{MyWhite}{0\hspace{18pt}}&\colorbox{MyOrangeBayes}{0\hspace{18pt}}\\[1pt] \hline
&&&&&&&&&&&\\[-10pt]
0.360&1.000&\colorbox{MyWhite}{0.020}&\colorbox{MyWhite}{0.020}&\colorbox{MyOrangeBayes}{0.005}&\colorbox{MyOrangeBayes}{0.005}&\colorbox{MyWhite}{0.200}&\colorbox{MyWhite}{0.200}&\colorbox{MyWhite}{0.200}&\colorbox{MyOrangeBayes}{0.075}&\colorbox{MyOrangeBayes}{0.075}&\colorbox{MyOrangeBayes}{0.200}\\[1pt] \hline
&&&&&&&&&&&\\[-10pt]
\end{tabular}
\caption{The believability $\left\{b_{x/\!\!/\frak X_{\!\mathscr M}}\colon x\in\frak X_{\!\mathscr M}\right\}$, probability $\left\{p(X/\!\!/\frak
X_{\!\mathscr M})\colon X\in\protect\rsS^{\frak X_{\!\mathscr M}}\right\}$, and  certainty $\left\{\varphi_{x}(X/\!\!/\frak X_{\!\mathscr M})\colon
x\in\frak X_{\!\mathscr M}, X\in\protect\rsS^{\frak X_{\!\mathscr M}}\right\}$  distributions of the co$\sim$event \colorbox{MyOrangeBayes}{$\mathscr
M$} for Example 4. $\bfPhi(\mathscr M)= 0.300$, and $\bfPhi^{\mbox{\tiny final}}(\mathscr M)= 0.360$. Initial  distributions (in the top table) and
final distributions (in the bottom table) for the 20-th iterative calculation by the co$\sim$event-based Bayes theorem.\label{tabEx4}} \end{table}

\clearpage
\begin{figure}[ht!]
\scriptsize
\centering
\includegraphics[width=3.10in]{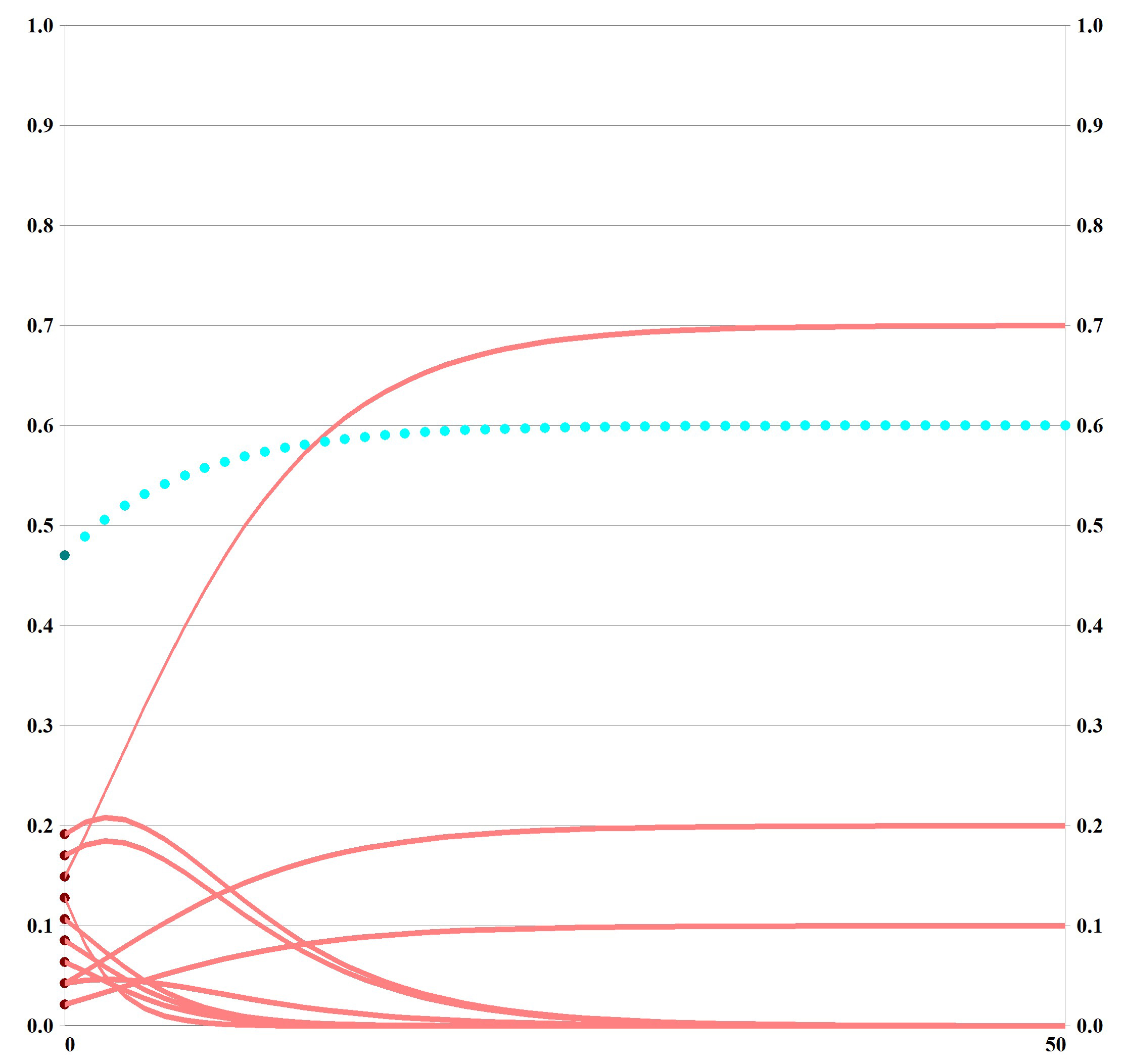}

\caption{The recurrent sequence of calculations by the co$\sim$event-based Bayes' formula: \emph{a posteriori} believability distributions
(\colorbox{MyOrangeBayes}{orange}) and \emph{a posteriori} certainty of $\mathscr M$ (\colorbox{MyAqua}{aqua}). Example 5.\label{Ex5}}
\end{figure}

\begin{table}[!h]\small\centering\begin{tabular}{ll|l|l|l|l|l|l|l|l|l|l|}&$p(X/\!\!/\frak X_{\!\mathscr M})$
&0.100&0.100&0.100&0.100&0.100&0.100&0.100&0.100&0.100&0.100\\ $p_{x/\!\!/\frak X_{\!\mathscr M}}$&$b_{x/\!\!/\frak X_{\!\mathscr M}}$  &  &  &  &  &  &  &  &  &  &\\[1pt] \hline
&&&&&&&&&&&\\[-10pt]
0.600&0.021&\colorbox{MyWhite}{0.002}&\colorbox{MyOrangeBayes}{0.002}&\colorbox{MyOrangeBayes}{0.002}&\colorbox{MyWhite}{0.002}&\colorbox{MyOrangeBayes}{0.002}&\colorbox{MyWhite}{0.002}&\colorbox{MyOrangeBayes}{0.002}&\colorbox{MyOrangeBayes}{0.002}&\colorbox{MyOrangeBayes}{0.002}&\colorbox{MyWhite}{0.002}\\[1pt] \hline
&&&&&&&&&&&\\[-10pt]
0.500&0.043&\colorbox{MyWhite}{0.004}&\colorbox{MyWhite}{0.004}&\colorbox{MyOrangeBayes}{0.004}&\colorbox{MyOrangeBayes}{0.004}&\colorbox{MyOrangeBayes}{0.004}&\colorbox{MyWhite}{0.004}&\colorbox{MyWhite}{0.004}&\colorbox{MyWhite}{0.004}&\colorbox{MyOrangeBayes}{0.004}&\colorbox{MyOrangeBayes}{0.004}\\[1pt] \hline
&&&&&&&&&&&\\[-10pt]
0.400&0.064&\colorbox{MyWhite}{0.006}&\colorbox{MyWhite}{0.006}&\colorbox{MyWhite}{0.006}&\colorbox{MyOrangeBayes}{0.006}&\colorbox{MyOrangeBayes}{0.006}&\colorbox{MyOrangeBayes}{0.006}&\colorbox{MyWhite}{0.006}&\colorbox{MyWhite}{0.006}&\colorbox{MyWhite}{0.006}&\colorbox{MyOrangeBayes}{0.006}\\[1pt] \hline
&&&&&&&&&&&\\[-10pt]
0.400&0.085&\colorbox{MyWhite}{0.009}&\colorbox{MyWhite}{0.009}&\colorbox{MyWhite}{0.009}&\colorbox{MyOrangeBayes}{0.009}&\colorbox{MyWhite}{0.009}&\colorbox{MyWhite}{0.009}&\colorbox{MyWhite}{0.009}&\colorbox{MyOrangeBayes}{0.009}&\colorbox{MyOrangeBayes}{0.009}&\colorbox{MyOrangeBayes}{0.009}\\[1pt] \hline
&&&&&&&&&&&\\[-10pt]
0.400&0.106&\colorbox{MyWhite}{0.011}&\colorbox{MyWhite}{0.011}&\colorbox{MyOrangeBayes}{0.011}&\colorbox{MyWhite}{0.011}&\colorbox{MyOrangeBayes}{0.011}&\colorbox{MyOrangeBayes}{0.011}&\colorbox{MyWhite}{0.011}&\colorbox{MyOrangeBayes}{0.011}&\colorbox{MyWhite}{0.011}&\colorbox{MyWhite}{0.011}\\[1pt] \hline
&&&&&&&&&&&\\[-10pt]
0.300&0.128&\colorbox{MyOrangeBayes}{0.013}&\colorbox{MyOrangeBayes}{0.013}&\colorbox{MyWhite}{0.013}&\colorbox{MyWhite}{0.013}&\colorbox{MyWhite}{0.013}&\colorbox{MyOrangeBayes}{0.013}&\colorbox{MyWhite}{0.013}&\colorbox{MyWhite}{0.013}&\colorbox{MyWhite}{0.013}&\colorbox{MyWhite}{0.013}\\[1pt] \hline
&&&&&&&&&&&\\[-10pt]
0.600&0.149&\colorbox{MyOrangeBayes}{0.015}&\colorbox{MyWhite}{0.015}&\colorbox{MyWhite}{0.015}&\colorbox{MyWhite}{0.015}&\colorbox{MyOrangeBayes}{0.015}&\colorbox{MyOrangeBayes}{0.015}&\colorbox{MyOrangeBayes}{0.015}&\colorbox{MyWhite}{0.015}&\colorbox{MyOrangeBayes}{0.015}&\colorbox{MyOrangeBayes}{0.015}\\[1pt] \hline
&&&&&&&&&&&\\[-10pt]
0.500&0.170&\colorbox{MyOrangeBayes}{0.017}&\colorbox{MyWhite}{0.017}&\colorbox{MyOrangeBayes}{0.017}&\colorbox{MyWhite}{0.017}&\colorbox{MyWhite}{0.017}&\colorbox{MyOrangeBayes}{0.017}&\colorbox{MyWhite}{0.017}&\colorbox{MyOrangeBayes}{0.017}&\colorbox{MyOrangeBayes}{0.017}&\colorbox{MyWhite}{0.017}\\[1pt] \hline
&&&&&&&&&&&\\[-10pt]
0.500&0.191&\colorbox{MyWhite}{0.019}&\colorbox{MyOrangeBayes}{0.019}&\colorbox{MyWhite}{0.019}&\colorbox{MyOrangeBayes}{0.019}&\colorbox{MyWhite}{0.019}&\colorbox{MyOrangeBayes}{0.019}&\colorbox{MyOrangeBayes}{0.019}&\colorbox{MyWhite}{0.019}&\colorbox{MyOrangeBayes}{0.019}&\colorbox{MyWhite}{0.019}\\[1pt] \hline
&&&&&&&&&&&\\[-10pt]
0.600&0.043&\colorbox{MyWhite}{0.004}&\colorbox{MyOrangeBayes}{0.004}&\colorbox{MyOrangeBayes}{0.004}&\colorbox{MyOrangeBayes}{0.004}&\colorbox{MyOrangeBayes}{0.004}&\colorbox{MyWhite}{0.004}&\colorbox{MyOrangeBayes}{0.004}&\colorbox{MyWhite}{0.004}&\colorbox{MyOrangeBayes}{0.004}&\colorbox{MyWhite}{0.004}\\[1pt] \hline
&&&&&&&&&&&\\[-10pt]
\end{tabular}

\vspace{10pt}

\begin{tabular}{ll|l|l|l|l|l|l|l|l|l|l|} &$p(X/\!\!/\frak X_{\!\mathscr M})$
&0.100&0.100&0.100&0.100&0.100&0.100&0.100&0.100&0.100&0.100\\ $p_{x/\!\!/\frak X_{\!\mathscr M}}$&$b^{\mbox{\tiny final}}_{x/\!\!/\frak X_{\!\mathscr M}}$  &  &  &  &  &  &  &  &  &  &\\[1pt] \hline
&&&&&&&&&&&\\[-10pt]
0.600&0.100&\colorbox{MyWhite}{0.010}&\colorbox{MyOrangeBayes}{0.010}&\colorbox{MyOrangeBayes}{0.010}&\colorbox{MyWhite}{0.010}&\colorbox{MyOrangeBayes}{0.010}&\colorbox{MyWhite}{0.010}&\colorbox{MyOrangeBayes}{0.010}&\colorbox{MyOrangeBayes}{0.010}&\colorbox{MyOrangeBayes}{0.010}&\colorbox{MyWhite}{0.010}\\[1pt] \hline
&&&&&&&&&&&\\[-10pt]
0.500&0\hspace{18pt}&\colorbox{MyWhite}{0\hspace{18pt}}&\colorbox{MyWhite}{0\hspace{18pt}}&\colorbox{MyOrangeBayes}{0\hspace{18pt}}&\colorbox{MyOrangeBayes}{0\hspace{18pt}}&\colorbox{MyOrangeBayes}{0\hspace{18pt}}&\colorbox{MyWhite}{0\hspace{18pt}}&\colorbox{MyWhite}{0\hspace{18pt}}&\colorbox{MyWhite}{0\hspace{18pt}}&\colorbox{MyOrangeBayes}{0\hspace{18pt}}&\colorbox{MyOrangeBayes}{0\hspace{18pt}}\\[1pt] \hline
&&&&&&&&&&&\\[-10pt]
0.400&0\hspace{18pt}&\colorbox{MyWhite}{0\hspace{18pt}}&\colorbox{MyWhite}{0\hspace{18pt}}&\colorbox{MyWhite}{0\hspace{18pt}}&\colorbox{MyOrangeBayes}{0\hspace{18pt}}&\colorbox{MyOrangeBayes}{0\hspace{18pt}}&\colorbox{MyOrangeBayes}{0\hspace{18pt}}&\colorbox{MyWhite}{0\hspace{18pt}}&\colorbox{MyWhite}{0\hspace{18pt}}&\colorbox{MyWhite}{0\hspace{18pt}}&\colorbox{MyOrangeBayes}{0\hspace{18pt}}\\[1pt] \hline
&&&&&&&&&&&\\[-10pt]
0.400&0\hspace{18pt}&\colorbox{MyWhite}{0\hspace{18pt}}&\colorbox{MyWhite}{0\hspace{18pt}}&\colorbox{MyWhite}{0\hspace{18pt}}&\colorbox{MyOrangeBayes}{0\hspace{18pt}}&\colorbox{MyWhite}{0\hspace{18pt}}&\colorbox{MyWhite}{0\hspace{18pt}}&\colorbox{MyWhite}{0\hspace{18pt}}&\colorbox{MyOrangeBayes}{0\hspace{18pt}}&\colorbox{MyOrangeBayes}{0\hspace{18pt}}&\colorbox{MyOrangeBayes}{0\hspace{18pt}}\\[1pt] \hline
&&&&&&&&&&&\\[-10pt]
0.400&0\hspace{18pt}&\colorbox{MyWhite}{0\hspace{18pt}}&\colorbox{MyWhite}{0\hspace{18pt}}&\colorbox{MyOrangeBayes}{0\hspace{18pt}}&\colorbox{MyWhite}{0\hspace{18pt}}&\colorbox{MyOrangeBayes}{0\hspace{18pt}}&\colorbox{MyOrangeBayes}{0\hspace{18pt}}&\colorbox{MyWhite}{0\hspace{18pt}}&\colorbox{MyOrangeBayes}{0\hspace{18pt}}&\colorbox{MyWhite}{0\hspace{18pt}}&\colorbox{MyWhite}{0\hspace{18pt}}\\[1pt] \hline
&&&&&&&&&&&\\[-10pt]
0.300&0\hspace{18pt}&\colorbox{MyOrangeBayes}{0\hspace{18pt}}&\colorbox{MyOrangeBayes}{0\hspace{18pt}}&\colorbox{MyWhite}{0\hspace{18pt}}&\colorbox{MyWhite}{0\hspace{18pt}}&\colorbox{MyWhite}{0\hspace{18pt}}&\colorbox{MyOrangeBayes}{0\hspace{18pt}}&\colorbox{MyWhite}{0\hspace{18pt}}&\colorbox{MyWhite}{0\hspace{18pt}}&\colorbox{MyWhite}{0\hspace{18pt}}&\colorbox{MyWhite}{0\hspace{18pt}}\\[1pt] \hline
&&&&&&&&&&&\\[-10pt]
0.600&0.700&\colorbox{MyOrangeBayes}{0.070}&\colorbox{MyWhite}{0.070}&\colorbox{MyWhite}{0.070}&\colorbox{MyWhite}{0.070}&\colorbox{MyOrangeBayes}{0.070}&\colorbox{MyOrangeBayes}{0.070}&\colorbox{MyOrangeBayes}{0.070}&\colorbox{MyWhite}{0.070}&\colorbox{MyOrangeBayes}{0.070}&\colorbox{MyOrangeBayes}{0.070}\\[1pt] \hline
&&&&&&&&&&&\\[-10pt]
0.500&0\hspace{18pt}&\colorbox{MyOrangeBayes}{0\hspace{18pt}}&\colorbox{MyWhite}{0\hspace{18pt}}&\colorbox{MyOrangeBayes}{0\hspace{18pt}}&\colorbox{MyWhite}{0\hspace{18pt}}&\colorbox{MyWhite}{0\hspace{18pt}}&\colorbox{MyOrangeBayes}{0\hspace{18pt}}&\colorbox{MyWhite}{0\hspace{18pt}}&\colorbox{MyOrangeBayes}{0\hspace{18pt}}&\colorbox{MyOrangeBayes}{0\hspace{18pt}}&\colorbox{MyWhite}{0\hspace{18pt}}\\[1pt] \hline
&&&&&&&&&&&\\[-10pt]
0.500&0\hspace{18pt}&\colorbox{MyWhite}{0\hspace{18pt}}&\colorbox{MyOrangeBayes}{0\hspace{18pt}}&\colorbox{MyWhite}{0\hspace{18pt}}&\colorbox{MyOrangeBayes}{0\hspace{18pt}}&\colorbox{MyWhite}{0\hspace{18pt}}&\colorbox{MyOrangeBayes}{0\hspace{18pt}}&\colorbox{MyOrangeBayes}{0\hspace{18pt}}&\colorbox{MyWhite}{0\hspace{18pt}}&\colorbox{MyOrangeBayes}{0\hspace{18pt}}&\colorbox{MyWhite}{0\hspace{18pt}}\\[1pt] \hline
&&&&&&&&&&&\\[-10pt]
0.600&0.200&\colorbox{MyWhite}{0.020}&\colorbox{MyOrangeBayes}{0.020}&\colorbox{MyOrangeBayes}{0.020}&\colorbox{MyOrangeBayes}{0.020}&\colorbox{MyOrangeBayes}{0.020}&\colorbox{MyWhite}{0.020}&\colorbox{MyOrangeBayes}{0.020}&\colorbox{MyWhite}{0.020}&\colorbox{MyOrangeBayes}{0.020}&\colorbox{MyWhite}{0.020}\\[1pt] \hline
&&&&&&&&&&&\\[-10pt]
\end{tabular}
\caption{The believability $\left\{b_{x/\!\!/\frak X_{\!\mathscr M}}\colon x\in\frak X_{\!\mathscr M}\right\}$, probability $\left\{p(X/\!\!/\frak
X_{\!\mathscr M})\colon X\in\protect\rsS^{\frak X_{\!\mathscr M}}\right\}$, and  certainty $\left\{\varphi_{x}(X/\!\!/\frak X_{\!\mathscr M})\colon
x\in\frak X_{\!\mathscr M}, X\in\protect\rsS^{\frak X_{\!\mathscr M}}\right\}$  distributions of the co$\sim$event \colorbox{MyOrangeBayes}{$\mathscr
M$} for Example 5. $\bfPhi(\mathscr M)= 0.470$, and $\bfPhi^{\mbox{\tiny final}}(\mathscr M)= 0.600$. Initial  distributions (in the top table) and
final distributions (in the bottom table) for the 50-th iterative calculation by the co$\sim$event-based Bayes theorem.\label{tabEx5}} \end{table}

\texttt{\indent Corollary \!\refstepcounter{ctrcor}\arabic{ctrcor}\,\label{cor-recurrent}\itshape\footnotesize (the limit certainty in a recurrent
co$\sim$event-based Bayes' formula)\!.}
\begin{eqnarray}
\label{Phi-limit}
\lim_{n\to\infty} \bfPhi^{(n)}(\mathscr M)&=&
\frac{1}{\displaystyle \sum_{x\in X_{\max}}\!\!\!\! b_{x}}
\sum_{x\in X_{\max}}\sum_{x\in X\in \rsS^{\frak X_{\!\mathscr M}}}
b_{x}p(X/\!\!/\frak X_{\!\mathscr M}).
\end{eqnarray}
\textsf{Proof}  is obvious.

Figures \ref{Ex3Ex4}, and \ref{Ex5} and Tables \ref{tabEx3}, \ref{tabEx4}, and \ref{tabEx5} illustrate Theorem and Corollary.

The recurrent application of the co$\sim$event-based Bayes' theorem can be interpreted as a repetition of the experienced-random experiment as a
result of which the same co$\sim$events $\mathscr H$ and $\mathscr R$ occur every time. Theorem \ref{Th-recurrent} asserts that if the
experienced-random experiment is repeated many times, the sequence of believability distributions tends to the limit believability distribution. And
the value of this limit distribution can be considered as the same characteristic of the believabilities of co$\sim$events $\mathscr H$ and $\mathscr
R$, which 1/2 is for the probabilistic distribution of a fair coin. Theorem 2 can serve as the basis for the formulation and proof of the
co$\sim$event law of large numbers.

\section{Notes in conclusion}

\textbf{\emph{We exist in the world of uncertainties.}} Any uncertainty always arises from a conflict of experience and chance, more precisely, from
a conflict between the observer's experience and the chance observation. In other words, this indivisible pair, ``experience and chance'', is the
source of any uncertainty. Other sources of uncertainty simply do not exist. A theory that describes this conflict strictly mathematically is the new
theory, called \emph{the theory of co$\sim$events, \emph{or} the theory of experience and chance} \cite{Vorobyev2016famems2}. Now we can say that
this theory is a theory of uncertainty in its broadest sense\footnote{Although the paper \cite{Vorobyev2016famems2} is devoted to the axioms of the
new theory of experience and chance and is primarily a mathematical text, there are enough preliminary considerations in it that can help
non-mathematicians understand the philosophy of the new theory and get used to new terminology, which sometimes contradicts everyday meaning. The
work \cite{Vorobyev2016famems1} is of purely technical significance for the new theory, mainly devoted to its mathematical apparatus and is intended
for mathematicians. The work \cite{Vorobyev2016famems3} tells about some applications of the new theory, which gave a decisive impetus to its
development. In the paper \cite{Vorobyev2017famems4} we consider a very simple example of the interpretation of the basic concepts of the theory in
applications.}.

\textbf{\emph{The eventology approach}} \cite{Vorobyev2007} gave impetus to the development of a theory that turned out to be broader than the theory
of probabilities. This new theory is a dual combination of two theories --- Kolmogorov's theory of probabilities of \emph{ket-events (k-e.'s)} and
its dual reflection --- a new theory of believabilities of \emph{bra-events (b-e.'s)}. Today, \emph{the theory of co$\sim$events} has a strict
axiomatics \cite{Vorobyev2016famems2,Vorobyev2016famems1,Vorobyev2016famems3}, in which Kolmogorov k-e.'s describing the \textbf{\emph{future
chance}} of observation are dually reflected in b-e.'s describing the \textbf{\emph{past experience}} of the observer. The co$\sim$events are defined
as the measurable binary relations on the Cartesian product: \emph{``a set of b-e.'s $\times$ a set of terraced k-e.'s''} and describe the
\textbf{\emph{present uncertainty}} that is generated by a pair of ``experience and chance''.

\textbf{\emph{Axiomatics of the theory of co$\sim$events \cite{Vorobyev2016famems2} terminates all the debate between different philosophical
interpretations of probabilities: frequentist
\cite{Bridgman1927,Church1940,Cramer1946,Feller1957,Jeffreys1939,Lindley1957,Martin-Lof1966,Mises1928,Neyman1950,Reichenbach1935,Russell1948},
propensity \cite{Peirce1958,Burks1978,Popper1977,Popper1957,Popper1959,Popper1967,Gillies2000,Giere1973,Lewis1980}, or Bayesian (subjective)
\cite{Berger1985,Bernardo1994,Davidson1957,Finetti1931,Finetti1937,Finetti1974,DeGroot1970,Hacking1967,Hajek2010,Hald1998,Jaynes1996,Jeffreys1939,Lindley1957,Morgenstern1978,Peirce1885,Pfanzagl1967,Pfanzagl1968,Ramsey1926,Stigler1990,Stigler1999}
probabilities and also other philosophical interpretation of probabilities.}} This new theory is not one of the branches of probability theory. It
offers a new axiomatic foundation for all those theories that are fundamentally based on mathematical probability. For example, for the Bayesian
(subjective), propensity, or frequentist probability. Until now, all these approaches relied on the same mathematical probability Kolmogorov
axiomatized. The new theory offers a new system of axioms, which contains Kolmogorov's axiomatics as one of the dual halves. Therefore the new theory
of experience and chance as a rigorous axiomatic mathematical basis allows one to unite all existing acceptable interpretations of uncertainty into
one general co$\sim$event-based approach.

\emph{\textbf{A past experience and a future chance are described by a unified theory.}} In the theory of co$\sim$events \cite{Vorobyev2016famems2}
the logic of experience is directed to the past and the logic of chance is directed to the future. So the logic of experience reverses the logic of
the chance with respect to the time direction. This is a simple consequence of the co$\sim$event-based axiomatics where the experience is described
by the b-e's, and the chance is described by the k-e's. And if some bra-event is experienced, then all the b-e's that are contained in it are
experienced, i.e. which might be its possible causes in the past. And when some ket-event happens, then all the k-e's happen that contain it, i.e.
which may be its consequences in the future. So the logic of experience is always the logic of the \emph{past} experience, and the logic of chance is
always the logic of the \emph{future} chance. This theory introduces a new \emph{believability} measure for measuring past experience and uses the
\emph{Kolmogorov probability} measure to measure a future chance.

\textbf{\emph{Only a measuring the believability of all the probabilistic distribution of the ket-event makes sense in the theory of
co$\sim$events.}} I think everyone agrees that when tossing a fair coin, one typically can measure her/his believability only in that the outcomes
``head'' and ``tail'' are equally likely to happen. This obvious statement agrees with the co$\sim$event-based axiomatics of the theory of experience
and chance. If you toss a fair coin many times, then the new theory allows you to measure your believability only in all the probabilistic
distribution of this coin. In the framework of this theory, you can measure your believability only in that the probability of a head happens and the
probability of a tail happens are somewhere near 1/2. In this theory, a measuring the believability in the fall of the head, or believability in the
fall of the tail does not make sense. Only a measuring the believability of all the probabilistic distribution of the coin makes sense here.

\textbf{\emph{A measuring the believability of a set of b-e's is a measuring its believability distribution.}} Also I think everyone agrees that when
tossing a given set of unfair coins, one typically can measure her/his believabilities only in a set of unfair coin probability distributions, i.e.,
one can measure only a set of believabilities which form the believability distribution of the set of unfair coins. In other words, one can measure
only a set of believabilities the sum of which is equal to one. Thus, if you are interested in several believabilities in results of some
experience-random experiment, then you can get the answer only in the form of some believability distribution.

\textbf{\emph{The measuring a co$\sim$event in the theory of experience and chance means the measuring three its measures simultaneously:
probability, believability and certainty.}} In this theory the probability measure is defined on a set of observations (k-e's), and the believability
measure is defined on a set of observers (b-e's). The probability measures the chances of observations, and the believability measures the
experiences of observers. It is said that with certainty $\bfPhi(\mathscr M)$ the co$\sim$event $\mathscr M$ has the probability distribution of
k-e's (observations) $\mathbf p_{\frak X_{\!\mathscr M}}$ and the believability distribution of b-e's (observers) $\breve{b}_{\frak X_{\!\mathscr
M}}$. Thus, the certainty of a co$\sim$event is a measure of the combination of its probabilistic and believabilistic measures, which is defined by
this co$\sim$event. In other words, we can say the certainty is a measure of conflict between observer' experiences with their believabilities and
chances of observations with their probabilities. This new semantics of the old familiar terms in an unusual combination represents the main
difficulty in understanding the new theory of experience and chance.

{\footnotesize
\bibliography{vorobyev}
}

\end{document}